\documentclass[apj]{emulateapj}

\newcommand{\myion}[2]{\ensuremath{\mathrm{#1}^{#2}}}
\newcommand{\um}{\,\micron}
\newcommand{\arII}{[Ar\,\textsc{ii}]}
\newcommand{\arIII}{[Ar\,\textsc{iii}]}
\newcommand{\neII}{[Ne\,\textsc{ii}]}
\newcommand{\neIII}{[Ne\,\textsc{iii}]}
\newcommand{\neV}{[Ne\,\textsc{v}]}
\newcommand{\sIII}{[S\,\textsc{iii}]}
\newcommand{\sIV}{[S\,\textsc{iv}]}
\newcommand{\siII}{[Si\,\textsc{ii}]}
\newcommand{\oIII}{[O\,\textsc{iii}]}
\newcommand{\oIV}{[O\,\textsc{iv}]}
\newcommand{\feII}{[Fe\,\textsc{ii}]}
\newcommand{\hII}{H\,\textsc{ii}}
\newcommand{\hs}[1]{H$_2$\,S(#1)}
\newcommand{\pahfit}{\textsc{Pahfit}}
\newcommand{\PAH}{\ensuremath{\textrm{\scriptsize PAH}}}
\newcommand{\TIR}{\ensuremath{\textrm{\scriptsize TIR}}}

\shortauthors{Smith \textit{et al.}}  
\shorttitle{PAH Emission in Star-Forming Galaxies}

\begin{document}

\title{The Mid-Infrared Spectrum of Star-Forming Galaxies: Global
  Properties of PAH Emission}

\author{%
J.D.T. Smith\altaffilmark{1}, 
B.T. Draine\altaffilmark{2},
D.A. Dale\altaffilmark{3}, 
J. Moustakas\altaffilmark{1},
R.C. Kennicutt, Jr.\altaffilmark{5,1}, 
G. Helou\altaffilmark{4}, 
L. Armus\altaffilmark{5}, 
H. Roussel\altaffilmark{6}, 
K. Sheth\altaffilmark{3}, 
G.J. Bendo\altaffilmark{7}, 
B.A. Buckalew\altaffilmark{5},
D. Calzetti\altaffilmark{8}
C.W. Engelbracht\altaffilmark{1}
K.D. Gordon\altaffilmark{1},
D.J. Hollenbach\altaffilmark{9},
A. Li\altaffilmark{10},
S. Malhotra\altaffilmark{11},
E.J. Murphy\altaffilmark{12}, and
F. Walter\altaffilmark{6}
}

\altaffiltext{1}{Steward Observatory, University of Arizona, Tucson, AZ
  85721} 
\altaffiltext{2}{Dept. of Astrophysical Sciences, Princeton
  University, Princeton, NJ 08544} 
\altaffiltext{3}{Dept. of Physics \&
  Astron., Univ. of Wyoming, Laramie, WY}
\altaffiltext{4}{Caltech, Pasadena, CA} 
\altaffiltext{5}{Spitzer Science Center, Caltech, Pasadena, CA}
\altaffiltext{5}{Institute of Astronomy, Cambridge University, Cambridge, UK}
\altaffiltext{6}{Max-Planck-Institut f\"{u}r Astronomie, Heidelberg, Germany}
\altaffiltext{7}{Imperial College, Blackett Laboratory, London, UK}
\altaffiltext{8}{STSci, Baltimore, MD }
\altaffiltext{9}{NASA Ames Research Center, Moffett Field, CA}
\altaffiltext{10}{ Dept. of Physics \& Astron., Univ.
  Missouri, Columbia, MO}
\altaffiltext{11}{Dept. of Physics \& Astron., Arizona State Univ.,
  Tempe, AZ}
\altaffiltext{12}{Dept. of Astronomy, Yale University, New Haven, CT}

\email{jdsmith@as.arizona.edu}

\begin{abstract}
  We present a sample of low-resolution 5--38\um\ Spitzer IRS spectra of
  the inner few square kiloparsecs of 59 nearby galaxies spanning a
  large range of star formation properties.  A robust method for
  decomposing mid-infrared galaxy spectra is described, and used to
  explore the behavior of PAH emission and the prevalence of silicate
  dust extinction.  Evidence for silicate extinction is found in
  $\sim$1/8 of the sample, at strengths which indicate most normal
  galaxies undergo $A_V\!\lesssim$\,3 magnitudes averaged over their
  centers.  The contribution of PAH emission to the total infrared power
  is found to peak near 10\% and extend up to $\sim$20\%, and is
  suppressed at metallicities $Z\lesssim Z_\sun/4$, as well as in
  low-luminosity AGN environments.  Strong inter-band PAH feature
  strength variations (2--5$\times$) are observed, with the presence of
  a weak AGN and, to a lesser degree, increasing metallicity shifting
  power to the longer wavelength bands.  A peculiar PAH emission
  spectrum with markedly diminished 5--8\um\ features arises among the
  sample solely in systems with relatively hard radiation fields
  harboring low-luminosity AGN.  The AGN may modify the emitting grain
  distribution and provide the direct excitation source of the unusual
  PAH emission, which cautions against using absolute PAH strength to
  estimate star formation rates in systems harboring active nuclei.
  Alternatively, the low star formation intensity often associated with
  weak AGN may affect the spectrum.  The effect of variations in the
  mid-infrared spectrum on broadband infrared surveys is modeled, and
  points to more than a factor of two uncertainty in results which
  assume a fixed PAH emission spectrum, for redshifts $z=0-2.5$.
\end{abstract}


\section{Introduction}

The mid-infrared (MIR) spectrum of star-forming galaxies is dominated by
strong emission features generally attributed to Polycyclic Aromatic
Hydrocarbons (PAHs).  PAHs are aromatic molecules ubiquitous in the
interstellar medium (ISM) of our own Galaxy, and found in most nearby
galaxies with ongoing or recent star formation.  These large molecules,
typically a few \AA\ in diameter and harboring up to a few hundred
carbon atoms, are now commonly believed to be responsible for the
prominent suite of broad emission bands in the MIR between 3--19\um\
\citep{Leger1984,Allamandola1985}.  PAH molecules efficiently absorb
ultraviolet and optical photons, and in many environments supply the
bulk of the photo-electrons responsible for heating interstellar gas
\citep[e.g.][]{Weingartner2001}.

The mid-infrared PAH bands can be very strong, both compared to their
underlying dust grain continuum, and even to the reprocessed bolometric
emission of larger dust grains.  In galaxies with intense star
formation, up to 20\% of the total infrared luminosity is emitted in the
PAH bands alone, and, longward of the Lyman and Balmer breaks, they are
among the most prominent spectral features with sufficient contrast for
photometric redshift determination.  In deep infrared surveys, they can
induce peaks in the flux-limited number density of 15--24\um-detected
galaxies at redshifts $z\gtrsim$1, as they move through fixed filter
passbands \citep[e.g.][]{Barvainis1999,LeFloch2005,Caputi2006}.  Thus
they are of fundamental importance in modeling deep infrared survey
results, and especially in constraining the role of dust-obscured star
formation at early epochs.  Yet the behavior or even presence of PAHs in
the distant universe is largely unexplored.  Despite tantalizing
indications of PAH emission at redshift beyond z$\gtrsim$2
\citep{Yan2005,Lutz2005,Huang2006}, other surveys probing the
mid-infrared properties of optically faint distant galaxy samples have
uncovered mid-infrared spectra dominated by continuum emission, cut by
deep silicate absorption \citep{Houck2005}.

A prevailing difficulty in studying PAH emission at high redshifts is
the relative ambiguity of the 7--13\um\ spectrum of star-forming
galaxies, with the broad PAH features at 7.7\um\ and 11.3\um\ almost
perfectly bracketing the 10\um\ silicate absorption trough.  Silicate
absorption can be pronounced in some galaxies' spectra, including high
redshift objects dominated by AGN emission.  At low signal, or in broad
photometric bands, discriminating between these two physically distinct
but observationally similar signatures can be challenging.
Interpretation of these spectral signatures, and of broadband properties
of even fainter and more distant systems, has relied predominantly on a
small set of local template spectra developed primarily from Infrared
Space Observatory (ISO) observations of bright nearby systems.

The \emph{Spitzer Infrared Nearby Galaxies Survey}
\citep[SINGS,][]{Kennicutt2003} provides an exceptional sample in which
to explore the global and spatially resolved behavior of these
fundamental emission bands over a much broader range of physical
environments than has been possible before.  We focus here on the global
behavior of the PAH emission spectrum and its underlying continuum.  A
future paper will focus on the spatially resolved behavior of PAH
emission in the centers and extended disks of galaxies in the SINGS
sample.

\section{Background on the Emission Processes}
\label{sec:backgr-emiss-proc}

The ``unidentified infrared bands'' were first discovered in
\citeyear{Gillett1973} in the spectra of Galactic planetary nebulae by
\citeauthor{Gillett1973} PAHs were later proposed as the origin of these
features \citep{Leger1984, Allamandola1985}, and invoked to explain the
12\um\ infrared excess seen by IRAS in the diffuse cirrus of the Milky
way \citep{Puget1985}.  Although the carriers of the mid-infrared
emission bands are now commonly attributed to the broad class of PAH
molecules, other interpretations remain consistent with at least some
aspects of the bands' strength, shape, and location \citep[see][and
references therein]{Li2001}.  Although the only certainty is that they
arise primarily in material with $sp^2$ hybridized carbon-carbon bonds
by infrared fluorescence, we adopt the convention of referring to them
as PAH emission bands; it should be kept in mind that the emitters could
encompass a larger group of individual species.  In this same spirit, we
refer to PAHs interchangeably as both molecules and small grains.

The ISO Key Project on Nearby Galaxies \citep{Helou2000} explored the
2--11\um\ spectrum of a sample of normal galaxies.  ISO was limited to
relatively high surface brightness sources, whereas SINGS extends to
much lower luminosity and surface brightness, and is exploring a much
broader range of local interstellar environment.  A surprising result of
the Key Project was the relative uniformity of the PAH emission spectrum
over a range of different galaxy environments.  \citet{Lu2003} found
band to band variations of at most 15\% out to 11\um.  A simple physical
argument which can be invoked to explain this apparent uniformity
relates to the underlying emission mechanism of small grains.  Since PAH
grains are transiently excited by single photons to high temperatures
\citep[higher than their equilibrium temperatures;
see][]{Leger1984,Draine1985,Allamandola1985,Draine2001a}, the emission
spectrum produced by a single grain population is relatively insensitive
to the strength and hardness of the exciting radiation field
\citep{Leger1984}, and the resulting PAH spectrum might therefore be
expected to be quite uniform.  However, a variety of mechanisms are
known to affect to various degrees the strength of the individual PAH
bands, including ionization state, distribution of grain sizes,
de-hydrogenation, and multi-photon heating in intense radiation fields
\citep[see][for a review]{Li2004}.  The possible destruction mechanisms
of PAH grains are varied as well, including photo-thermo
\citep{Leger1989,Guhathakurta1989} and Coulomb dissociation, sputtering,
and dissociation in shocks.

Evidence in our own Galaxy
\citep[e.g.][]{Joblin1996,Verstraete1996,Hony2001,Peeters2004b,Bregman2005}
and resolved regions in nearby galaxies like the LMC and SMC
\citep{Contursi2000,Vermeij2002}, indicate that the PAH emission
spectrum is variable on small scales, and can depend sensitively on the
local gas density and details of the incident radiation field.
Variations can occur in the relative strengths, widths, and wavelength
centers of individual emission bands \citep[e.g.][]{Peeters2002}.

It is not yet known whether the primary processes regulating PAH
emission are chemical, mechanical, or radiative, or whether they are
expressed through inhibited or enhanced grain formation rates, selective
destruction in differing environments, or varying efficiencies of
photo-excitation.  Given the predicted and observed diversity in the
form of the PAH emission spectrum within individual objects in the
Galaxy, the question which remains, and which will be explored here, is
whether the natural variations found on small scales vanish when
averaged over real distributions of emitting environments in the disks
and nuclei of galaxies.

\section{Observations and Reduction}
\label{sec:observ-reduct}

The SINGS survey includes uniform 5--38\um\ spectral mapping with the
Infrared Spectrograph \citep[IRS,][]{Houck2004a}.  Full observational
details regarding the SINGS spectral mapping program can be found in
\citet{Kennicutt2003}.  Briefly, a low-resolution ($R\sim50-100$),
5--15\um\ spectral map of approximate size 55\arcsec$\times$34\arcsec\
is centered on the nucleus of each target.  Larger, 15--38\um\
\emph{radial strip} maps extending to roughly half the optical radius
and covering 50\arcsec$\times$3.5\arcmin --15.4\arcmin\ are also
obtained.  The total effective integration time per pixel was uniform,
at 28 seconds in the short wavelength module, and 2 minutes in the long
wavelength module.  The IRS spectra were supplemented for all sample
targets with scan-map images at 24, 70, and 160\um\ obtained with the
MIPS instrument \citep{Rieke2004}.

\subsection{Sample}


\begin{deluxetable*}{rccrrr@{\,$\times$\,}lccccc}    
\tablecaption{Sample Parameters\label{tab:sample}}
\tablecolumns{12}
\tabletypesize{\small}
\tablehead{%
  \colhead{Galaxy} &
  \colhead{Nuc. Type} &
  \colhead{Distance} &
  \multicolumn{6}{c}{Extraction Aperture} &
  \colhead{$L_{\TIR}$} &
  \colhead{$f_{\TIR}$} &
  \colhead{$L_{TIR}/\nu L_\nu(B)$}\\
  \colhead{} &
  \colhead{} &
  \colhead{Mpc} &
  \multicolumn{2}{c}{Center} &
  \multicolumn{2}{c}{Size ($\arcsec$)} &
  \colhead{A(kpc$^2$)} &
  \colhead{PA($\degr$)} &
  \colhead{$10^{8}\,L_\sun$} &
  \colhead{} &
  \colhead{}\\
  \colhead{(1)} & 
  \colhead{(2)} & 
  \colhead{(3)} & 
  \multicolumn{2}{c}{(4)} &
  \multicolumn{2}{c}{(5)} &
  \colhead{(6)} &
  \colhead{(7)} &
  \colhead{(8)} &
  \colhead{(9)} & 
  \colhead{(10)}}
\startdata
     HoII &       \hII &    3.4 & 08:19:12.86 & +70:43:09.6 &  14.8 &   9.3 &   0.04 & -155.6 &   0.1 &   0.06 &   1.94\\
   IC4710 &       \hII &    9.0 & 18:28:39.55 & -66:58:20.8 &  11.1 &   9.2 &   0.19 &  176.9 &   0.3 &   0.06 &   3.22\\
    Mrk33 &       \hII &   22.9 & 10:32:31.82 & +54:24:02.5 &  38.8 &  31.4 &  15.03 &   41.9 &  51.8 &   0.78 &   1.88\\
  NGC0024 &       \hII &    7.3 & 00:09:56.37 & -24:57:51.2 &  35.1 &  29.6 &   1.32 &  -39.5 &   0.9 &   0.17 &   0.76\\
  NGC0337 &       \hII &   22.4 & 00:59:50.20 & -07:34:45.8 &  42.6 &  25.9 &  12.99 &  149.0 &  60.4 &   0.37 &   3.40\\
  NGC0584 &      LINER &   20.0 & 01:31:20.90 & -06:52:05.1 &  18.5 &  18.5 &   3.20 &  -32.5 &   0.7 &   0.12 &   0.03\\
  NGC0628 &       \hII &    7.3 & 01:36:41.60 & +15:47:00.0 &  53.6 &  29.6 &   1.99 &  151.8 &   2.7 &   0.03 &   0.90\\
  NGC0855 &       \hII &    9.6 & 02:14:03.70 & +27:52:38.4 &  18.5 &  24.1 &   0.95 &  156.1 &   2.1 &   0.55 &\nodata\\
  NGC0925 &       \hII &    9.1 & 02:27:17.25 & +33:34:41.6 &  27.8 &  29.6 &   1.61 &  156.2 &   2.3 &   0.05 &   1.17\\
  NGC1097 &      LINER &   17.1 & 02:46:18.86 & -30:16:27.2 &  37.0 &  31.4 &   7.99 &  146.7 & 249.3 &   0.38 &   6.19\\
  NGC1266 &      LINER &   30.0 & 03:16:00.71 & -02:25:36.9 &  22.2 &  20.4 &   9.58 &  157.7 & 141.9 &   0.54 &  16.24\\
  NGC1291 &      LINER &   10.8 & 03:17:18.59 & -41:06:28.0 &  38.8 &  27.7 &   2.97 &  -78.5 &   4.2 &   0.11 &   0.17\\
  NGC1316 &      LINER &   24.3 & 03:22:41.68 & -37:12:29.4 &  22.2 &  25.9 &   7.99 &  149.0 &  27.7 &   0.28 &   0.27\\
  NGC1404 &      LINER &   18.5 & 03:38:51.95 & -35:35:39.1 &  25.9 &  22.2 &   4.60 &  -79.7 &   0.6 &   0.18 &   0.02\\
  NGC1482 &       \hII &   23.2 & 03:54:38.88 & -20:30:07.1 &  27.7 &  29.6 &  10.40 & -171.4 & 299.2 &   0.68 &  38.04\\
  NGC1512 &      LINER &   11.8 & 04:03:54.17 & -43:20:54.4 &  22.2 &  25.9 &   1.89 &  141.9 &   8.7 &   0.21 &   1.62\\
  NGC1566 &    Seyfert &   20.3 & 04:20:00.33 & -54:56:16.6 &  25.9 &  24.1 &   6.06 &  129.0 &  47.2 &   0.08 &   2.06\\
  NGC1705 &       \hII &    5.1 & 04:54:14.50 & -53:21:36.4 &  14.8 &  14.8 &   0.13 &  120.9 &   0.1 &   0.19 &   0.41\\
  NGC2403 &       \hII &    3.2 & 07:36:49.95 & +65:36:03.5 &  51.8 &  33.3 &   0.42 &   30.4 &   0.8 &   0.02 &   0.62\\
  NGC2798 &       \hII &   26.2 & 09:17:22.80 & +41:59:59.4 &  25.9 &  22.2 &   9.26 &   23.2 & 246.4 &   0.66 &   6.69\\
  NGC2841 &    Seyfert &   14.1 & 09:22:02.50 & +50:58:34.1 &  22.2 &  22.2 &   2.30 &   38.6 &   4.0 &   0.03 &   0.11\\
  NGC2915 &       \hII &    3.8 & 09:26:10.03 & -76:37:32.2 &  18.5 &  18.5 &   0.11 &    8.4 &   0.1 &   0.33 &\nodata\\
  NGC2976 &       \hII &    3.6 & 09:47:15.22 & +67:55:00.3 &  38.8 &  27.8 &   0.32 &   32.5 &   0.7 &   0.08 &   0.93\\
  NGC3049 &       \hII &   23.9 & 09:54:49.59 & +09:16:18.1 &  37.0 &  20.4 &  10.11 &   15.1 &  32.7 &   0.61 &   2.18\\
  NGC3184 &       \hII &   11.1 & 10:18:16.90 & +41:25:24.7 &  14.8 &  13.0 &   0.56 &   33.9 &   2.8 &   0.03 &   1.28\\
  NGC3190 &      LINER &   20.9 & 10:18:05.63 & +21:49:54.2 &  27.7 &  31.5 &   8.96 &   18.5 &  30.9 &   0.37 &   0.58\\
  NGC3198 &      LINER &   13.7 & 10:19:54.84 & +45:32:58.7 &  22.2 &  22.2 &   2.17 &   36.0 &  15.6 &   0.17 &   5.46\\
  NGC3265 &       \hII &   23.2 & 10:31:06.80 & +28:47:45.6 &  27.8 &  22.2 &   7.76 &   22.1 &  25.2 &   0.71 &   1.81\\
  NGC3351 &       \hII &    9.3 & 10:43:57.72 & +11:42:13.5 &  40.7 &  29.6 &   2.46 & -162.3 &  28.5 &   0.36 &   3.20\\
  NGC3521 &      LINER &   10.1 & 11:05:48.58 & -00:02:07.3 &  53.7 &  31.4 &   4.01 & -162.1 &  28.3 &   0.10 &   1.21\\
  NGC3621 &      LINER &    6.6 & 11:18:16.51 & -32:48:49.3 &  53.6 &  31.4 &   1.75 & -145.9 &   6.1 &   0.07 &   1.71\\
  NGC3627 &    Seyfert &    9.4 & 11:20:15.04 & +12:59:29.0 &  44.4 &  31.5 &   2.89 &   22.8 &  23.5 &   0.09 &   0.82\\
  NGC3773 &       \hII &   11.9 & 11:38:12.98 & +12:06:45.8 &  24.1 &  24.0 &   1.92 & -159.6 &   4.1 &   0.53 &   0.82\\
  NGC3938 &      LINER &   13.3 & 11:52:49.32 & +44:07:13.6 &  49.9 &  29.6 &   6.19 &   36.6 &  11.2 &   0.10 &   0.75\\
  NGC4125 &      LINER &   22.9 & 12:08:05.84 & +65:10:29.5 &  24.0 &  24.1 &   7.13 &  173.4 &   5.1 &   0.32 &   0.06\\
  NGC4254 &       \hII &   16.6 & 12:18:49.57 & +14:24:57.5 &  40.7 &  24.1 &   6.33 &   18.9 &  56.7 &   0.11 &   1.59\\
  NGC4321 &      LINER &   14.3 & 12:22:54.87 & +15:49:19.2 &  31.4 &  25.9 &   3.92 &   24.6 &  54.6 &   0.16 &   3.39\\
  NGC4450 &      LINER &   16.6 & 12:28:29.71 & +17:05:08.7 &  48.1 &  29.6 &   9.20 &   31.4 &   9.8 &   0.20 &   0.23\\
  NGC4536 &       \hII &   14.4 & 12:34:27.03 & +02:11:16.5 &  25.9 &  24.0 &   3.06 &   21.4 &  96.3 &   0.73 &   9.12\\
  NGC4552 &    Seyfert &   15.9 & 12:35:39.88 & +12:33:23.3 &  25.9 &  24.0 &   3.71 & -160.9 &   0.7 &   0.25 &   0.01\\
  NGC4559 &       \hII &   10.3 & 12:35:57.58 & +27:57:34.2 &  51.8 &  33.3 &   4.29 &   40.9 &   8.8 &   0.13 &   0.79\\
  NGC4569 &    Seyfert &   16.6 & 12:36:49.76 & +13:09:45.5 &  25.9 &  22.2 &   3.72 &   25.4 &  45.3 &   0.31 &\nodata\\
  NGC4579 &    Seyfert &   16.6 & 12:37:43.53 & +11:49:03.8 &  22.2 &  18.5 &   2.65 &   19.6 &  17.1 &   0.12 &   0.45\\
  NGC4594 &    Seyfert &    9.3 & 12:39:59.56 & -11:37:23.2 &  46.3 &  31.4 &   2.98 &   17.4 &   4.3 &   0.10 &   0.08\\
  NGC4625 &       \hII &    9.2 & 12:41:52.68 & +41:16:26.9 &  37.0 &  29.6 &   2.16 &   46.3 &   2.4 &   0.39 &   0.68\\
  NGC4631 &       \hII &    8.1 & 12:42:07.80 & +32:32:34.6 &  50.0 &  33.3 &   2.54 & -175.9 &  32.6 &   0.12 &   5.21\\
  NGC4725 &    Seyfert &   11.9 & 12:50:26.59 & +25:30:01.2 &  31.4 &  20.4 &   2.13 &   31.2 &   3.6 &   0.04 &   0.28\\
  NGC4736 &      LINER &    5.0 & 12:50:53.15 & +41:07:14.4 &  53.7 &  33.3 &   1.05 & -161.9 &  12.2 &   0.18 &   0.35\\
  NGC4826 &    Seyfert &    5.0 & 12:56:43.59 & +21:40:58.0 &  29.6 &  31.5 &   0.54 &   30.2 &  12.3 &   0.34 &   2.38\\
  NGC5033 &    Seyfert &   14.8 & 13:13:27.32 & +36:35:35.2 &  49.9 &  31.5 &   8.11 &   53.1 &  61.4 &   0.24 &   1.95\\
  NGC5055 &      LINER &    7.8 & 13:15:49.35 & +42:01:45.7 &  38.8 &  29.6 &   1.65 &   22.9 &  16.9 &   0.08 &\nodata\\
  NGC5194 &    Seyfert &    7.8 & 13:29:52.80 & +47:11:43.5 &  55.5 &  31.5 &   2.48 &   61.9 &  32.8 &   0.08 &   2.36\\
  NGC5195 &    Seyfert &    8.0 & 13:29:59.50 & +47:15:56.7 &  33.3 &  25.9 &   1.30 &   24.3 &  16.4 &   0.75 &   2.66\\
  NGC5713 &       \hII &   29.4 & 14:40:11.38 & -00:17:24.2 &  44.4 &  29.6 &  26.72 &   18.7 & 318.9 &   0.47 &   5.99\\
  NGC5866 &      LINER &   15.1 & 15:06:29.48 & +55:45:45.0 &  22.2 &  29.6 &   3.54 &   29.4 &  22.9 &   0.41 &   0.59\\
  NGC6946 &       \hII &    6.8 & 20:34:52.23 & +60:09:14.4 &  44.4 &  25.9 &   1.25 &  120.7 &  41.2 &   0.11 &  39.07\\
  NGC7331 &      LINER &   14.5 & 22:37:04.15 & +34:24:55.3 &  55.5 &  12.9 &   3.56 &  -31.1 &  26.5 &   0.05 &   1.22\\
  NGC7552 &      LINER &   21.0 & 23:16:10.83 & -42:35:05.5 &  37.0 &  29.6 &  11.36 &  -30.1 & 537.4 &   0.58 &  12.97\\
  NGC7793 &       \hII &    3.8 & 23:57:49.84 & -32:35:27.1 &  53.6 &  31.4 &   0.58 &  165.5 &   1.0 &   0.04 &   1.08%
\enddata

\tablecomments{Col. (2): Based on the optical classification methodology
  of \citet{Kauffmann2003} applied to the compiled spectroscopic sample
  of Moustakas et~al., (2006, in prep.).  See also \citet{Dale2006}.
  Col.  (3): Adopted distances from the literature compilation and local
  flow model of Masters, et al.  (2006, in prep.).  Col.  (4): Center of
  extraction in J2000 coordinates.  Col.  (6): Physical extraction
  aperture area.  Col.  (7): Position angle in degrees E of N of first
  (slit-aligned) axis of extraction rectangle.  Col.  (8): Total
  infrared luminosity (3--1100\um) within the extraction aperture.  Col.
  (9): Fraction of global TIR contained in the aperture.  Col. (10):
  Total infrared to B-band luminosity ratio in aperture.}

\end{deluxetable*}


A subset of 59 galaxies from the SINGS sample was selected, and their
relevant parameters are given in Table~\ref{tab:sample}.  The sub-sample
spans a range in luminosity and infrared/optical ratio similar to the
complete sample, and includes galaxies from all represented
morphological types.  The only requirements for inclusion in the
sub-sample were full 5--38\um\ spectral coverage, and sufficient
signal-to-noise to put useful upper limits on PAH emission (the
continuum must be detectable).  Among the galaxies not included were an
individual source with anomalous silicate extinction
\citep{Roussel2006}, a source with strong silicate emission, and several
undetected low-luminosity dwarf galaxies.

\subsection{MIPS Data Reduction and Matched Fluxes}
\label{sec:mips-data-reduction}
All MIPS images were reduced as described in \citet{Dale2005}, and
global MIPS fluxes were taken from that work, modified slightly
($<$10\%) to account for the most recent MIPS calibration factors.
Obtaining total infrared luminosities ($L_\TIR$) matched to the
apertures listed in Table~\ref{tab:sample}, which are in most cases
smaller than the 38\arcsec\ diffraction beam full width of the MIPS
160\um\ instrument, required a method to overcome the inherent
resolution limits of the two long-wavelength channels.  We adopted a
consistent approach to estimate the 70\um\ and 160\um\ surface
brightness.  Representative 24\um:70\um:160\um\ band ratios were
obtained in 25\arcsec\ radius regions centered on the spectral
extraction apertures, from 24\um\ and 70\um\ images convolved to the
MIPS 160\um\ resolution, and from 160\um\ images at native resolution.
These central color ratios were assumed to hold in the smaller spectral
apertures, and were combined with matched 24\um\ photometry (which does
not suffer resolution effects) to predict absolute 70\um\ and 160\um\
surface brightness.  From these, the matched total infrared intensity,
$I_{\TIR}$, was calculated according to the prescription of
\citet{Dale2002}.  Except where otherwise noted, all references to the
total infrared luminosity are matched in this way to the spectral
apertures.

Since the 70\um/24\um, and 160\um/24\um\ ratios in galaxies tend to
\emph{increase} as the strength of the radiation field diminishes
\citep[e.g.][]{Li2001}, this method can overestimate $I_{\TIR}$ if the
innermost regions have warmer FIR colors than the surrounding material.
To estimate the magnitude of this effect, we considered the nuclear and
inner disk MIPS photometry of \citet{Dale2005}, from nearby SINGS
galaxies M81 and M51.  If the luminosity weighted ratio of nuclear to
off-nuclear emission in the 25\arcsec\ beam is 3:1, the maximum amount
by which $I_{\TIR}$ is overestimated using this technique is 25\%; most
galaxies should show no bias in this estimate.

\subsection{IRS Data Reduction and Spectra}

All IRS data were reduced with \textsc{Cubism} (Smith et al., 2006, in
prep.), a custom tool created for the assembly and analysis of
spatially-resolved spectral cubes from IRS spectral maps.  The flux
calibration was achieved using appropriate extended-source corrections
(see Smith et al., 2006, in prep).  All spectral data were processed
with IRS pipeline version S14.  Spectra were extracted from IRS modules
long-low (LL; 15-37\um) and short-low (SL; 5--15\um) in matched
extraction apertures of varying size, chosen to encompass the largest
useful region of circumnuclear and inner disk emission, including
star-forming rings and central star formation knots.  Details of the
extraction apertures are given in Table~\ref{tab:sample}.  The enclosed
fraction of the total infrared emission illustrates the range of
aperture coverage --- from 5--10\% in large nearby galaxies, to nearly
the entire system for smaller and more distant targets.  Each spectrum
is composed of data from the four low-resolution orders of IRS: SL2
(5.25\um-7.6\um), SL1 (7.5\um-14.5\um), LL2 (14.5\um--20.75\um), and LL1
(20.5\um--38.5\um).  Slight mismatches between order segments, resulting
from small residual photometric and astrometric uncertainties, were
addressed by fitting the continuum near the overlap region with
low-order polynomials, then scaling SL2 to match SL1, LL1 to match LL2,
and finally SL to match LL --- a method based on cross-calibration with
IRAC and MIPS.  The scaled spectra were then interpolated onto a common
wavelength grid and averaged in the overlap region.  Typical scaling
adjustments were 10\%.

Statistical errors are propagated from the ramp-fitting uncertainties
produced by the IRS pipeline.  Full flux uncertainty cubes are extracted
alongside the flux cubes, and the resulting spectral uncertainties are
used consistently throughout the analysis.  At signal-to-noise above
$\sim$10, systematic errors exceed the statistical uncertainties in
background-subtracted IRS data, particularly in certain wavelength
regimes, e.g. 20--22\um, where LL1 residual fringing is strong, and at
the ends of the spectral orders, where throughput declines sharply.  The
magnitude of these systematic errors is typically 15\% or less.  All
spectral mapping BCD exposures were background-subtracted, using
background frames taken from dedicated offset observations or from the
periphery of the spectral maps themselves.  In a small number of cases,
when suitable SL background observations were not available, off-source
exposures from the archive, observed nearby-in-time ($\Delta t<\!3$
days) and ecliptic latitude ($|l-l_0|\!<\!10\degr$), were substituted.

All 59 low-resolution IRS spectra are shown in Fig.~\ref{fig:spectra_0}.
The error bars include only statistical uncertainty, though fixed
uncertainties are important at $S/N\!\!\gtrsim$10.

\begin{figure*}[ht]
\centering
\leavevmode
\includegraphics[width={.88\linewidth}]{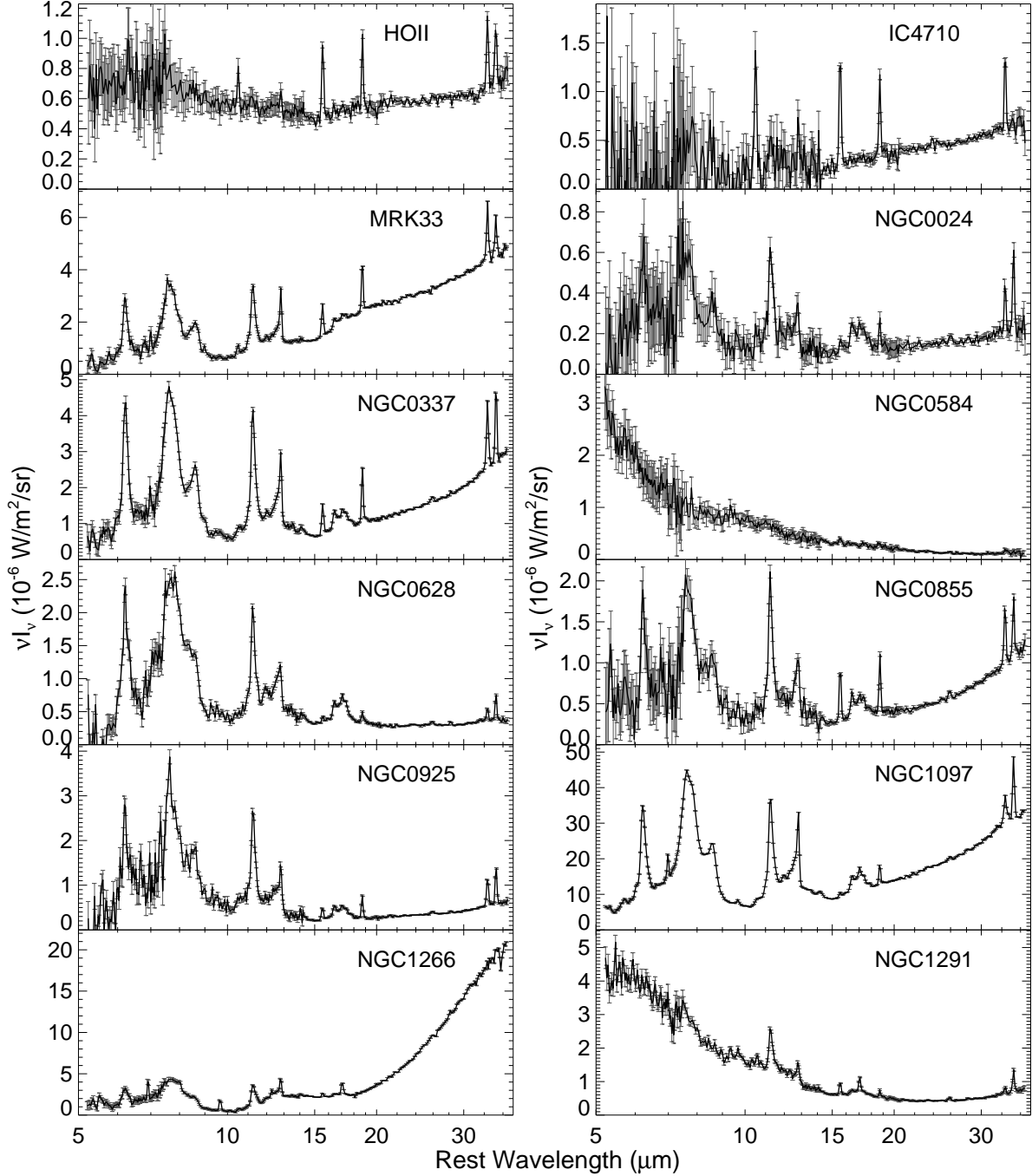}
\caption{Full low-resolution IRS central spectra with logarithmic
  wavelength axis.  Data points are connected.}
\label{fig:spectra_0}
\end{figure*}

\begin{figure*}[ht]
\figurenum{\ref{fig:spectra_0}}
\centering
\leavevmode
\includegraphics[width={.88\linewidth}]{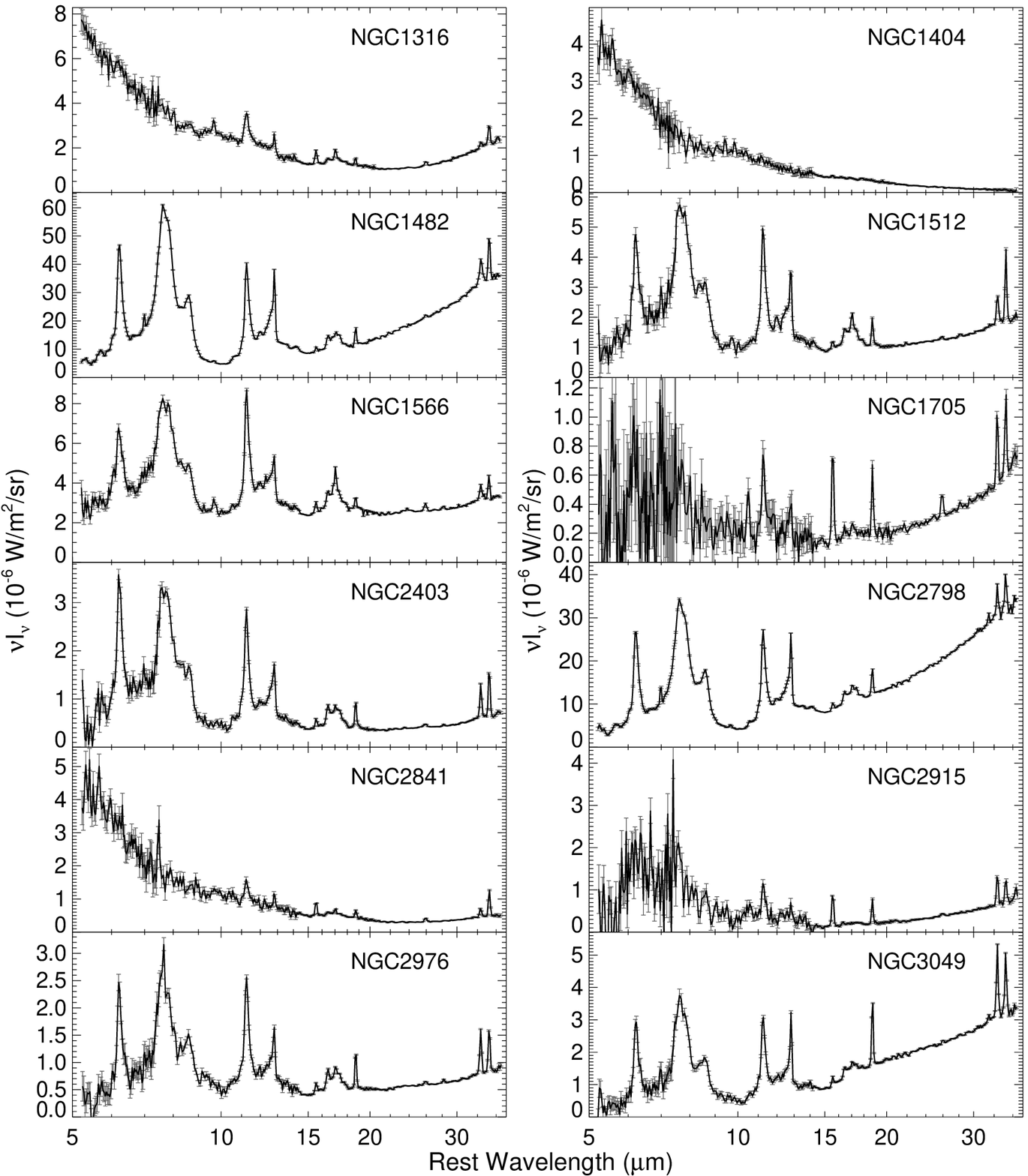}
\caption{ \textit{continued}}
\end{figure*}

\begin{figure*}[ht]
\figurenum{\ref{fig:spectra_0}}
\centering
\leavevmode
\includegraphics[width={.88\linewidth}]{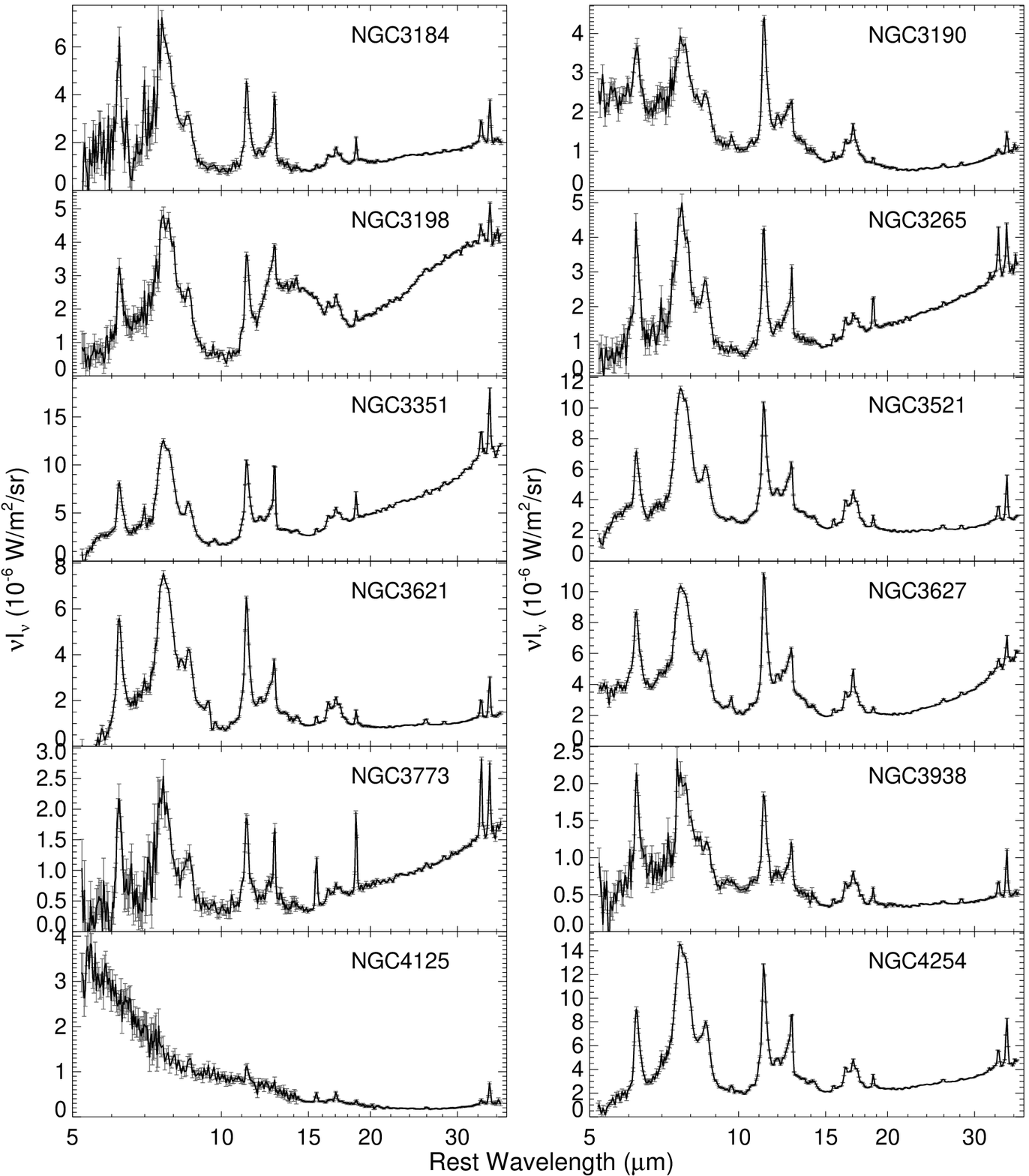}
\caption{ \textit{continued}}
\end{figure*}

\begin{figure*}[ht]
\figurenum{\ref{fig:spectra_0}}
\centering
\leavevmode
\includegraphics[width={.88\linewidth}]{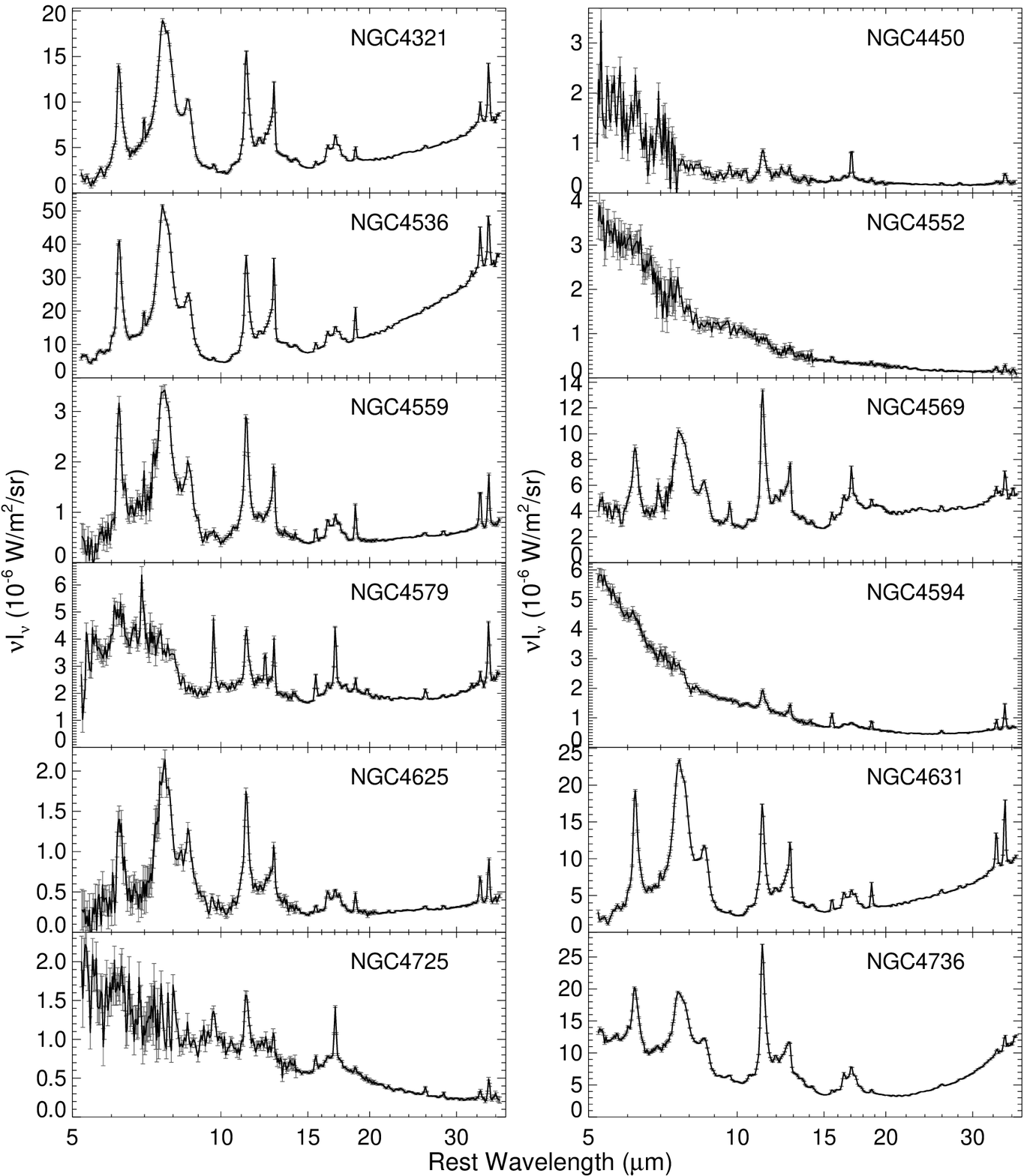}
\caption{ \textit{continued}}
\end{figure*}

\begin{figure*}[ht]
\figurenum{\ref{fig:spectra_0}}
\centering
\leavevmode
\includegraphics[width={.88\linewidth}]{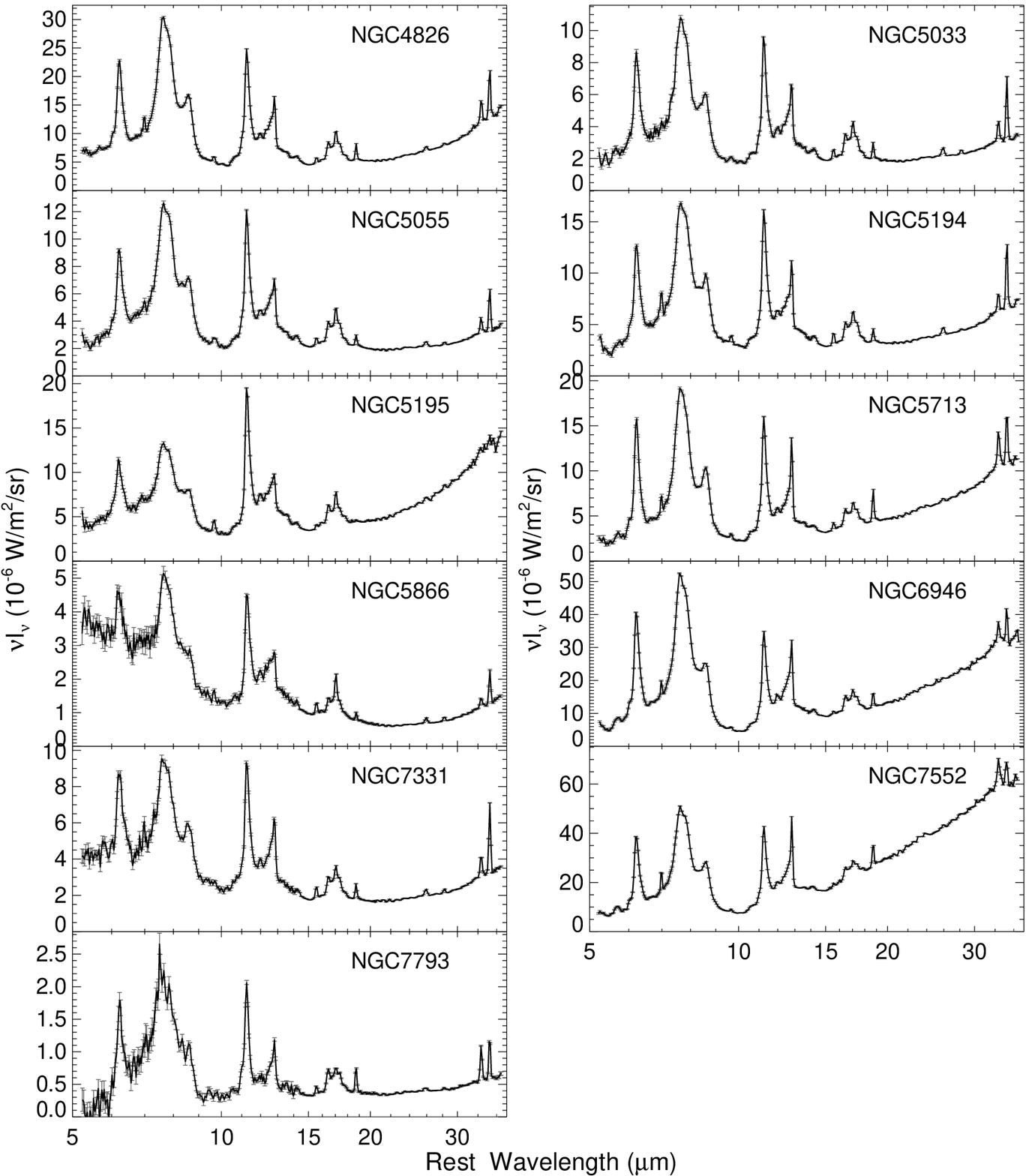}
\caption{ \textit{continued}}
\end{figure*}

\section{Spectral Decomposition of the MIR Spectra of Galaxies}
\label{sec:decomp-mir-galaxy}

In star-forming galaxies, the mid-infrared spectral region hosts among
the greatest diversity of emission components of any wavelength regime.
The broad, strong PAH features are set against a background continuum
transitioning from the rapidly declining direct photospheric emission of
old, cool stars at the shorter mid-infrared wavelengths, to intermediate
wavelengths dominated by hot, stochastically-heated small dust grains,
settling finally on the steadily rising emission of the energetically
important cool dust component, radiating thermally at longer
wavelengths.  Set against this diversity of continuum and emission
features are broad absorption bands of silicate (at 9.7 and 18\um) and,
in extreme cases, water and other ices \citep[e.g.][]{Spoon2004}.  Many
of the PAH features are blended with prominent forbidden spectral lines,
with \neII\ 12.8\um, blended with a band at 12.7\um, being the most
severe example.  Especially at low spectral resolution, separating the
lines, features, multiple continuum components, and absorption features
in a consistent way requires a coherent approach.

To decompose low-resolution IRS spectra, we construct a model consisting
of a small number of physically motivated components: dust continuum in
fixed equilibrium temperature bins, starlight, prominent emission lines,
individual and blended PAH emission bands, and extinction by silicate
grains mixed uniformly with all emitting components.  The main purpose
of this simple yet physically realistic decomposition method is to
estimate robustly the strengths of weak features, features blended with
line emission, and features affected by moderate silicate absorption ---
both in absolute terms, and relative to the local continuum.

\subsection{The Model}
\label{sec:model}

The surface brightness in the spectral extraction aperture is modeled as
a sum of (1) starlight continuum; (2) featureless thermal dust
continuum; (3) pure rotational lines of H$_2$; (4) fine-structure lines;
and (5) dust emission features.  The model also allows for dust
extinction, characterized by an optical depth $\tau_\lambda$ through the
emitting region.  If the absorbing dust is assumed to be uniformly mixed
with the emitting material, the emergent flux intensity is then:

\begin{equation}
  \label{eq:1}
  I_\nu=\left[ \tau_\star B_\nu(T_\star) + 
\sum_{m=1}^M \tau_m \frac{B_\nu(T_m)}{(\lambda/\lambda_0)^2} + 
  \sum_{r=1}^R I_r(\nu) \right]
\frac{(1-e^{-\tau_\lambda})}{\tau_\lambda}
\end{equation}

\noindent where $B_\nu$ is the blackbody function, $T_\star$ is the
temperature of the stellar continuum, $T_m$ are the $n=M$ thermal dust
continuum temperatures, the $I_r(\nu)$ are the $n=R$ resolved (dust) and
unresolved (line) emission features, and $\tau_\lambda$ is the dust
opacity.  These features are described in detail below.

\subsubsection{Starlight}
\label{sec:starlight}

The infrared emission from older stellar populations is represented by
blackbody emission at fixed temperature $T_\star=5000$K, similar to the
temperature of the stars which dominate stellar emission redward of
3--5\um.  This temperature was found to best represent the stellar
continuum in average Starburst99 spectral energy distributions
\citep[SEDs,][]{Leitherer1999} for stellar populations older than
100Myr, and wavelength longward of 3\um.  For shorter rest-frame
wavelengths, or in fits involving near-infrared (NIR) photometry,
including a more realistic suite of stellar SEDs would be appropriate,
as would considering the very hot ($10^3$K) small grain NIR dust
emission component inferred in ISO-Phot spectra of normal galaxies
\citep{Lu2003}.  These components are unimportant beyond rest frame
wavelengths of 3--5\um.

\subsubsection{Dust Continuum}
\label{sec:dust-continuum}

We allow up to $M=8$ thermal dust continuum components represented by
modified blackbodies at fixed temperatures $T_m=\{35, 40, 50, 65, 90,
135, 200, 300\mathrm{K}\}$, with $\nu^2$ emissivity normalized at
$\lambda_0=9.7\um$.  Although the bulk of the dust in galaxies is heated
to relatively low equilibrium temperatures ($\sim$15--20\,K; see Draine
et al., 2006, in prep.), these grains are too cold to make any
contribution at $\lambda<40$\um, and are not considered here.
Contributions from a hot grain continuum, which would be important in
strong AGN sources, are also not included; if present, e.g. for strong
SINGS AGN like NGC\,1566 and NGC\,5194, these continuum components would
be absorbed by a combination of warm thermal and stellar continuum.  The
dust continuum is thus determined by the $M$ parameters $\tau_m$, each
required to be non-negative (but not non-zero).  The number of
components was chosen to reproduce smooth distributions of grain
temperature, though reducing the number (but not temperature range) of
these components would not significantly impact the resulting
decomposition.

No special significance is attached to the strength of individual
temperature components; rather, they are chosen to produce a physically
realistic dust continuum underlying the discrete lines and dust
features.  Including 70\um\ and 160\um\ MIPs photometry, along with IRAC
and near-infrared photometry in matched apertures, would permit placing
more significance on the resulting non-stochastic dust temperature
distribution; this is not pursued here.  For extracting PAH, silicate,
and line strengths, such additional short and long wavelength
constraints have little impact, and changes in the number and exact
placement of the temperature bins also do not strongly affect the
resulting feature parameters.  On average, only 5--6 of the maximum
allowed 8 thermal dust components have non-negligible power in a given
model fit, though all components are well populated when considering the
sample as a whole.

\subsubsection{Spectral Line Features}
\label{sec:line-features}

\begin{deluxetable}{ccc}
  \tablecaption{Unresolved Line Parameters\label{tab:fitted_lines}}
    \tablecolumns{3}
    \tablehead{
      \colhead{Line} &
      \colhead{$\lambda_r(\um)$} &
      \colhead{FWHM(\um)}\\
      \colhead{(1)} &
      \colhead{(2)} &
      \colhead{(3)}}
\startdata
\hs{7} &   5.511 &  0.053\\
\hs{6} &   6.109 &  0.053\\
\hs{5} &   6.909 &  0.053\\
\arII  &   6.985 &  0.053\\
\hs{4} &   8.026 &  0.100\\
\arIII &   8.991 &  0.100\\
\hs{3} &   9.665 &  0.100\\
\sIV   &  10.511 &  0.100\\
\hs{2} &  12.278 &  0.100\\
\neII  &  12.813 &  0.100\\
\neIII &  15.555 &  0.140\\
\hs{1} &  17.035 &  0.140\\
\sIII  &  18.713 &  0.140\\
\oIV   &  25.910 &  0.340\\
\feII  &  25.989 &  0.340\\
\hs{0} &  28.221 &  0.340\\
\sIII  &  33.480 &  0.340\\
\siII  &  34.815 &  0.340%
\enddata
\tablecomments{Col. (1): line name.  Col. (2): central wavelength.  Col.
  (3): full-width at half maximum.}
\end{deluxetable}

A diverse collection of spectral lines is emitted in the mid-infrared
range probed by the IRS, including the \hs{0}--\hs{6} pure rotational
lines of molecular hydrogen, strong \hII-region and photo-dissociation
region (PDR) cooling lines from low-ionization species such as
\myion{Si}{+}, with an ionization potential of 8.2 eV, to massive star
and AGN tracers from \myion{O}{+++} (55eV) and \myion{Ne}{4+} (97 eV).
These lines are modeled primarily to allow realistic decomposition of
blended features and the underlying continuum.  The strong line pair
ratio \neIII/\neII\ provides an indicator of radiation hardness
unaffected by all but the heaviest dust extinction \citep{Thornley2000}.

Spectral lines in the model are represented by Gaussian profiles, with
widths set per instrument module based on direct measurements of
unresolved features, and shifted into the rest frame.  The lines which
are fitted to the spectra are listed in Table~\ref{tab:fitted_lines}.
Line wavelengths are allowed to vary by 0.05\um\ (roughly 1/2 of a
spectral pixel, approximately the wavelength calibration uncertainty at
the blue ends of each IRS order), and line widths are allowed to vary by
10\% from their initial default values.  The central intensity of all
lines is required to be non-negative.

\subsubsection{Dust Features}
\label{sec:dust-features}

Dust features are represented by individual and blended Drude profiles:

\begin{equation}
  \label{eq:2}
  I_\nu^{(r)}=\frac{b_r\gamma_r^2}{(\lambda/\lambda_r -
    \lambda_r/\lambda)^2 + \gamma_r^2}
\end{equation}

\noindent where $\lambda_r$ is the central wavelength of the feature,
$\gamma_r$ is the fractional FWHM, and $b_r$ is the central intensity,
which is required to be non-negative.  The Drude profile is the
theoretical frequency profile for a classical damped harmonic
oscillator, and is a natural choice to model PAH emission features
\citep[and other emission and absorption bands, e.g., the 2175\AA\
extinction feature,][]{Fitzpatrick1986}.  The Drude profile, like the
Lorentzian, has more power in the extended wings than a Gaussian.  Power
in the overlapping wings of multiple bands can dominate the apparent
continuum, especially in the 5--12\um\ region.  The integrated intensity
of the Drude profile is

\begin{equation}
  \label{eq:3}
  I^{(r)}\equiv\int I_\nu^{(r)} d\nu=\frac{\pi c}{2}\frac{b_r\gamma_r}{\lambda_r}
\end{equation}

Table~\ref{tab:dust_features} lists all modeled dust features, some of
which are represented by up to 3 sub-features, grouped as blended
complexes.  The 17\um\ complex has two resolvable components, at
16.4\um\ \citep{Moutou2000} and 17.4\um\
\citep{2004ApJS..154..199S,Werner2004} --- see
Fig.~\ref{fig:examp_pahfit_ngc1482}.  For the other blends (7.7\um,
11.3\um, and 12.7\um), sub-features were adopted to best reproduce the
band shape.  Unresolved complexes of this type are considered only in
aggregate, and the sub-features comprising them are not intended to
represent distinct physical emission bands. The number and placement of
these sub-features was determined by training the model on a subset of
25 high signal galaxies from the sample, by fixing all other fitting
components, and allowing the width and central wavelengths to vary.  A
similar optimization was performed for the weaker individual features.
Once determined, feature positions and widths were held fixed.  Related
fitting approaches have been used with ISO spectra
\citep{Boulanger1998,Verstraete2001}.

\begin{deluxetable}{clc}
  \tablecaption{Dust Feature Parameters\label{tab:dust_features}}
    \tablecolumns{3}
    \tablehead{
      \colhead{$\lambda_r$(\um)} &
      \colhead{$\gamma_r$} & 
      \colhead{FWHM(\um)} \\
      \colhead{(1)} &
      \colhead{(2)} &
      \colhead{(3)}}

\startdata
  5.27 &  0.034 &  0.179\\
  5.70 &  0.035 &  0.200\\
  6.22 &  0.030 &  0.187\\
  6.69 &  0.070 &  0.468\\
  7.42\tablenotemark{a} &  0.126 &  0.935\\
  7.60\tablenotemark{a} &  0.044 &  0.334\\
  7.85\tablenotemark{a} &  0.053 &  0.416\\
  8.33 &  0.050 &  0.417\\
  8.61 &  0.039 &  0.336\\
 10.68 &  0.020 &  0.214\\
 11.23\tablenotemark{b} &  0.012 &  0.135\\
 11.33\tablenotemark{b} &  0.032 &  0.363\\
 11.99 &  0.045 &  0.540\\
 12.62\tablenotemark{c} &  0.042 &  0.530\\
 12.69\tablenotemark{c} &  0.013 &  0.165\\
 13.48 &  0.040 &  0.539\\
 14.04 &  0.016 &  0.225\\
 14.19 &  0.025 &  0.355\\
 15.90 &  0.020 &  0.318\\
 16.45\tablenotemark{d}&  0.014 &  0.230\\
 17.04\tablenotemark{d}&  0.065 &  1.108\\
 17.375\tablenotemark{d}&  0.012 &  0.209\\
 17.87\tablenotemark{d}&  0.016 &  0.286\\
 18.92&  0.019 &  0.359\\
 33.10 &  0.050 &  1.655%
\enddata 

\tablecomments{Col. (1): central wavelength.  Col. (2): fractional FWHM.
  Col. (3): full-width at half maximum.}

\tablenotetext{a}{7.7\um\  Complex}
\tablenotetext{b}{11.3\um\ Complex}
\tablenotetext{c}{12.7\um\ Complex}
\tablenotetext{d}{17\um\ Complex}
\end{deluxetable}

\paragraph{Features from 5--10\um}

The bulk of the PAH emission between 5--12\um\ is comprised of the
well-studied set of features discovered in planetary nebulae by
\citet{Gillett1973}, and probed in detail in Galactic objects with ISO
\citep[e.g. reflection nebula NGC 7023,][]{Cesarsky1996}.  The weak
8.33\um\ band, which sits between the prominent 7.7\um\ and 8.6\um\
features, was tentatively detected in the ISO SWS spectra of M82
\citep{Sturm2000}, and is present in many of our spectra, serving in
most cases only to fill in the saddle between the two stronger features.
A weak new feature in external galaxies occurs at 6.69\um\ \citep[see
also][]{Werner2004}.

A small number of stellar-dominated spectra from elliptical and other
early type systems in the sample show broad 10\um\ excess
(e.g. NGC\,4552), which has been attributed in similar systems to
silicate emission from the windos of AGB populations
\citep{Bressan2006}.  These features are not considered in our fit, and
are weak enough not to influence the recovery of stellar continuum in
these cases, though they can induce spurious dust features (especially
at 7.7\um), and for this reason are excluded from results pertaining to
these features.

\paragraph{Features from 10--20\um}

\begin{figure*}[ht]
\centering
\leavevmode
\includegraphics[width={.8\linewidth}]{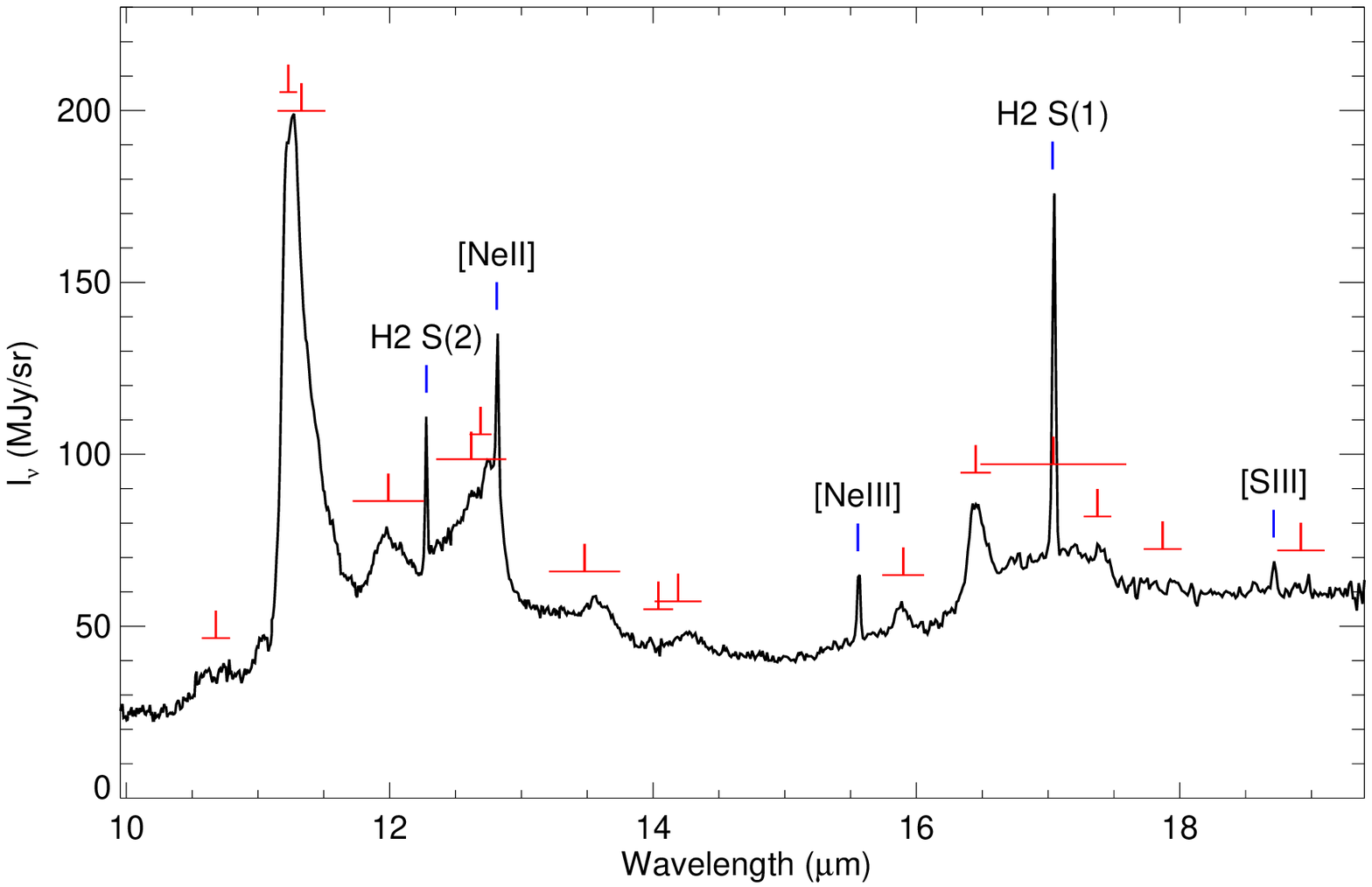}
\caption{High-resolution spectrum of NGC\,5195 (M51b) illustrating the
  sub-set of features from 10--19\um\ identified in the full
  low-resolution sample. The PAH feature central wavelength and FWHM for
  features from Table~\ref{tab:dust_features} with central wavelengths
  $>$10\um\ are indicated above the spectra, and unresolved emission
  lines in the spectrum are also labeled.}
\label{fig:ngc5195}
\end{figure*}

Along with the strong bands at 11.3\um, 12.7\um, and 17.0\um, a variety
of weak features are seen in most low-resolution spectra between
10--20\um.  The 10.7\um\ band, which falls in the (potentially
silicate-extincted) trough between the main 7.7 and 11.3\um\ features
was tentatively identified in the ISO spectra of M82 \citep{Sturm2000},
and is not uncommon in the sample, occasionally blending with nearby
\sIV\ at 10.51\um.  Other weak bands are found at 12.0\um, 12.6\um,
13.6\um, 14.3\um, 15.9\um, 16.4\um, and 17.4\um.  \citet{Sturm2000}
identified all of these features in ISO SWS spectra of the bright
starburst galaxies NGC\,253 and M82, and many of the individual
sub-features were first discovered in Galactic sources
\citep[e.g.][]{Moutou2000}.  The 14.3\um\ dust feature may be blended
with \neV\ 14.32\um\ in sources with AGN nuclei.  Another very weak
feature at 18.92\um\ is likely the same as the 19\um\ feature found in
Spitzer spectra of reflection nebula NGC\,7023 by \citet{Werner2004}.
It is blended with \sIII\ at 18.7\um.  Dust features $>$10\um\ are
marked in Fig.~\ref{fig:ngc5195} on the high-resolution IRS nuclear
spectrum of NGC\,5195 (M51b).  Note in particular that the 12.7\um\ PAH
band, which is blended with \neII\ at low-resolution, cleanly separates
with \neII\ to the red wing of the blend at higher resolution.

Interestingly, the strongest feature at $\lambda>12\um$, the broad band
centered at 17.04\um, was not seen in the ISO spectra of the brightest
starbursts, and was first identified in an extragalactic source in the
SINGS galaxy NGC\,7331 \citep{2004ApJS..154..199S}.  Despite the
strength of the 17\um\ complex, which can carry power comparable to the
prominent 11.3\um\ band (see \S\,\ref{sec:pah-energetics}), its large
width ($\sim$1.1\um), and the lack of available low-resolution
spectroscopy at wavelengths above 16\um\ combined to make this prominent
band difficult to detect with ISO.  The 17\um\ feature was reported
simultaneously in Spitzer spectra of reflection nebulae NGC\,7023
\citep{Werner2004}.  Evidence for a similar band complex was found in
ISO SWS spectra of Galactic sources
\citep{Beintema1996,VanKerckhoven2000}.

\paragraph{Features beyond 20\um}
\label{sec:features-beyond-20um}

A single feature was included beyond 20\um: a broad band centered at
33.1\um.  Though not present in all sources, it underlies the \siII\ and
\sIII\ lines, and is required to obtain accurate line fluxes for these
lines.  \citet{Sturm2000} notice a broad plateau from 33--34\um\ in 3
ISO SWS galaxy spectra, and attribute it to crystalline silicates.  A
similar plateau was noted in the SWS spectra of planetary nebulae
\citep{Waters1998}.  This feature is strong in $\sim$1/6 of our sample
(e.g., see NGC\,4321), with no correlation of the feature's strength
with silicate extinction (see \S\,\ref{sec:prev-silic-extinct}).

\subsubsection{Dust Extinction}
\label{sec:dust-extinction}

\begin{figure}
\plotone{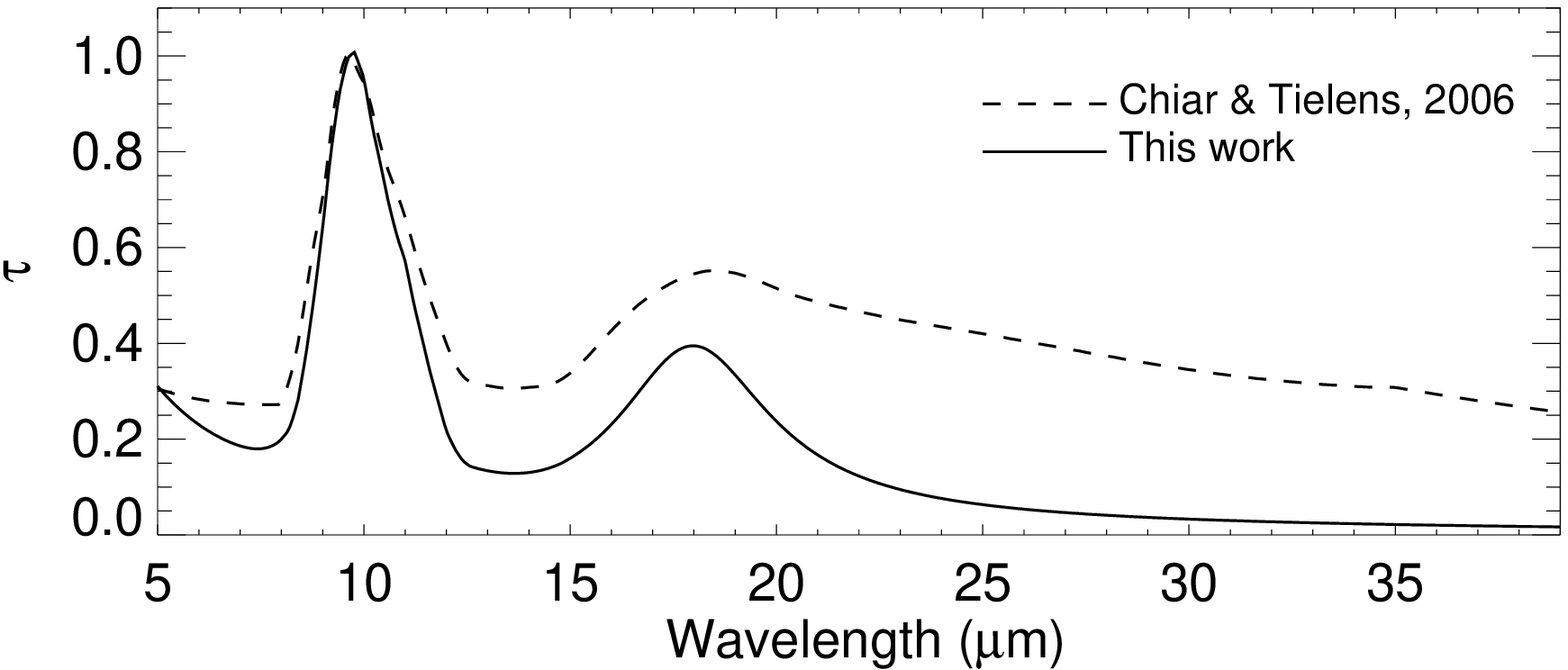}
\caption{The adopted extinction curve, along with the curve of
  \citet{Chiar2006}.}
\label{fig:ext_curve}
\end{figure}

We allow for absorption by dust with extinction properties similar to
Miky Way dust.  We take the infrared extinction to consist of a
power-law plus silicate features peaking at 9.7\um\ and 18\um:

\begin{equation}
  \label{eq:4}
  \tau(\lambda)=\tau_{9.7}\left[ (1-\beta) P_{\mathrm{Si}}(\lambda) + 
    \beta (9.7\um/\lambda)^{1.7}\right]
\end{equation}

\noindent where $\tau_{9.7}$ is the total extinction at 9.7\um, and
$P_{\mathrm{Si}}(\lambda)$ is the normalized silicate profile function,
with $P_{\mathrm{Si}}(9.7\um)\!=\!1$.  We adopt $\beta\!=\!0.1$; with
this choice, $\tau(3.6\um)\!=\!0.54\,\tau(9.7\um)$, which is within the
current uncertainty range regarding extinction from 3--8\um\ \citep[see,
e.g., ][]{Draine2001a}.  The 9.7\um\ component of
$P_{\mathrm{Si}}(\lambda)$ is taken from the Galactic center profile of
\citet{Kemper2004} between 8.0\um\ and 12.5\um, with
$P_{\mathrm{Si}}(8\um)=0.06$.  For $\lambda<8\um$, we smoothly extend
this profile by assuming
$P_{\mathrm{Si}}(\lambda)=P_{\mathrm{Si}}(8\um)\:e^{2.03(\lambda-8\mu
  m)}$.  For $\lambda>12.7\um$, we add a Drude profile extinction
component peaking at 18\um, with FWHM=4.45\um\ ($\gamma_r$=0.247),
smoothly joining the two profiles between 12.5\um\ and 12.7\um.  The
fitted silicate depth $\tau_{9.7}$ is required to be non-negative; no
silicate emission was included in the model.  We also explored using the
Galactic center extinction curve of \citet{Chiar2006}, with similar
values of $\tau_{9.7}$ derived, but the much larger extinction at long
wavelengths ($A_{30\mu m}\!\sim\! A_{8\mu m}$) produced poorer overall
fits in absorbed glaxies with our choice of dust continuum temperatures.
The adopted extinction curve is shown in Fig.~\ref{fig:ext_curve}.

The default extinction geometry is taken to be absorbing dust fully
mixed with emitting stars and grains; a uniform foreground screen of
dust is also considered.  Since all components of the model are presumed
to be equally affected by dust extinction, line and feature strengths
are automatically corrected to their extinction-free values.  This
implicit correction may be unphysical if the material emitting the
mid-infared continuum, from which the dust absorption is primarily
determined, and the PAH emitting material undergo differential
extinction.  The TIR luminosity estimates described in
\S\,\ref{sec:observ-reduct} are not extinction-corrected, but the
correction for $\lambda>25\um$, where the bulk of the TIR power emerges,
is negligible.

\subsection{Fitting and Uncertainties}
\label{sec:fitting}

The model in Eq.~\ref{eq:1} is fitted to the observed flux intensity by
minimizing the global $\chi^2$ using the Levenberg-Marquardt algorithm,
modified to allow the values of individual fit parameters to be fixed
within allowed limits.  Formal statistical uncertainties, derived from
the IRS pipeline ramp fitting uncertainties, are returned for each fit
parameter, and full uncertainties in combined quantities such as
integrated intensities are formulated using the full covariance matrix.
In particular, for estimating uncertainties in the integrated intensity
of blended dust features, correlations between components can introduce
significant cross-term corrections to the propagation of error. For
upper limits, additional uncertainty estimates, based on the residuals
of the observed flux intensity from the best fitting model, are adopted.
The response of the model to a variety of simulated changes to the
continuum was tested, and the strengths of dust features recovered were
found to be robust against such variations.  In particular, adding
strong stellar continuum to PAH-dominated spectra and re-fitting
resulted in feature fluxes which changed by less than 2\%.

\subsection{Example Decomposition  and Tool Availability}
\label{sec:example-fits}

\begin{figure*}[ht]
\plotone{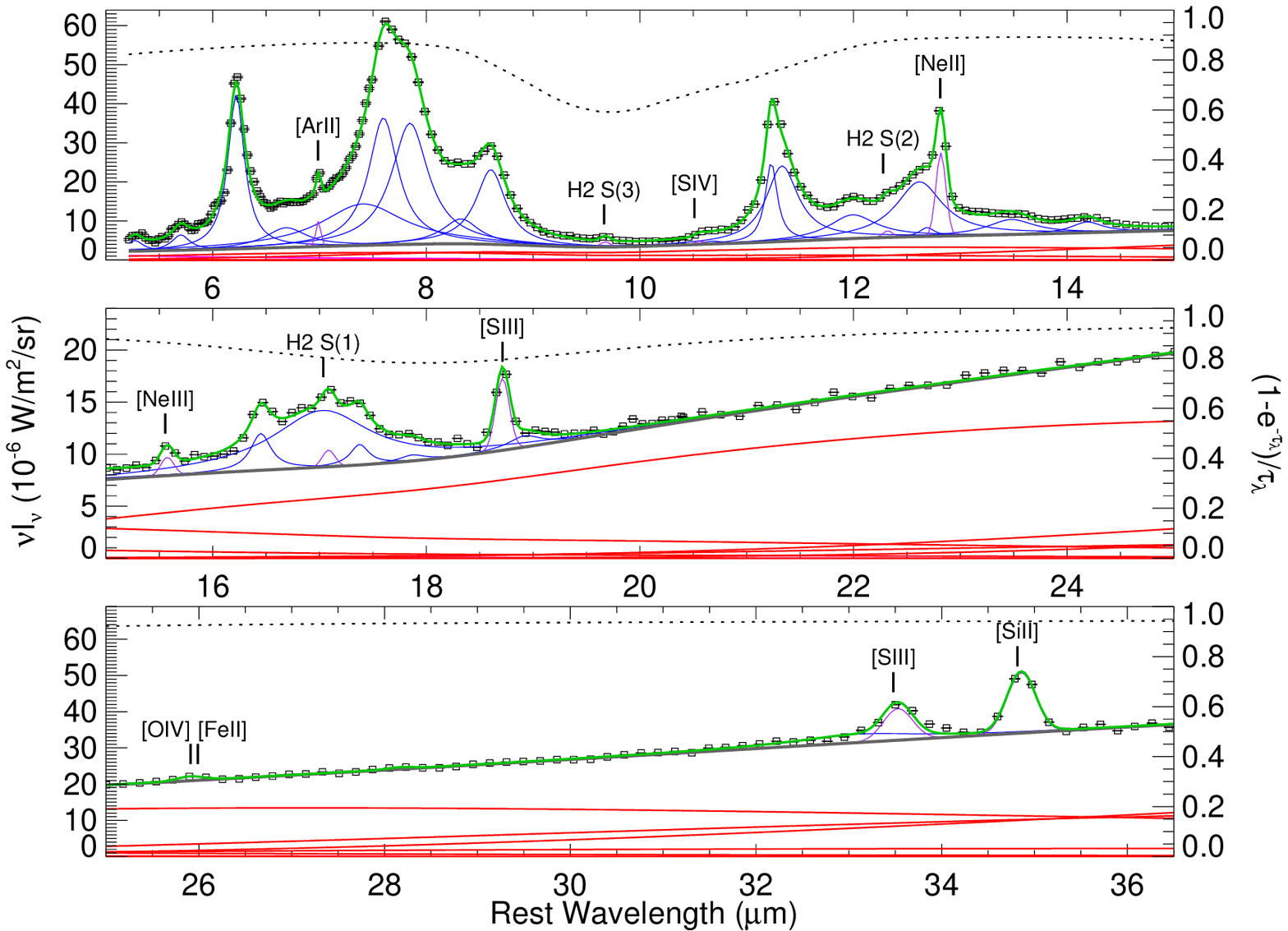}
\caption{Detailed decomposition of NGC\,1482 from 5--37\um.  Red solid
  lines represent the thermal dust continuum components, the magenta
  line shows the stellar continuum (weak in this case), and the thick
  gray line shows the total (dust + stellar) continuum.  Blue lines set
  above the total continuum are dust features, while the narrower violet
  peaks are unresolved atomic and molecular spectral lines.  All
  components are diminished by the fully mixed extinction, indicated by
  the dotted black line, with axis at right.  The solid green line is
  the full fitted model, plotted on the observed flux intensities and
  uncertainties.}
\label{fig:examp_pahfit_ngc1482}
\end{figure*}

\begin{figure*}[ht]
\centering
\leavevmode
\includegraphics[width={.95\linewidth}]{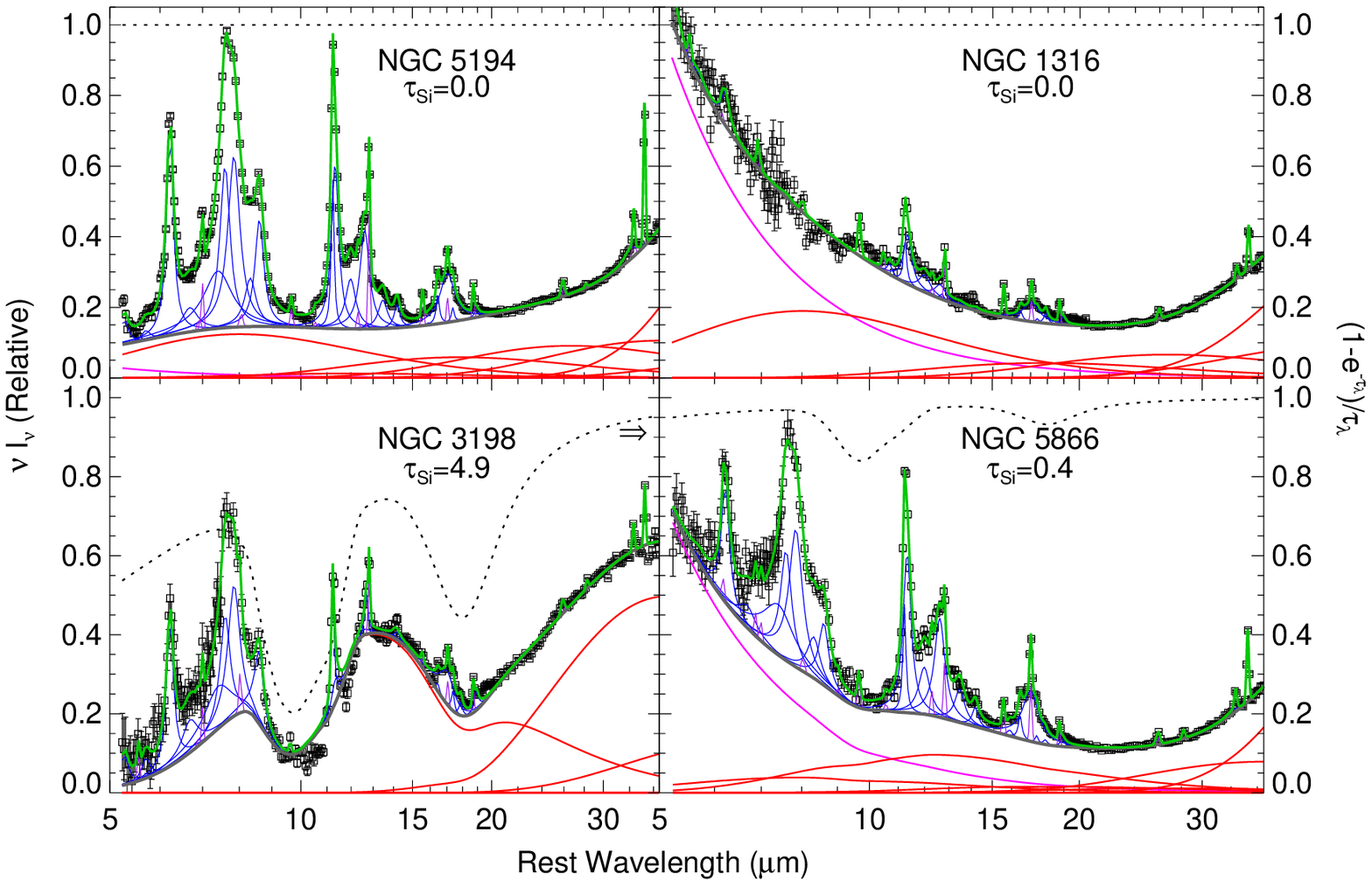}
\caption{Example decomposition of four SINGS spectra, for sources with
  and without significant silicate extinction, and dominated by either
  stellar or thermal dust continuum at short wavelengths.  All plot
  symbols as in Fig.~\ref{fig:examp_pahfit_ngc1482}.  Logarithmic
  wavelength axis.}
\label{fig:examp_pahfit}
\end{figure*}

The detailed components of the spectral decomposition method are
illustrated in Fig.~\ref{fig:examp_pahfit_ngc1482} for NGC\,1482, and in
Fig.~\ref{fig:examp_pahfit} for four additional examples --- two sources
with significant dust extinction, and two sources with starlight
dominating their short-wavelength continuum.  All galaxies in the sample
are well reproduced by the modeled components.  A tool implementing the
decomposition method is available in \pahfit, an IDL package developed
for the decomposition of low resolution IRS spectra comprised of stellar
and thermal dust continuum (not including very hot dust in strong AGN),
dust features, line emission, and silicate absorption.  Source code and
documentation for \pahfit\ are
available\footnote{\url{http://sings.stsci.edu/pahfit}}.

\section{Results}
\label{sec:results}

The spectral decomposition of all 59 sample galaxies was used to
investigate silicate extinction, the energetics of PAH emission,
variations in the large number of individual and blended PAH features
recovered, and the impact of these variations on studies of
high-redshift galaxies in deep broadband surveys.  Recovered strengths
for all important dust features, and the \neII\ and \neIII\ lines are
given in Table~\ref{tab:feature_strength}.

\begin{deluxetable}{c}
  \tablecolumns{1} 
  \tablecaption{Recovered Dust Feature and Line Strengths
    \label{tab:feature_strength}} 
  \tablehead{
    \colhead{This table is available only on-line as a machine-readable
      table.}}
  \startdata
  \enddata
\end{deluxetable}

\subsection{The Nature of the 17\um\ Emission Band}
\label{sec:nature-17um-emission}

\begin{figure}[ht]
\plotone{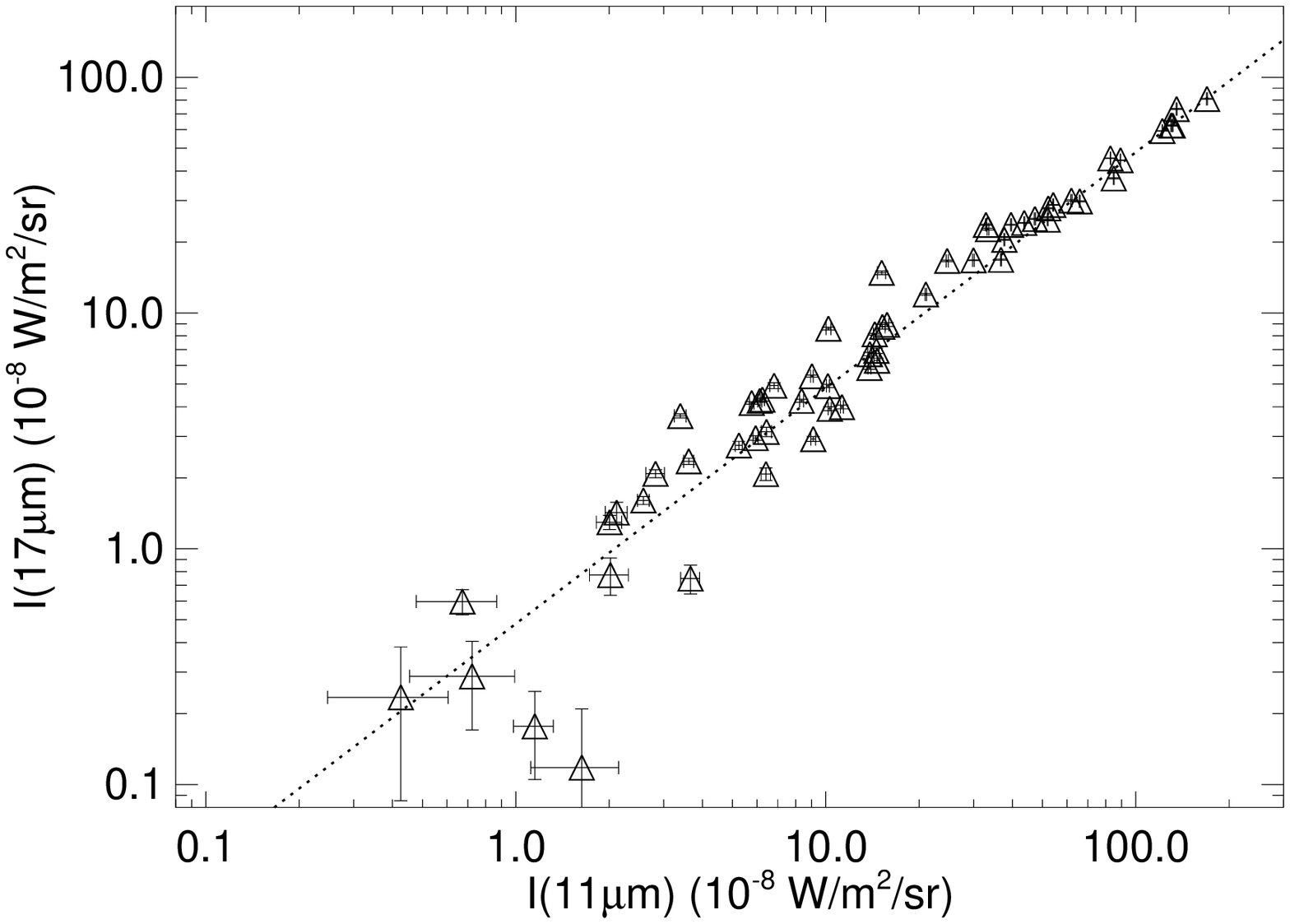}
\caption{The correlation between the intensity of the 11.3\um\ and
  17\um\ PAH complexes, shown with a slope unity line.}
\label{fig:pahs_11_17}
\end{figure}

The 17\um\ PAH complex is the strongest band longward of 12\um.  In
contrast to the highly variable plateau reported by
\citet{Peeters2004b}, all SINGS spectra exhibit a dominant component of
the 17\um\ complex which varies very little in position and width.  This
uniformity could be a result of beam-averaging over a range of physical
conditions in PDRs drawn from similar environmental distributions, or
perhaps an indication that the Galactic sources considered by
\citeauthor{Peeters2004b} were dominated by atypical band carriers or
radiation environments.  The distinct flanking sub-features at 16.4\um\
and 17.4\um\ contribute only $\sim$20\% of the total power in the
complex, with the remainder contained in the broad (FWHM$\sim$1\um)
feature centered at 17.04\um.  A small number of galaxy spectra in the
sample do exhibit an apparent plateau-like extension of the 17\um\ band
to $\sim$19\um\ (e.g.  NGC\,1566, NGC\,4569), but it does not correlate
with the strength of the 17\um\ feature or any other PAH band, and may
arise from a different carrier.

Although laboratory spectra of PAH mixtures contain many features
between 15--20\um\ presumed due to C-C-C out-of-plane bending modes
\citep{Peeters2004b}, it is of interest to compare the general behavior
of the 17\um\ band with the other more traditional PAH bands.
Fig.~\ref{fig:pahs_11_17} compares the integrated feature power of the
17\um\ and 11.3\um\ complexes.  The features correlate very well with
each other over 2 orders of magnitude, strongly supporting the
association of the 17\um\ complex with PAH emission.  Some variations in
the 17\um/11\um\ intensities ratio is evident; these will be explored in
\S\,\ref{sec:pah-strength-vari}.

\subsection{The Incidence of Silicate Extinction}
\label{sec:prev-silic-extinct}

\citet{Helou2000} found little evidence for silicate absorption in the
5.9--11.7\um\ spectra of the ISO Key Project sample of 45 normal
galaxies, and \citet{Sturm2000} also concluded that high-resolution ISO
spectra of bright starbursts like M82 were consistent with little or no
silicates.  With a more varied sample, and access to the full 9.7\um\
and associated 18\um\ features, the SINGS spectra are uniquely suited to
the investigation of moderate silicate absorption in a range of galaxy
types.  

Attempting to measure weak to moderate silicate absorption in spectra
with strong PAH emission features can lead to ambiguous measures of both
the feature strengths and the silicate optical depth.  In systems
dominated by deeply embedded infrared sources (e.g. many ULIRGs) with
strong silicate absorption, strong to weak or even absent PAH bands, and
accompanying shorter wavelength ice absorption features, the
identification and measurement of the silicate and other absorption
optical depths is straightforward
\citep[e.g.][]{Armus2004,Spoon2005,Roussel2006}.  However, as the
silicate optical depth decreases, and the PAH strength relative to the
local continuum increases, the PAH bands at 7.7/8.6\um\ and 11.3\um,
which closely flank the main silicate absorption trough centered at
9.7\um, can mask or mimic the true silicate absorption.  With this
absorption feature alone it is difficult to discriminate between
moderate PAH emission superposed on a silicate-absorbed continuum, and
strong PAH features with a relatively weak underlying continuum.
However, since the dust continuum at 15--20\um\ is fairly regular in the
spectra of star-forming galaxies, increasing silicate absorption in the
associated 18\um\ band results in a measurable change in the shape of
the 15--20\um\ continuum.

A compelling example of this degeneracy in recovering silicate
extinction from limited wavelengths around the 10\um\ band is found
NGC\,5866.  \citet{Lu2003} noted an unusually high value of
$L(6.2\um)/L(7.7\um)$ for this compact, edge-on disk galaxy.  With
access to the full 5--38\um\ spectrum, and identification of the
non-negligible silicate extinction $\tau_{9.7}\!=\!0.36$, the band ratio
$L(6.2\um)/L(7.7\um)$ is placed on the \emph{low} end of the
distribution (see Fig.~\ref{fig:main_band_rat}).  In this case, the
strong photospheric continuum which dominates NGC\,5866's spectrum out
to 10\um\ may have further complicated accurate measurement of the band
strengths.

\begin{deluxetable}{lccc}
  \tablecaption{Silicate Dust Extinction Sources\label{tab:silicate}}
  \tablecolumns{4}
  \tablehead{
    \colhead{} & 
    \multicolumn{2}{c}{$\tau_{9.7\um}$} &
    \colhead{}\\
    \colhead{Galaxy} & 
    \colhead{Mixed} &
    \colhead{Screen} &
    \colhead{$\chi^2(\tau_{\mathrm{Si}}=0)/\chi^2(\tau_{\mathrm{Si}})$}}
  \startdata
      NGC\,1266 &  2.05 &  0.91 &  1.868\\
      NGC\,1482 &  1.04 &  0.37 &  1.346\\
      NGC\,3198 &  4.88 &  1.90 &  4.978\\
      NGC\,4536 &  0.37 &  0.18 &  1.086\\
      NGC\,4631 &  0.40 &  0.20 &  1.031\\
      NGC\,5033 &  0.41 &  0.20 &  1.058\\
      NGC\,5866 &  0.36 &  0.17 &  1.060\\
      NGC\,6946 &  0.76 &  0.37 &  1.258%
   \enddata
\end{deluxetable}

Only 13.6\%, or 8 of the 59 sample spectra shown in
Fig.~\ref{fig:spectra_0}, require non-negligible absorption by silicate
dust to reproduce their spectra.  These objects, and their inferred
silicate optical depths $\tau_{9.7}$, are listed in
Table~\ref{tab:silicate}, for the default fully mixed dust geometry, as
well as a uniform foreground screen.  The silicate depths are lower for
the screen geometry, by roughly a factor of two.  The mixed and screen
models provide equally acceptable fits.  Also given is
$\chi^2(\tau_{\mathrm{Si}}\!=\!0)/\chi^2(\tau_{\mathrm{Si}})$ --- the
ratio of $\chi^2(\tau_{\mathrm{Si}}\!=\!0)$, formed by explicitly
forcing the silicate depth to be zero, to the normal
$\chi^2(\tau_{\mathrm{Si}})$ including absorption.  This ratio
demonstrates the degree to which the fit is improved by adding silicate
extinction, and indicates that the decomposition method can robustly
recover modest extinction with mixed silicate optical depths
$\tau_{\mathrm{Si}}\gtrsim0.35$, or screen depths half of that value.
Thus it appears that \emph{most} normal galaxies do not exhibit dust
extinction to the mid-infrared emitting material in their central few
square kiloparsecs above $A_V\sim5$ for fully mixed dust geometries, or
$A_V\sim3$ for foreground screens of dust \citep[assuming
$A_V/\tau_{\mathrm{Si}}=16.6$,][]{Rieke1985}.  Screen extinctions
derived separately from optical nuclear spectra for the SINGS sample
show $\langle A_V\rangle\sim1.3$, consistent with the inferred silicate
depths \citep{Dale2006}.  \citet{Brandl2006} found a larger range and
greater incidence of silicate absorption depths in the nuclear spectra
of a sample of starburst galaxies.

\subsection{Recovering PAH Strengths}
\label{sec:recov-pah-strengths}

\begin{figure*}[t]
\plotone{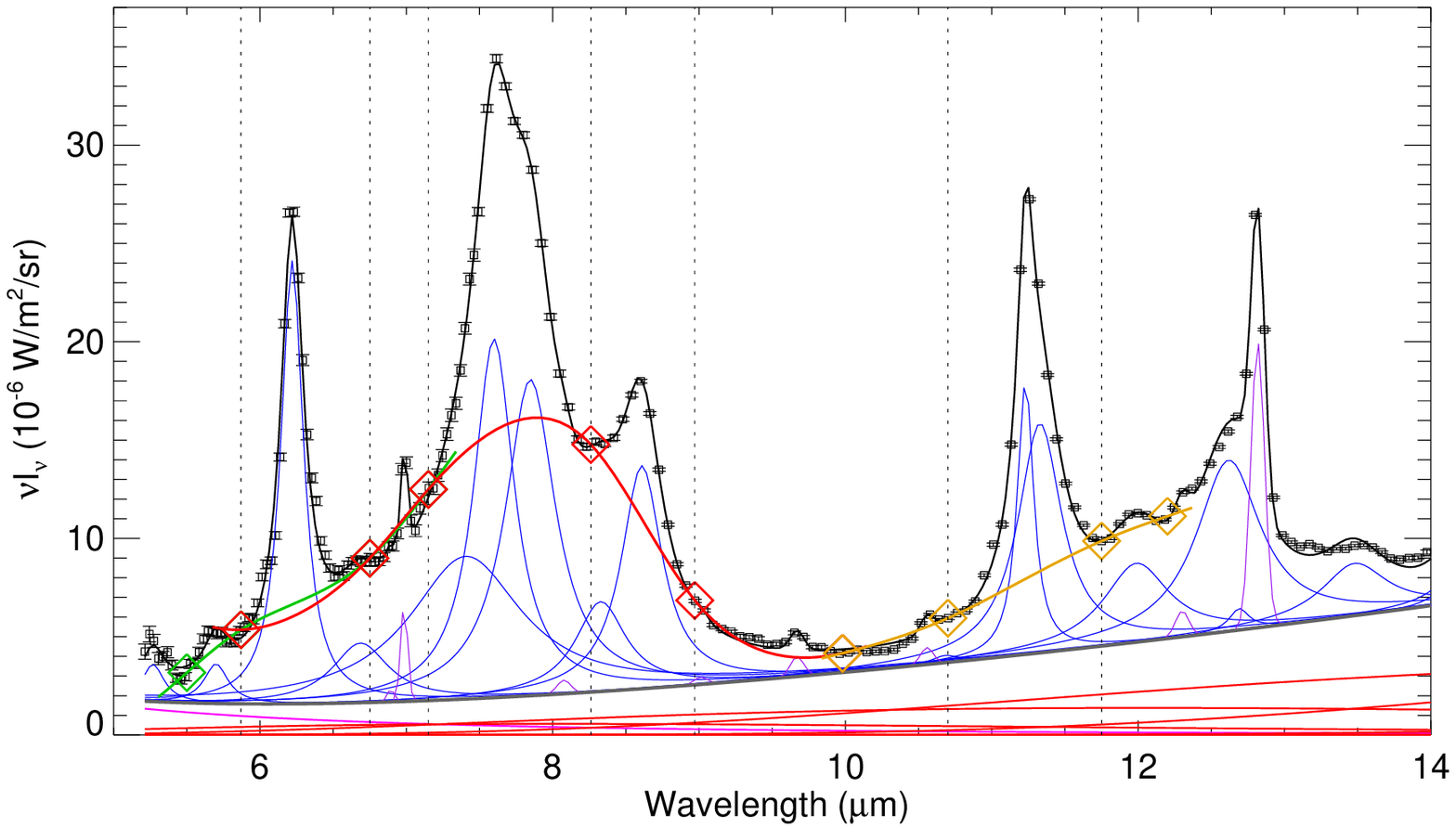}
\caption{An example comparing the spline-based and full
  feature+continuum decomposition in SINGS galaxy NGC\,2798.  Three
  different spline fits are shown with continuum anchor points: in green
  (leftmost), underlying the the 6.2\um\ feature, in red (center),
  underlying both the 7.7\um\ and 8.6\um\ features, and in orange
  (rightmost), under the 11.3\um\ feature.  Vertical dotted lines mark
  the ranges over which the spline-subtracted features are integrated.
  The fitted continuum, dust, and line features, are as in Fig.
  \ref{fig:examp_pahfit_ngc1482}.}
\label{fig:spline_comp_2798}
\end{figure*}

A widely used method for extracting PAH strength involves selecting
narrow continuum regions free from line or feature emission, forming
linear or cubic spline fits through these continuum points, and
integrating the spectrum above this pseudo-continuum
\citep[e.g.][]{Uchida2000,Peeters2002,Brandl2006}.  An application of
this method to the SINGS spectrum of NGC\,2798 is shown in
Fig.~\ref{fig:spline_comp_2798}.  The method's advantage is that it is
relatively straightforward to apply, and requires no specific
assumptions regarding the nature of the continuum.  The disadvantage is
that locating nearby continuum anchors free from line, feature, or
absorption contamination can prove difficult and to some degree
arbitrary, with the resulting feature strengths depending sensitively on
the exact choice of fiducial continuum points.  This problem is
compounded in low quality spectra, where continuum regions free from
these contaminating effects may be less readily identified.

\begin{figure}
\plotone{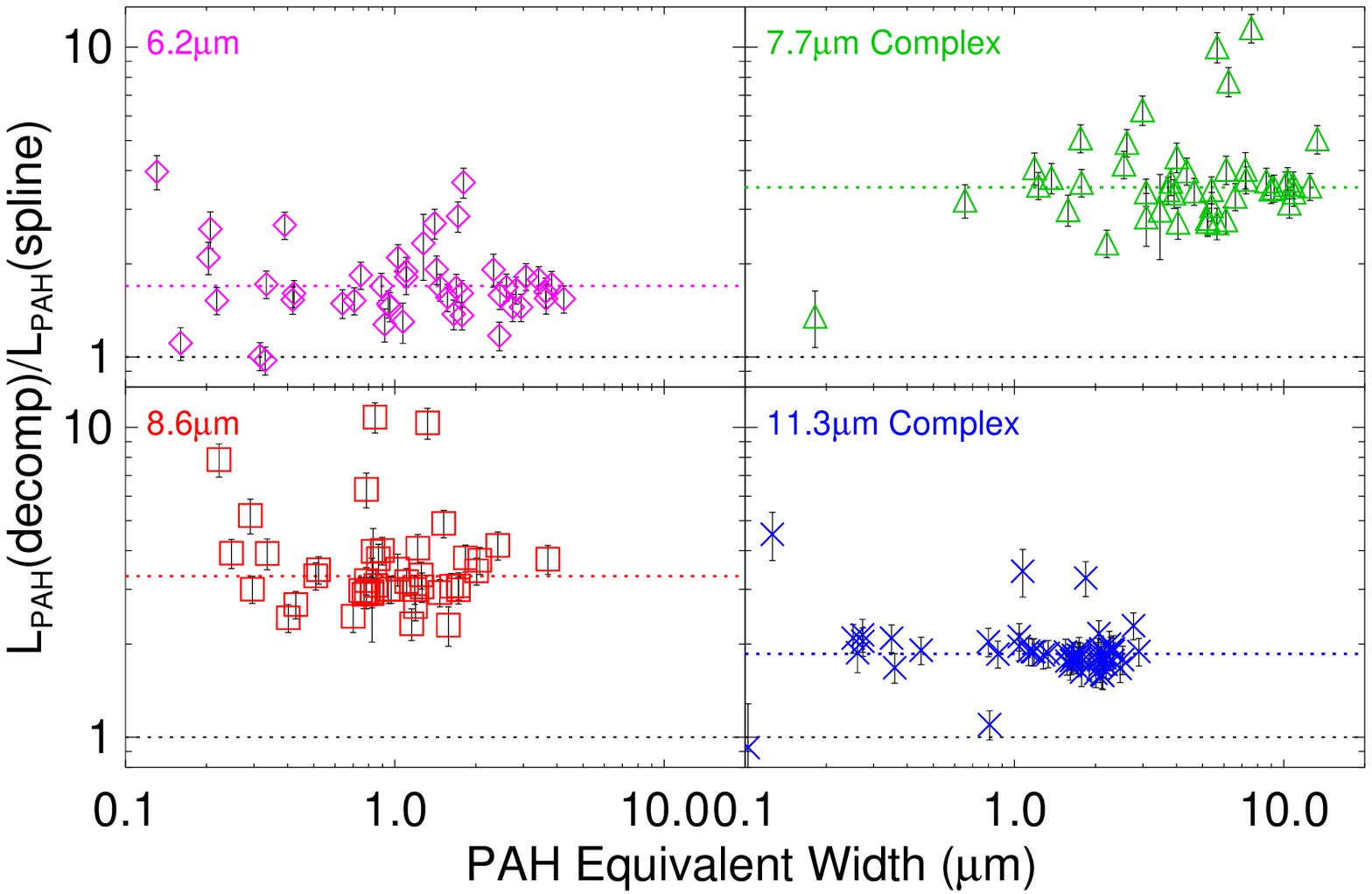}
\caption{The ratio between the PAH features strength from the full
  spectral decomposition, to the PAH strength recovered from
  spline-based continuum fitting vs. the equivalent width measured from
  the decomposition.  Mean and unity values are indicated with
  horizontal dotted lines.}
\label{fig:spline_comp}
\end{figure}

To compare the band strengths recovered by the spline-fitting and full
decomposition techniques, we have applied the cubic spline method for
the features at 6.2\um, 7.7\um, 8.6\um, and 11.3\um\ to the full sample,
fixing the rest wavelength continuum pivots following
\citet{Peeters2002}.  Figure~\ref{fig:spline_comp_2798} illustrates the
chosen fixed rest-frame continuum anchor points: 3 points for 6.2\um, 4
for 11.3\um, and 5 for the 7.7+8.6\um\ blend.  The PAH bands at 12.6\um\
and 17\um\ are blended with strong emission lines (\neII\ and \hs{1}),
and for this reason cannot be recovered with spline-based methods.

Fig.~\ref{fig:spline_comp} shows the PAH strengths recovered from full
feature decomposition vs. spline fitting as a function of the feature
equivalent width (defined as $\int (I_\nu-I^{cont}_\nu)/I^{cont}_\nu
d\lambda$).  Equivalent widths of the shortest wavelength features can
be undefined, since the underlying continuum can vanish.  In these
cases, the profile-weighted average continuum was substituted.  For the
two unblended features in clean continuum regions, at 6.2\um\ and
11.3\um, the decomposition and spline-based methods yield feature powers
differing by an approximately constant ratio,
$L(\textrm{decomp})/L(\textrm{spline})\sim$1.75, independent of the
equivalent width of the feature. Since a fraction $f_{wing}\sim59\%$ of
the power in a Drude profile lies outside its FWHM, it is not surprising
that the offset between the two is roughly $1/f_{wing}$.

For the blended features at 7.7\um\ and 8.6\um, however, the fraction of
the full feature strength recovered by the spline method is lower, with
larger scatter.  This results from the continuum anchor placement in the
filled trough between these features, located to ensure that the
de-blended peaks are independent.  The overall scatter is probably
somewhat larger than would have been achieved with hand-tuned spline
continuum placement in all spectra.  However, the large systematic
offset would remain, such that the spline-based methods underestimate by
a factor of 3--6 the full 7.7\um\ and 8.6\um\ feature strengths.  Table
\ref{tab:spline_comp} lists the average offsets for feature equivalent
widths larger than 0.1\um, after applying an iterative 3$\sigma$
trimming of outliers.  These $L(\textrm{decomp})/L(\textrm{spline})$
factors can be used to scale spline-recovered PAH fluxes to estimate
crudely the total power in the PAH features -- although, if possible, a
more accurate estimate will be obtained directly using a decomposition
method like the one described in \S\,\ref{sec:decomp-mir-galaxy}.  They
may have dependence on the form of underlying continuum and presence of
silicate or other absorption, and thus may not be appropriate for all
spectra.

\begin{deluxetable}{lr@{\,$\pm$\,}l}
  \tablecaption{PAH Feature Luminosity Estimates: Full Decomposition vs.
    Spline Ratios
    \label{tab:spline_comp}} 
  \tablecolumns{3} 
  \tablehead{
    \colhead{Feature} & 
    \multicolumn{2}{c}{$L(\textrm{decomp})/L(\textrm{spline})$}}
  \startdata
            6.2\um &   \hspace{3ex}1.70 &   0.42\\
    7.7\um\ Complex &   \hspace{3ex}3.53 &   0.62\\
            8.6\um &   \hspace{3ex}3.31 &   0.64\\
   11.3\um\ Complex &   \hspace{3ex}1.86 &   0.15%
   \enddata
\end{deluxetable}

\subsection{PAH Energetics}
\label{sec:pah-energetics}

\begin{figure*}
\centering
\leavevmode
\includegraphics[width={.8\linewidth}]{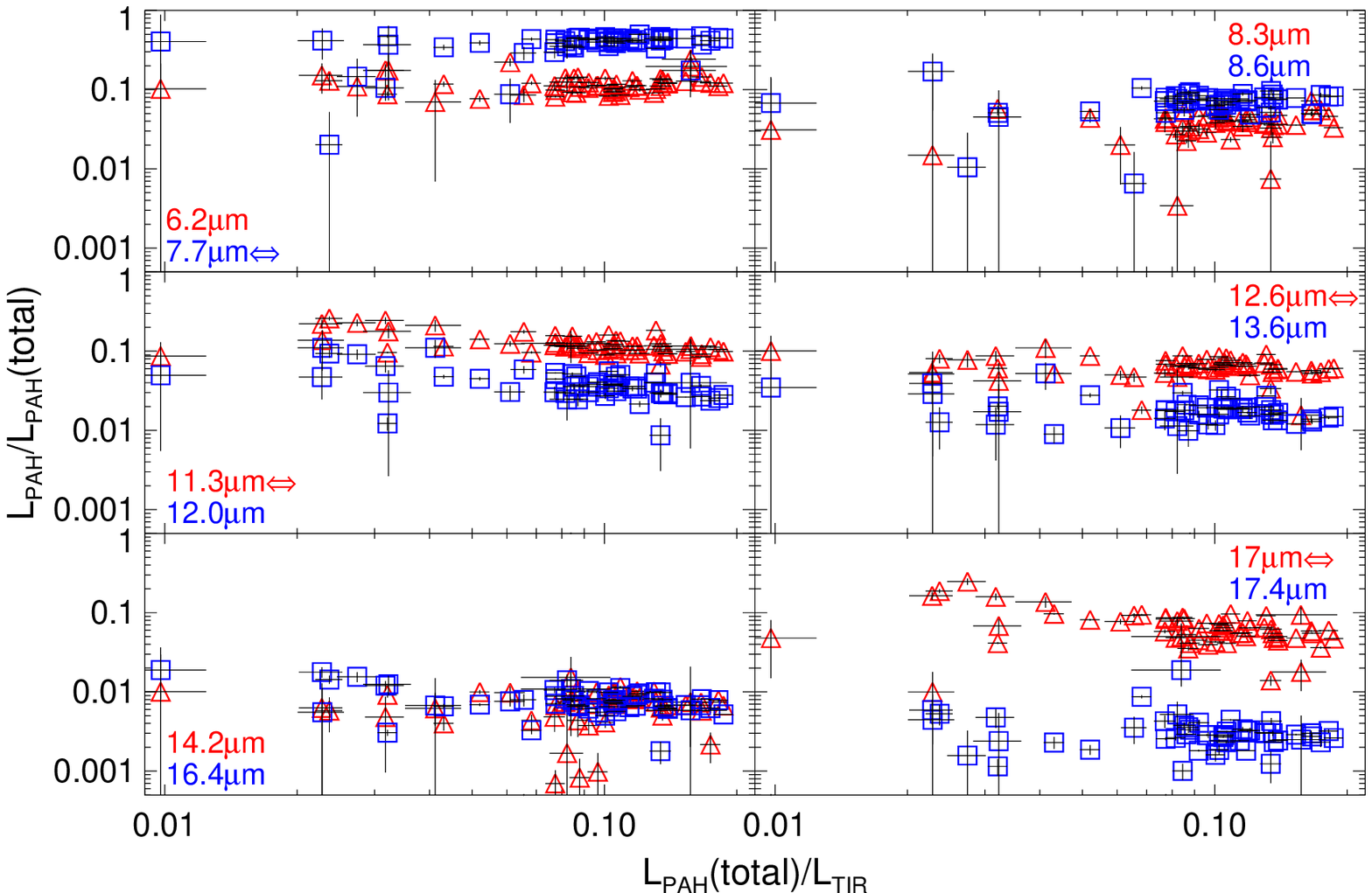}
\caption{The fractional PAH luminosity relative to the total PAH power
  for the main features identified in the spectra, against the
  integrated PAH contribution to the total infrared luminosity from
  3--1100\um, in the same region.  Blended complexes are marked by a
  double arrow in the legend; see Table~\ref{tab:dust_features}.}
\label{fig:pah_tir}
\end{figure*}

\begin{figure}
\plotone{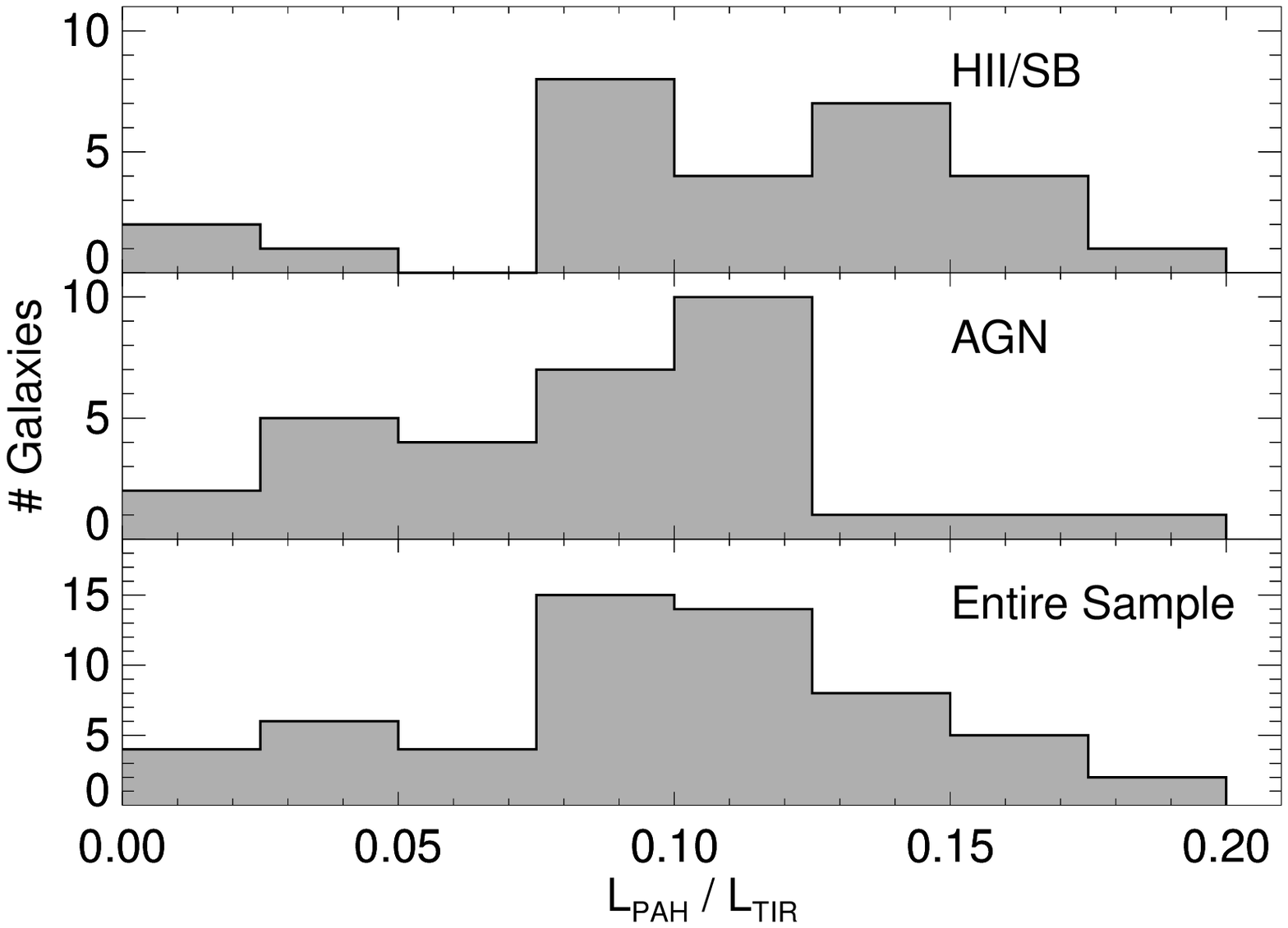}
\caption{The distribution of the integrated PAH luminosity relative to
  the total infrared, $L_\PAH/L_{\TIR}$, for the entire sample, and
  individually for galaxies with AGN and \hII/Starburst classifications}
\label{fig:pah_tir_hist}
\end{figure}

The total power emitted in the PAH bands can be large, and the current
sample provides an opportunity to form a census of the energetic
importance of individual features, independent of the underlying
mid-infrared continuum, which can be dominated by widely differing
emission processes and environments and depend on the details of the
stellar and dust grain populations.  In Fig.  \ref{fig:pah_tir}, the
fractional power relative to the total PAH luminosity is shown for all
significant features present in the sample.

The 7.7\um\ PAH complex can contribute nearly one-half of the total PAH
luminosity, and deliver up to 10\% of the total infrared luminosity
alone.  In aggregate, the combined power output of all PAH features
ranges from a few percent up to 20\% of the total infrared.  Fig.
\ref{fig:pah_tir_hist} shows the distribution of $L_\PAH/L_{\TIR}$ in
the apertures considered, for the full sample, and separately for
galaxies with evidence of nuclear Seyfert or LINER emission in their
nuclear optical spectra, and galaxies with only \hII-region or starburst
classifications.  The peak is near $L_\PAH/L_{\TIR}=0.1$ for both \hII\
and AGN galaxies.  At the highest relative PAH luminosities, systems
dominated by star formation predominate (at $L_\PAH/L_{\TIR}>0.12$, only
3/14 systems host AGN), despite the fact that the spectral apertures
include both nuclear and inner disk emission.  This may suggest a
natural limit on the absolute PAH strength in sources with weak AGN,
either due to partial destruction of the grains, or the contribution of
other continuum sources to the total infrared luminosity.

\begin{figure}
\plotone{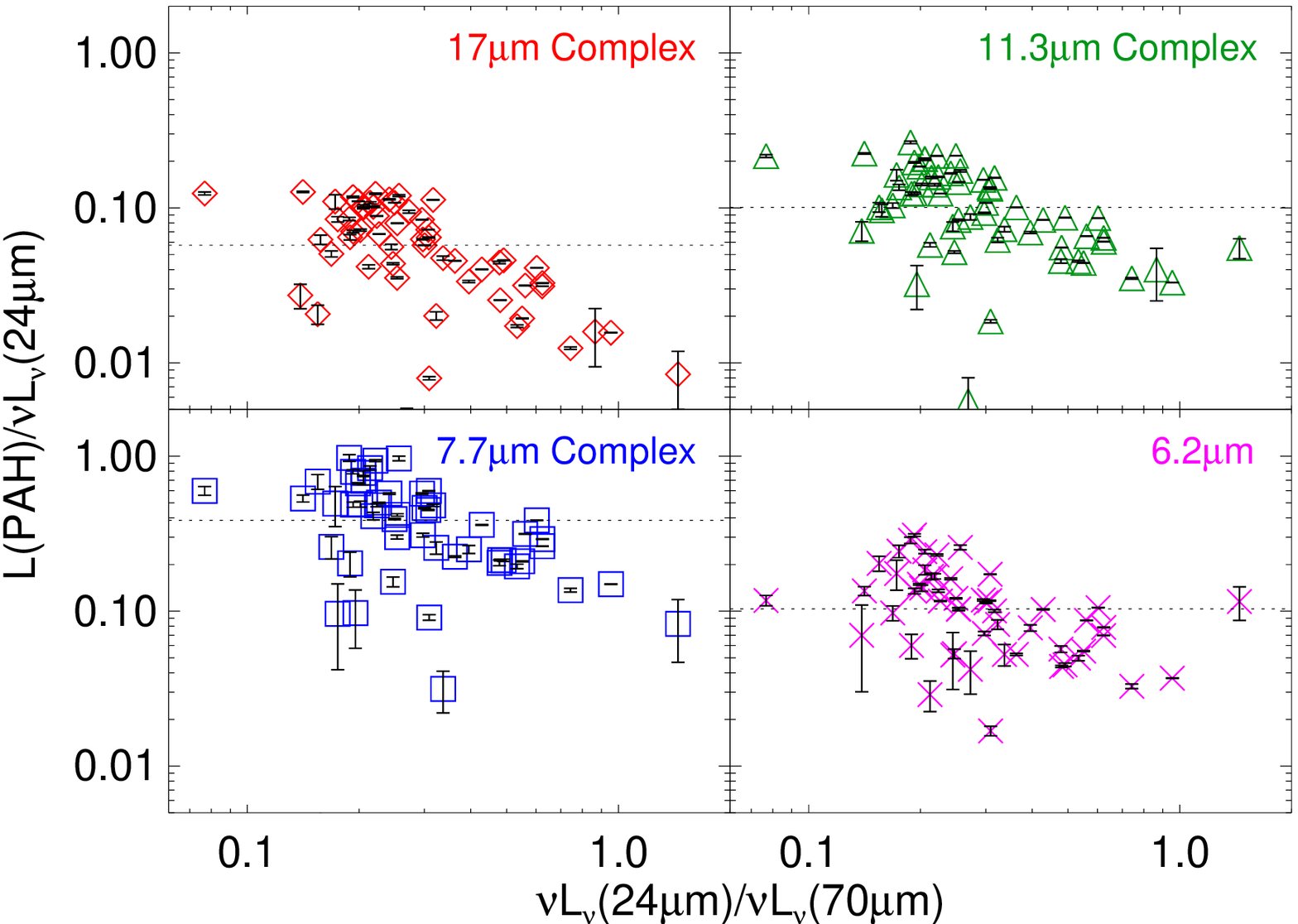}
\caption{The PAH to MIPS 24\um\ luminosity ratio vs. far-infrared color
  for the four strongest PAH bands, in matched apertures.}
\label{fig:pahs_24}
\end{figure}

Fig.~\ref{fig:pahs_24} demonstrates the large scatter in the ratio
$L(\textrm{PAH})/\nu L_\nu(24\um)$ for the four strongest PAH features
as a function of the local far-infrared color temperature $\nu
L_\nu(24\um)/\nu L_\nu(70\um)$.  The full range is typically over a
factor of 10 for each band, and may in fact be a lower limit on the true
scatter in the global emission of galaxies, for which diffuse scattered
light in the outer regions can account for a substantial fraction of the
24\um\ luminosity \citep{Calzetti2005}.  For all bands, there is a
definite negative trend of $L(\textrm{PAH})/\nu L_\nu(24\um)$, which
suggests that a mix of both small stochastically-heated and thermally
emitting grains contribute to the 24\um\ continuum, perhaps due to
varying contributions from PDR and \hII\ regions
\citep[e.g.][]{Laurent2000}.

\subsection{The PAH Contribution to IRAC 8\um\ and MIPS 24\um}
\label{sec:pah-contr-irac}

\begin{figure}[t]
\plotone{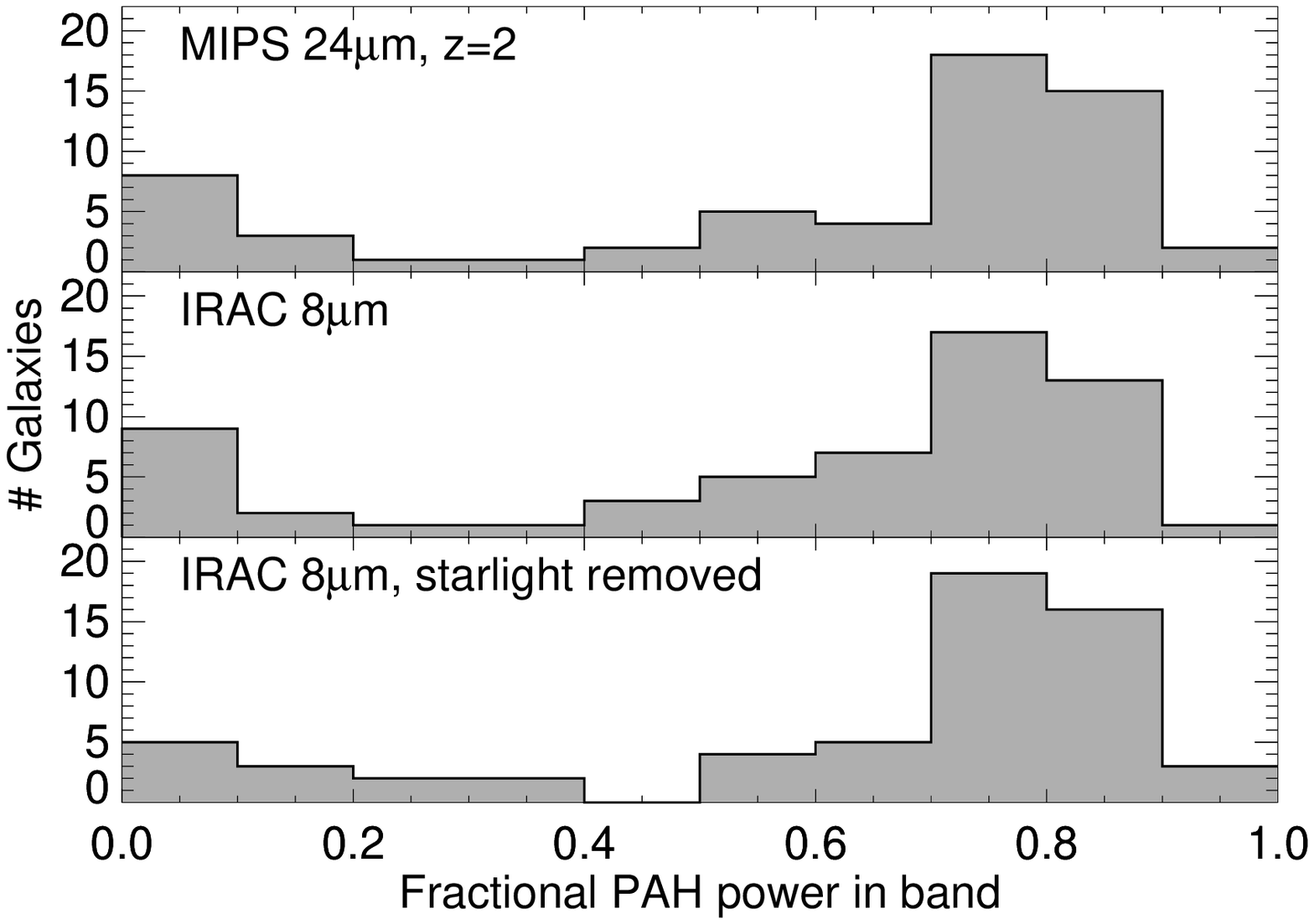}
\caption{The distribution among the sample of the fractional
  contribution of dust features to the Spitzer broad bands, for IRAC
  8\um\ at redshift $z=0$, with and without stellar continuum removed
  from the band, as well as MIPS 24\um\ at
  $z=2$.\label{fig:pah_iracmips}}
\end{figure}

With the full spectral decomposition of continuum and dust features, we
can examine the fractional contribution of PAH emission and stellar or
thermal dust continuum to the broad Spitzer bandpasses.  We consider
IRAC 8\um\ at $z=0$, and MIPS 24\um\ at $z=2$.  These broad bands are
commonly used as surrogates for PAH strength in nearby and distant
galaxy surveys \citep[e.g.][]{Hogg2005,Papovich2006,Reddy2006}.  Since
IRAC 8\um\ at low redshift, and MIPS 24\um\ at $z\sim2$ encompass the
7.7\um\ and 8.6\um\ PAH bands --- typically the strongest (although, see
\S\,\ref{sec:pah-strength-vari})--- a common assumption is that
broadband photometry at these rest wavelengths serves as a direct proxy
for PAH emission, total infrared or bolometric luminosity, and,
potentially, star formation rate.  For low redshift sources,
prescriptions exist for removing the stellar contribution using the
3--5\um\ channels of IRAC \citep[e.g.][]{Helou2004}.  The relatively
isolated 24\um\ band does not typically permit such starlight removal
(although the IRS Peak-Up Blue channel at 16\um\ could in principle
serve this purpose).  

Fig.~\ref{fig:pah_iracmips} highlights the large fractional contribution
of PAH feature emission to the uncorrected IRAC 8\um\ band at $z\!=\!0$
and MIPS 24\um\ band at $z\!=\!2$.  The median contribution for both
$\sim$\,0.7, with a small number of sample members (sources dominated by
stellar continuum) showing very little PAH contribution.  For over 75\%
of the sample, PAH emission contributes more than one half of the broad
band 8\um\ power.  If the stellar continuum is \emph{perfectly} removed,
the distribution shifts only slightly to higher fractional PAH
contribution.

The typical source in deep MIPS 24\um\ surveys has total infrared
luminosity $L_{\TIR}\gtrsim 10^{11}L_\sun$ \citep{Papovich2006,Yan2004},
and may therefore exibit a much different contribution of PAH emission
in the 8\um\ rest frame compared to star-forming galaxies of lower
luminosity.  The sample explored here has median global $L_{\TIR}$ of
$0.8\times10^{10}L_\sun$.  It should also be noted that the dominance of
PAH emission in the band, e.g. MIPS 24\um\ at z$\sim$2, does not imply
anything about the efficacy of using this single photometric point to
derive the total infrared luminosity or star formation rate; see
\citet{Dale2005}.

\subsection{PAH Strength Variations}
\label{sec:pah-strength-vari}

\begin{figure*}[t]
\centering \leavevmode
\includegraphics[width={.8\linewidth}]{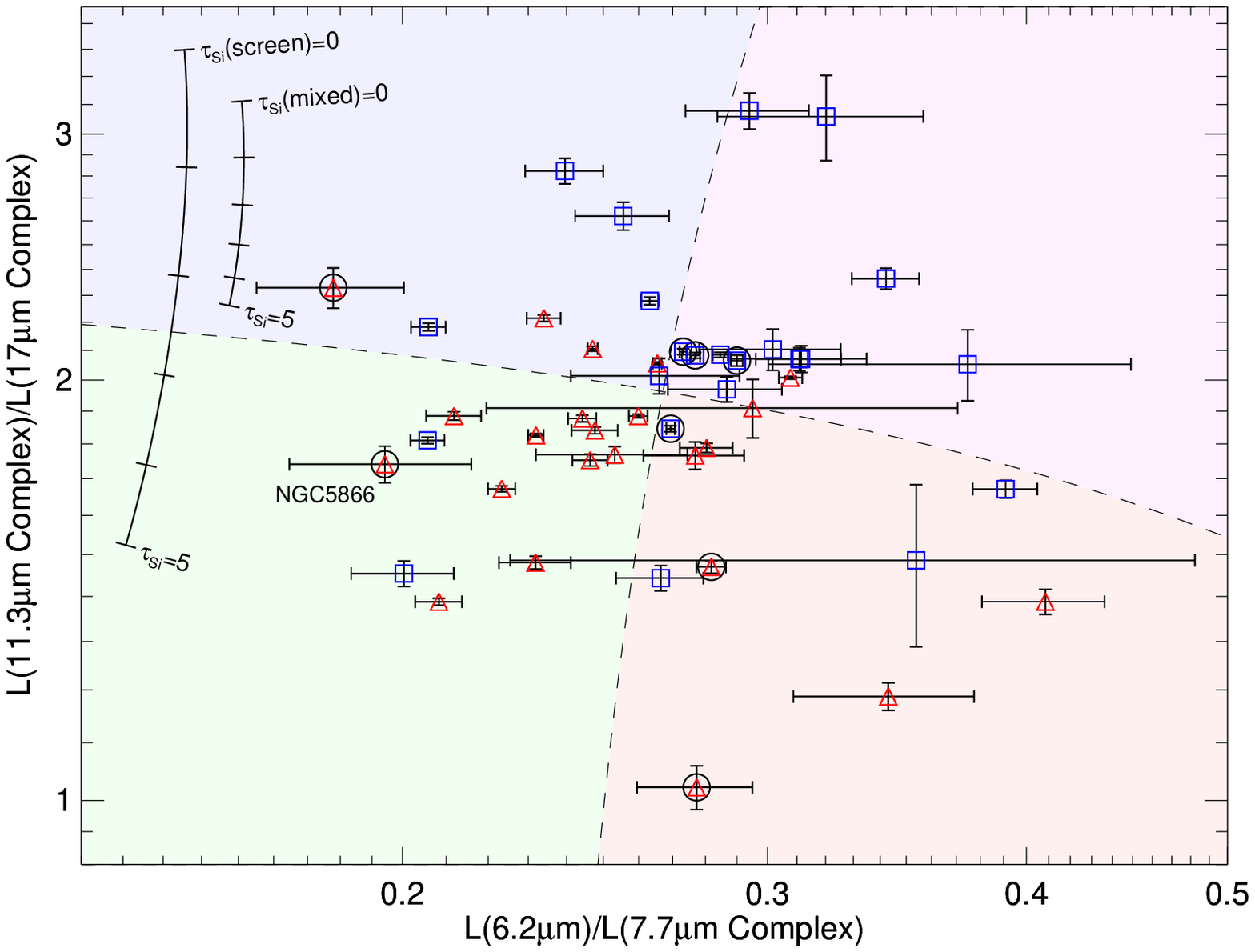}
\caption{The inter-band strength ratios of the four main PAH bands at
  6.2\um, 7.7\um, 11.3\um, and 17\um.  Triangles (red) indicate galaxies
  with AGN (Seyfert or LINER) types, whereas squares (blue) are galaxies
  with \hII-dominated nuclei.  Galaxies with measurable silicate
  absorption ($\tau_{\mathrm{Si}}>.01$) are circled.  Also shown are the
  extinction trends for these band ratios, parametrized by the silicate
  optical depth for both uniform foreground screen and fully mixed dust
  geometries.  The highlighted regions indicate zones used for the
  construction of template spectra (see \S\,\ref{sec:model-infr-seds}).}
\label{fig:main_band_rat}
\end{figure*}

Fig.~\ref{fig:main_band_rat} demonstrates the variation in inter-band
strength ratios over the full sample for the four strongest PAH bands at
6.2\um, 7.7\um, 11.3\um, and 17\um.  Both the 6.2\um/7.7\um\ and
17\um/11.3\um\ ratios vary by more than a factor of two.  Sources with
weak AGN are almost entirely offset from sources with \hII-like nuclear
optical spectra --- an effect which will be explored in the next
section.  Sources with non-negligible mixed silicate absorption optical
depths (circled) are offset slightly from the main locus of points for
their type (\hII\ or AGN), to lower values of $L(6.2\um)/L(7.7\um)$ and
$L(11.3\um)/L(17\um)$.  This is in the direction of increased reddening,
despite the implicit de-reddening performed by the spectral
decomposition method, and could be evidence of different absolute
extinctions for the underlying continuum and the PAH emitting regions,
or of limitations in the adopted extinction curve. 

The medians and 10\%--90\% range of variation for luminosity ratios
among the most important PAH bands, the integrated PAH luminosity, and
the total infrared luminosity are given in
Table~\ref{tab:pah_variations}.  Typical variations in the inter-band
luminosities are a factor of 2--5, with $1\sigma$ variations in these
ratios of $\pm$30\%.

\tabletypesize{\scriptsize}
\begin{deluxetable*}{rr@{\,(}r@{--}l@{)\ }r@{\,(}r@{--}l@{)\ }r@{\,(}r@{--}l@{)\ }r@{\,(}r@{--}l@{)\ }r@{\,(}r@{--}l@{)\ }r@{\,(}r@{--}l@{)\ }r@{\,(}r@{--}l@{)\ }r@{\,(}r@{--}l@{)\ }}
\tablecaption{PAH Band Luminosity Ratios L($\lambda_1$)/L($\lambda_2$)\label{tab:pah_variations}}
\tablecolumns{25}
\tablehead{
  \colhead{Band\,\raisebox{-1ex}{(1)}$\backslash$\raisebox{1ex}{(2)}} &
  \multicolumn{3}{c}{\textbf{6.2\um}} &
  \multicolumn{3}{c}{\textbf{7.7\um}\tablenotemark{$\Leftrightarrow$}} &
  \multicolumn{3}{c}{\textbf{8.6\um}} &
  \multicolumn{3}{c}{\textbf{11.3\um}\tablenotemark{$\Leftrightarrow$}} &
  \multicolumn{3}{c}{\textbf{12.6\um}\tablenotemark{$\Leftrightarrow$}} &
  \multicolumn{3}{c}{\textbf{17\um}\tablenotemark{$\Leftrightarrow$}} &
  \multicolumn{3}{c}{\textbf{$\sum$\,PAH}} &
  \multicolumn{3}{c}{\textbf{TIR/100}}}
\startdata
\textbf{                         6.2\um} & \multicolumn{3}{c}{\nodata}  &    0.28 &    0.21 &    0.75  &     1.5 &     1.2 &     3.0  &     1.1 &    0.52 &     1.5  &     1.8 &     1.1 &     2.8  &     1.9 &    0.93 &     2.9  &    0.11 &   0.086 &    0.15  &     1.1 &    0.35 &     2.1\\
\textbf{                         7.7\um} &     3.6 &     1.3 &     4.8  & \multicolumn{3}{c}{\nodata}  &     5.7 &     4.7 &     9.0  &     3.6 &     1.5 &     4.8  &     6.5 &     3.6 &     8.3  &     6.9 &     3.1 &      11  &    0.42 &    0.18 &    0.45  &     4.1 &    0.54 &     6.2\\
\textbf{                         8.6\um} &    0.66 &    0.38 &    0.86  &    0.18 &    0.11 &    0.21  & \multicolumn{3}{c}{\nodata}  &    0.68 &    0.35 &    0.81  &     1.1 &    0.66 &     1.5  &     1.2 &    0.65 &     2.0  &   0.073 &   0.049 &   0.088  &    0.72 &    0.16 &     1.2\\
\textbf{                        11.3\um} &    0.95 &    0.65 &     1.9  &    0.28 &    0.21 &    0.67  &     1.5 &     1.2 &     2.9  & \multicolumn{3}{c}{\nodata}  &     1.8 &     1.5 &     3.2  &     1.9 &     1.4 &     2.8  &    0.12 &   0.092 &    0.18  &     1.1 &    0.57 &     1.7\\
\textbf{                        12.6\um} &    0.55 &    0.36 &    0.93  &    0.16 &    0.12 &    0.27  &    0.93 &    0.65 &     1.5  &    0.57 &    0.31 &    0.65  & \multicolumn{3}{c}{\nodata}  &     1.1 &    0.54 &     1.6  &   0.065 &   0.047 &   0.087  &    0.60 &    0.19 &    0.92\\
\textbf{                          17\um} &    0.52 &    0.34 &     1.1  &    0.15 &   0.089 &    0.32  &    0.81 &    0.50 &     1.5  &    0.53 &    0.35 &    0.72  &    0.93 &    0.62 &     1.8  & \multicolumn{3}{c}{\nodata}  &   0.059 &   0.039 &   0.097  &    0.62 &    0.28 &    0.94\\
\textbf{                    $\sum$\,PAH} &     8.8 &     6.6 &      12  &     2.4 &     2.2 &     5.7  &      14 &      11 &      20  &     8.6 &     5.5 &      11  &      15 &      11 &      21  &      17 &     10. &      25  & \multicolumn{3}{c}{\nodata}  &     10. &     3.2 &      16\\
\textbf{                        TIR/100} &    0.88 &    0.48 &     2.9  &    0.25 &    0.16 &     1.9  &     1.4 &    0.85 &     6.1  &    0.88 &    0.60 &     1.7  &     1.7 &     1.1 &     5.2  &     1.6 &     1.1 &     3.6  &   0.100 &   0.064 &    0.31  & \multicolumn{3}{c}{\nodata}%
\enddata
\tablenotetext{$\Leftrightarrow$}{Indicates blended PAH Complexes; see Table~\ref{tab:dust_features} for components.}
\tablecomments{
 Each entry is encoded as \emph{med (low--high)} where \emph{med} 
 is the median of the ratio, and \emph{low--high} is the 10\%--90\% 
 range of variation.}
\end{deluxetable*}
\tabletypesize{\footnotesize}

\subsubsection{PAHs in AGN}
\label{sec:pahs-agn}

Nuclear black holes are now recognized as a fundamental property of many
local elliptical and spiral galaxies: $\gtrsim$1/3 have optical nuclear
spectra typical of low-luminosity active galactic nuclei \citep{Ho1995,
  Ho1997a}.  The SINGS sample was not defined to include powerful AGN
sources, yet roughly half of the galaxies considered here show some form
of AGN activity in their nuclear optical spectroscopy.  The SINGS AGN
sources are in the low-luminosity regime, with $L(\oIII\
5007\AA)<10^{6.5} L_\sun$ (compared to
$L(\oIII)\!\sim\!10^8$--$10^9L_\sun$ for traditional AGN, Kewley et al.,
2006, in prep.).  This sample makes it possible to test the behavior of
PAH emission in the presence of weak AGN.

\begin{figure*}[t]
\plotone{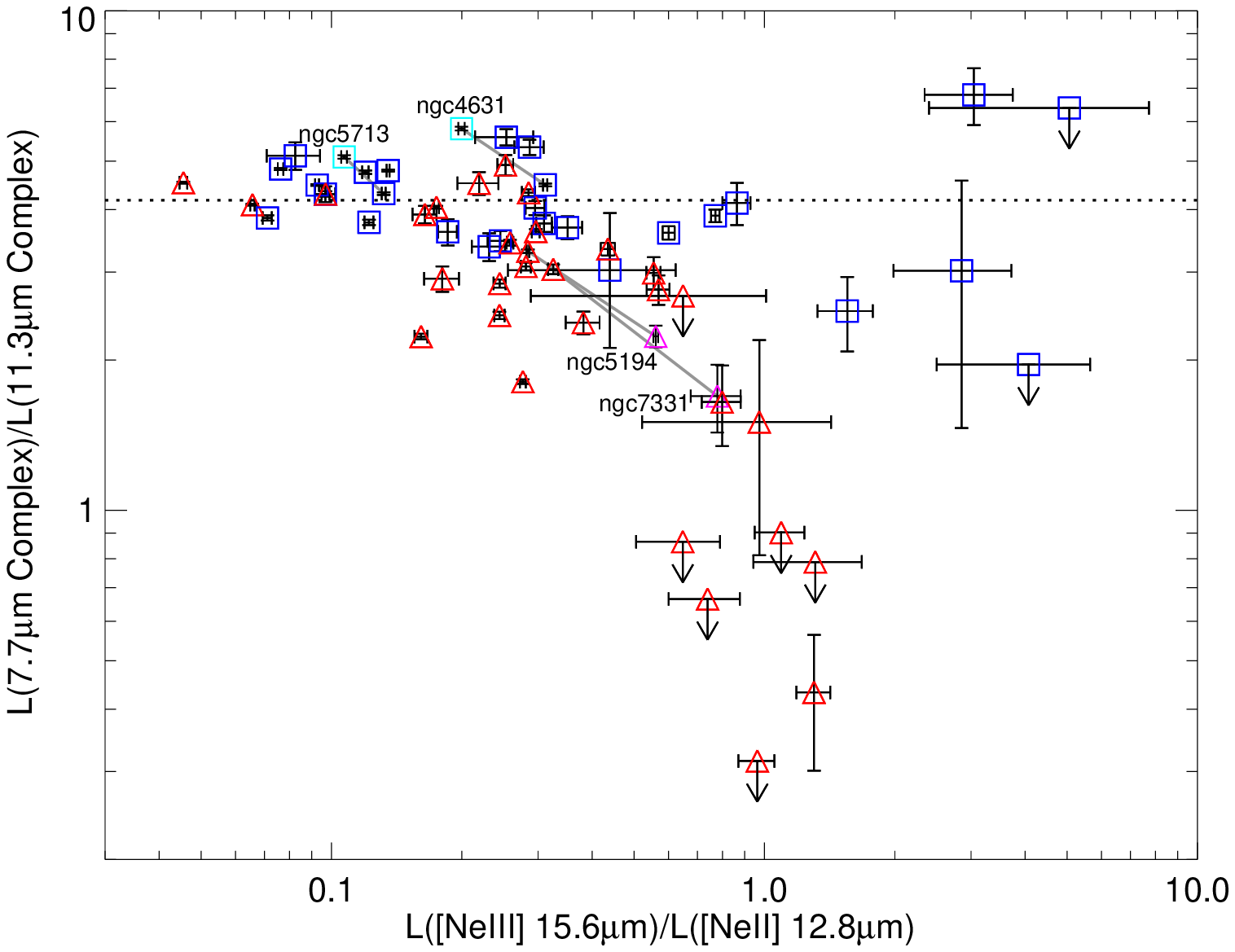} 
\caption{The variation of inter-band strength ratios of the PAH features
  at 7.7\um\ relative to the 11.3\um\ band with the line ratio
  \neIII/\neII, an indicator of the hardness of the radiation field.
  Triangles (red) indicate galaxies with AGN (Seyfert or LINER) types,
  whereas squares (blue) are galaxies with \hII-dominated nuclei.  Upper
  limits are 2$\sigma$.  The dotted line shows the median of the
  \hII-dominated sources.  Four sources (two AGN --- magenta triangles
  --- and two \hII\ nuclei --- cyan squares, labeled) are plotted with
  additional values obtained from spectra extracted over small
  9$\arcsec^2$ apertures centered on their nuclei, and connected by
  solid gray lines to their larger aperture values.}
\label{fig:pah_ne_rat}
\end{figure*}

\begin{figure}[t]
\plotone{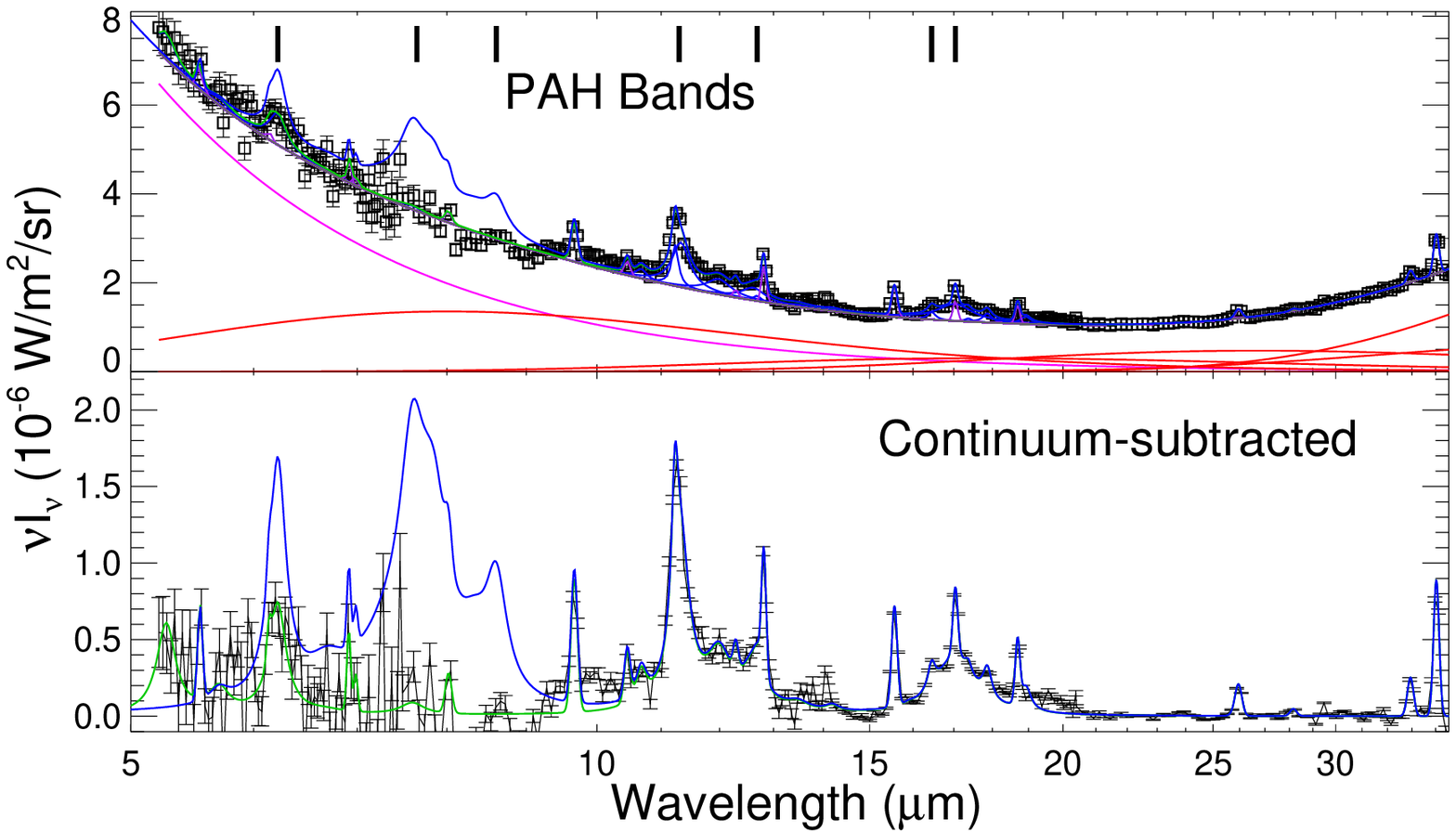}
\caption{An example of the peculiar low-luminosity AGN PAH emission
    spectrum in barred LINER galaxy NGC\,1316, with the full
    decomposition shown as the solid green line.  The PAH spectrum
    predicted from the strength of the 11.3\um features, and 
    adapted to the SINGS median band ratios, is shown in blue, with a
    continuum-subtracted version below.
    \label{fig:pah_ngc1316_peculiar}}
\end{figure}

In Fig.~\ref{fig:pah_ne_rat}, the ratio of two strong PAH bands is
compared to the hardness indicator \neIII/\neII, with an ionization
potential ratio of 41\,eV/21.6\,eV.  Galaxies with \hII\ region or
starburst-like optical spectra exhibit a nearly constant
$L(7.7\um)/L(11.3\um)$ ratio across the full range of radiation
hardness, including those low metallicity systems (when PAHs are
detected) with \neIII/\neII$\:\gtrsim\!2$.  In striking contrast, the
AGN as a group are offset below the locus of \hII-type galaxies, even at
moderate hardness ratios, and at \neIII/\neII$\sim1$, fall away rapidly.
Among the 10 galaxies in the sample with the lowest limiting band ratios
$L(7.7\um)/L(11.3\um)$, \emph{all} show evidence of low-luminosity LINER
or Seyfert activity.  In these systems, the relative strength of the
short wavelength PAH emission bands is suppressed by up to a factor of
10, compared to star formation dominated systems.  An example of such a
spectrum is shown in Fig.~\ref{fig:pah_ngc1316_peculiar}, for the LINER
galaxy NGC\,1316, with the predicted spectrum for the sample's median
band ratios overlayed.  

One effect which could lead to the observed behavior is ionization
state.  Since the 11.3\um\ feature is thought to be produced by neutral
PAHs, whereas the 7.7\um\ feature arises primarily from PAH cations
\citep[e.g.][]{Allamandola1999,Li2001}, a change in the neutral fraction
could impact their relative strengths.  Harder radiation fields are also
more efficient at destroying PAH grains, an effect which may dominate
the observed trend, if the grain or molecular carriers of the 7.7\um\
and 11.3\um\ bands are different, and undergo photo-destruction via
ionization or dissociation at substantially differing rates.  The
11.3\um\ feature can be produced by single-photon heating of PAHs that
are larger than those that are effective at producing 7.7\um\ emission
\citep[e.g.][]{Schutte1993,Draine2006}.  The observed decrease in
$L(7.7\um)/L(11.3\um)$ may then be the result of selective destruction
of those PAHs small enough to emit at 7.7\um.  Low-metallicity dwarfs
with the \emph{highest} \neIII/\neII\ ratios appear to have roughly
normal band ratios (when PAHs are detected), suggesting that AGN must
play a role.

Further evidence comes from a related trend in spatially resolved
spectral maps of large SINGS spiral galaxies with AGN nuclei.  The
7.7\um/11.3\um\ band ratio in many cases decreases noticeably in the
nucleus, compared to larger apertures which sample star-forming regions
in the inner disk.  An example of this effect for four galaxies chosen
at $d\!\lesssim\!8\,$Mpc with both bright inner disk and nuclear
emission is illustrated in Fig.~\ref{fig:pah_ne_rat}.  Ratios found in
the default apertures are connected by solid lines to those recovered in
small 9.25\arcsec$\times$9.25\arcsec\ (roughly 300\,pc) apertures
centered on the nucleus.  For the two systems with AGN (NGC\,5194 and
NGC\,7331), the short wavelength PAH features show further suppression
in their nuclear spectra, whereas the \hII\ galaxies (NGC\,4631 and
NGC\,5713) move to slightly \emph{higher} values of
$L(7.7\um)/L(11.3\um)$.  Such spatially-resolved variations will be
considered in more detail in a subsequent paper, but are consistent with
the interpretation that the hard radiation fields near AGN have a strong
impact on the strength of the short wavelength PAH bands.

\citet{Kaneda2005} note an unusually weak 7.7\um\ feature in IRS spectra
of three X-ray-bright elliptical galaxies, as well as SINGS elliptical
NGC\,4125 (which has the fourth lowest PAH band ratio in
Fig.~\ref{fig:pah_ne_rat}).  They suggest that unusual ISM properties in
evolved elliptical galaxies could be responsible.  However, two of these
three ellipticals have established optical AGN activity (NGC\,2974, a
Seyfert 2, and NGC\,4589, a LINER), and our own recently obtained
nuclear optical spectra of the third, NGC\,3962, indicates that it also
hosts an unambiguous AGN.  This suggests that the AGN environment,
rather than the global morphological type, leads to the unusual emission
characteristics.  Elliptical galaxies, with little contamination from
the PAH emission associated with ongoing star formation, may simply
offer the most favorable systems for detecting this unusual spectrum.
Consistent with this, \citet{Brandl2006} found a large scatter, but no
significant variation of $L(7.7\um)/L(11.3\um)$ over a wide range of
excitation in starburst nuclei.

An alternative explanation is that the AGN is unrelated to or only
partially responsible for the change in the PAH spectrum.  This would be
the case if AGN classification acts merely as a surrogate of low star
formation intensity, since strong nuclear star formation can overwhelm
the AGN diagnostic lines.  \citet{Cesarsky1998} found a depressed 7\um\
band in differential ISOCAM spectra within M31, suggesting this could be
due to the weak UV field.  In Spitzer samples of $\sim$50 passively
evolving early-type galaxies \citep{Bregman2006,Bressan2006}, only two
quiescent sources show PAH emission, one of which, NGC\,4550, hosts a
bright LINER nucleus and exhibits the same strong 11.3\um\ feature we
report.  The other, NGC\,4697, is without an apparent AGN, yet also
shows a moderately depressed 7\um\ PAH, though \citeauthor{Bregman2006}
conclude that after removing silicate emission, a spectrum similar to
the diffuse ISM of the Galaxy is recovered.  Future studies will better
constrain the role of AGN and weak star formation intensity in the
modification of the PAH emission spectrum.

If AGN dominate the effect, the absence of PAH emission in nearly all
quiescent galaxies suggests that, especially in systems with little
ongoing star formation, the AGN itself could excite PAH emission,
supplying UV radiation dominated by Lyman-$\alpha$ and H$_2$
Lyman-Werner bands in the extended X-Ray dominated regions
\citep{Meijerink2005}, with a hardness spectrum quite unlike that
typically found in environments harboring young stellar populations.  In
addition, an AGN may emit X-rays sufficient to destroy small PAHs
directly even at distances of kpc \citep{Voit1992}.  Because larger PAHs
are expected to be less susceptible to destruction by X-rays, the
presence of an AGN can alter the PAH emission spectrum, even if the PAH
excitation itself is dominated by starlight from an older stellar
population.  Clearly, the absolute luminosity density of the AGN source
is also an important factor regulating the PAH spectrum, since more
luminous AGN either destroy all PAH grains
\citep[e.g.][]{Voit1992,Lutz1998a}, or produce mid-infrared continuum
photons in such copious quantities that the PAH bands are diluted away
\citep{Siebenmorgen2005,Hao2005,Sturm2005}.  This effect was exploited
by \citet{Genzel1998} to differentiate between AGN and starbursts at
high luminosity. Although grain size modification may operate in many
contexts, the unusual band ratios observed in low-luminosity AGN nuclei
($L(7.7\um)/L(11.3\um)\lesssim2$) could offer an efficient way to detect
such weak active nuclei, when the UV field is not dominated by star
formation, and dust absorption hinders optical classification.  Though
it is not yet known over what regimes this effect operates, the
possibility that AGN sources could directly excite PAH emission suggests
considerable caution when using the PAH bands as direct indicators of
star formation in systems known to harbor active nuclei.

Care must be taken to quantify the band ratios which constitute an
unusual PAH spectrum of this type, since the 11.3\um\ feature, while not
the most luminous, has the highest specific luminosity per wavelength
interval, and does not typically suffer continuum dilution from
starlight as the shorter wavelength PAH features can.  At low signal
and/or with strong stellar continuum, the 11.3\um\ feature can remain
the most easily detectable PAH band, and appear to dominate despite
normal band ratios.  An example of this phenomenon is found in the SINGS
dwarf galaxy NGC\,1705 \citep{Cannon2006}.

\subsubsection{PAHs and Metallicity}
\label{sec:pahs-metallicity}

That the absolute strength of PAH emission can depend on metallicity has
recently been well established
\citep{Engelbracht2005,Madden2006,Wu2006,OHalloran2006,Rosenberg2006},
with indications of a threshold $12+\log(\mathrm{O/H})\sim8.1$, or
approximately $Z_\sun/4$, below which PAH emission is suppressed.  From
detailed SED fitting of the SINGS sample, Draine et al. (2006, in prep)
find that the fraction of the dust mass contributed by PAHs is
suppressed below a similar threshold.  Most studies rely on Spitzer
broadband photometry (e.g. 8\um/24\um) to assess absolute PAH strengths.
Although broadband 8\um\ flux does correlate well with PAH strength (see
\S\,\ref{sec:pah-contr-irac}), the range of mid-infrared continuum
properties can overwhelm even significant changes in the relative PAH
luminosity --- see, e.g., the large variation in 7.7\um\ PAH to 24\um\
luminosity in Fig.~\ref{fig:pahs_24}.  Broadband methods are also not
sensitive to changes in the form of the PAH emission spectrum.

\begin{figure}[t]
\plotone{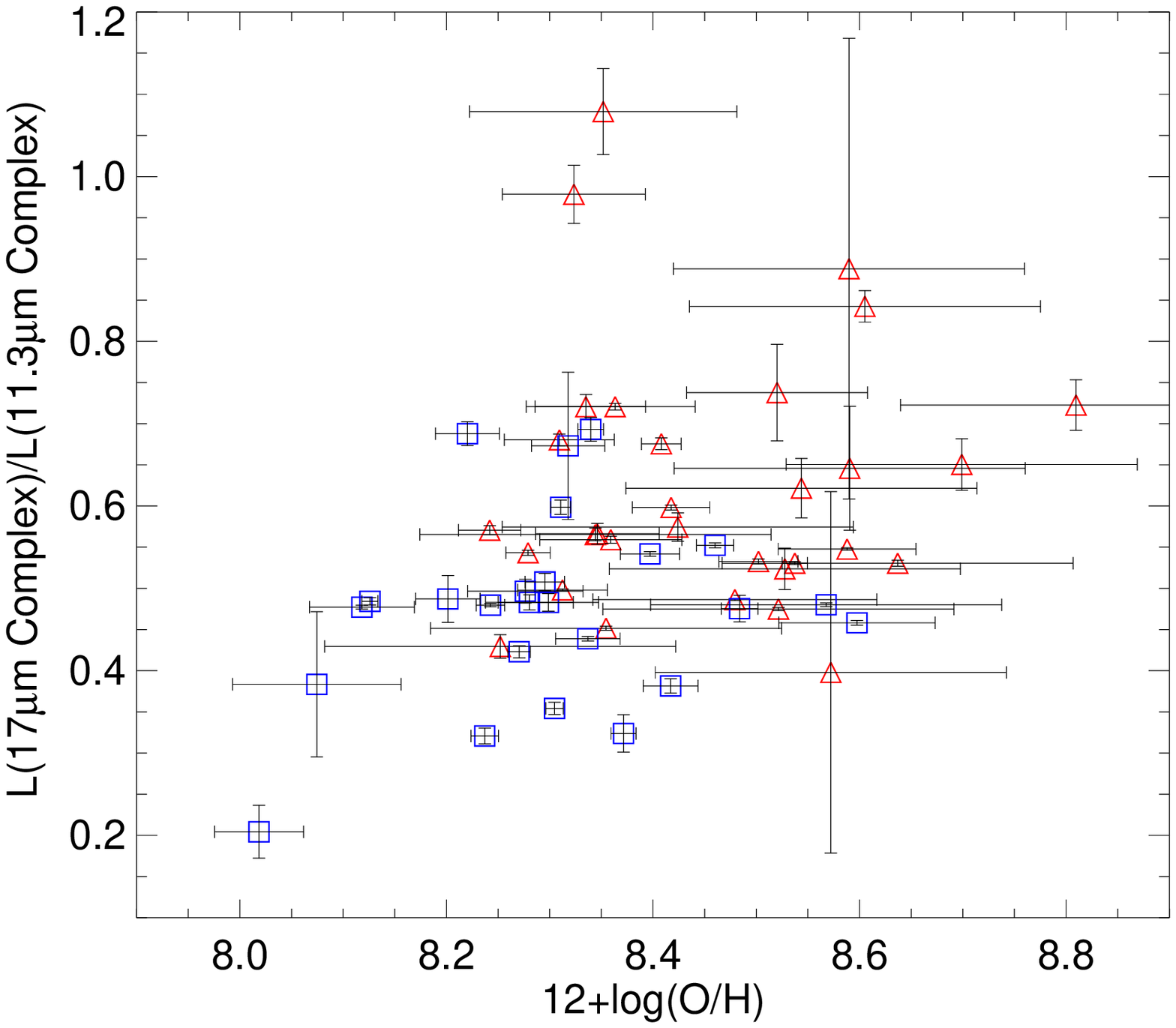}
\caption{The variation of the 17\um/11.3\um\ PAH luminosity ratio with
  oxygen abundance.  Red triangles are sources with LINER/Seyfert
  classifications, blue squares have \hII\ classifications.}
\label{fig:pah_metallicity}
\end{figure}

\begin{figure}
\plotone{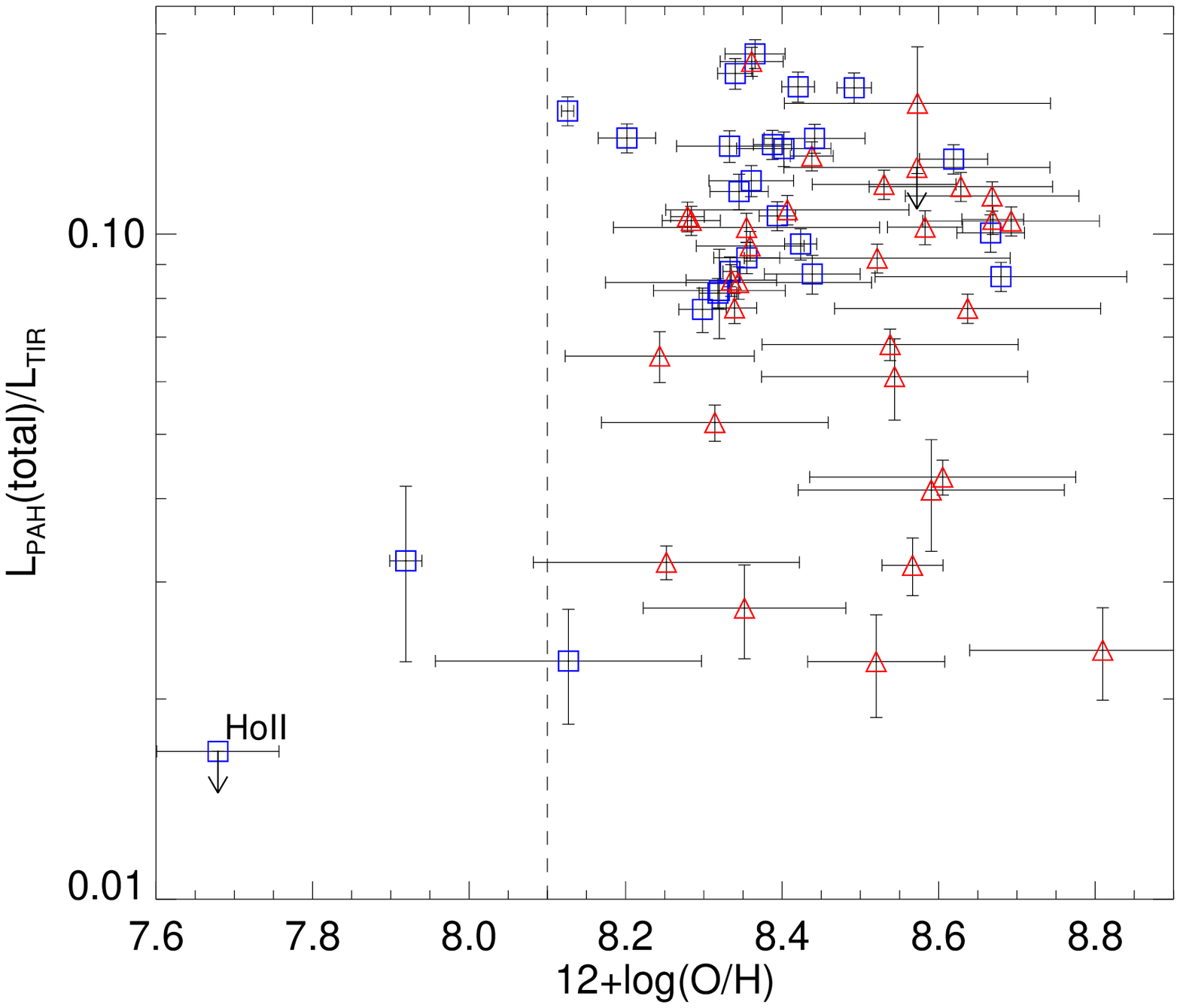}
\caption{The variation of the total PAH luminosity relative to the total
  infrared (3--1100\um) luminosity.  The threshold abundance
  $12\!+\!\log(\mathrm{O/H})=8.1$ is indicated.  Symbols are as in
  Fig.~\ref{fig:pah_metallicity}.}
\label{fig:pah_total_metallicity}
\end{figure}

Figs.~\ref{fig:pah_metallicity} and \ref{fig:pah_total_metallicity}
illustrate two examples of the behavior of PAH emission with changes in
nuclear oxygen abundance: the relative strengths of the 17\um\ and
11.3\um\ PAH bands, and the total integrated PAH luminosity relative to
the total infrared.  Oxygen abundances for the SINGS galaxies have been
determined by Moustakas et al. (2006, in prep.), based on a combination
of nuclear and circumnuclear optical spectroscopy and observations of
individual \hII\ regions culled from the literature, and placed on the
common $12\!+\!\log(\mathrm{O/H})$ strong-line calibration scale of
\citet{Pilyugin2005}.

The central spectra of SINGS galaxies probe much higher oxygen abundance
than the low-$Z$ sample of \citet{Engelbracht2005}, yet, even at these
relatively high metallicities, up to $\sim1.5Z_\sun$, there is a
definite trend in the inter-band strength ratios, with higher
metallicity leading to increased $L(17\um)/L(11\um)$.  Though low
metallicity systems do not tend to harbor AGN \citep{Kauffmann2003}, the
presence of a weak AGN in the nucleus clearly has a strong impact on the
behavior of $L(17\um)/L(11\um)$, with the highest ratios observed only
in systems with active nuclei.

Since the 15--20\um\ PAH bands are hosted by larger grains than the
shorter wavelength features, the trend with metallicity is suggestive of
a scenario in which formation of the large grains that contribute the
17\um\ band is enhanced in the presence of higher metal abundance.  This
effect could act in combination with the destruction of smaller PAH
grains in AGN environments, since it is likely that the mechanism that
suppresses the shortest wavelength PAH features in AGN has a similar, if
smaller effect on the 11.3\um\ feature.

Low metallicity systems also have harder radiation fields due to a
combination of two effects: 1) hotter massive stars form in low
metallicity environments \citep[an effect dominated by the metallicity
sensitivity of line-driven stellar winds,][]{Martin-Hernandez2002}, and
2) dust production is inhibited, leading to reduced shielding by dust
grains.  This could also impact the PAH band ratios of
Fig.~\ref{fig:pah_metallicity}.  However, since both the 11.3\um\ and
17\um\ band are contributed by neutral PAHs, the ionization state cannot
drive the trend in their ratios.  Likewise, selective destruction of the
smaller grains in the harder fields would result in the opposite trend
--- higher $L(17\um)/L(11.3\um)$ at lower metallicity.  This leaves the
inhibited formation of the largest grains at low metallicity as the most
likely dominant effect.  Clearly, however, metallicity variation alone
cannot explain the full range of $L(17\um)/L(11.3\um)$ in
Fig.~\ref{fig:pah_metallicity}, since at $Z\!\gtrsim\!Z_\sun/2$, AGN
consistently have the largest ratio.

The behavior of the total PAH luminosity, integrated over all bands from
5--19\um, relative to the total infrared luminosity in
Fig.~\ref{fig:pah_total_metallicity} shows a similar trend, with high
fractional PAH contribution to the total infrared ($\gtrsim$\,10\%)
occuring only at metallicities above $12\!+\!\log(\mathrm{O/H})\sim8.1$.
There is also a marked separation between galaxies with weak AGN and
\hII\ classifications.  At a given metallicity (even super-solar), AGN
sources emit a much smaller fraction of their central infrared
luminosity as PAH emission.  Coupled with the anomalous PAH band ratios
seen in the most extreme such sources (Fig.~\ref{fig:pah_ne_rat}), a
picture of weak AGN profoundly influencing the absolute strength and
shape of the PAH emission spectrum emerges.

\section{SED Models in the High-Redshift Context}
\label{sec:model-infr-seds}

Rather than directly modeling the physical processes and resulting
mid-infrared emission properties of star-forming galaxies, most
broadband infrared studies of moderate to high redshift galaxies have
utilized infrared \emph{templates}, constructed from small samples of
local galaxy spectra.  Since the PAH bands are so important
energetically, and impart the 3--19\um\ spectrum of star-forming
galaxies with features that contrast strongly with each other and with
the underlying dust and stellar continuum, they are a vital component of
any infrared SED template.  A small number of bright, very well studied
systems with high-quality ISO spectra \citep[e.g.  Arp\,220, M82,
NGC\,253, Mrk\,231;][]{Rigopoulou1999,Sturm1996,Sturm2000} have been
used extensively \citep[e.g.][]{Elbaz2002,Webb2006,Reddy2006}.  The
average ISOPhot spectrum of \citet{Lu2003} has also seen wide used,
having been integrated into the SED library of \citet{Dale2001} and
others.  Even physically realistic SED models \citep[e.g.][]{Silva1998}
are often used with the mid-infrared substituted by scaled template
spectra.

The importance of quantifying not just the average shape of the
mid-infrared spectrum of star-forming galaxies, but the range of
variation in that shape, is well illustrated by the source count models
of \citet{Lagache2004}.  To model deep MIPS 24\um\ source counts
\citep{Papovich2004}, \citeauthor{Lagache2004} applying a modest tilt to
the ISO PAH template spectrum to enhance the power at 15\um.  With this
modified SED, the source counts, which could not be reproduced by
pre-existing models, came into very good agreement.  This perturbation
played a large role in the interpretation of the redshift distribution
and star formation history implied by the broadband survey sample.  Such
sensitivity to the detailed form and shape of the mid-infrared emission
spectrum underlines the urgent need to quantify variations in this
spectrum, and establish the role of such natural variations in the
interpretation of resolved and unresolved survey results.

\begin{figure}[t]
  \plotone{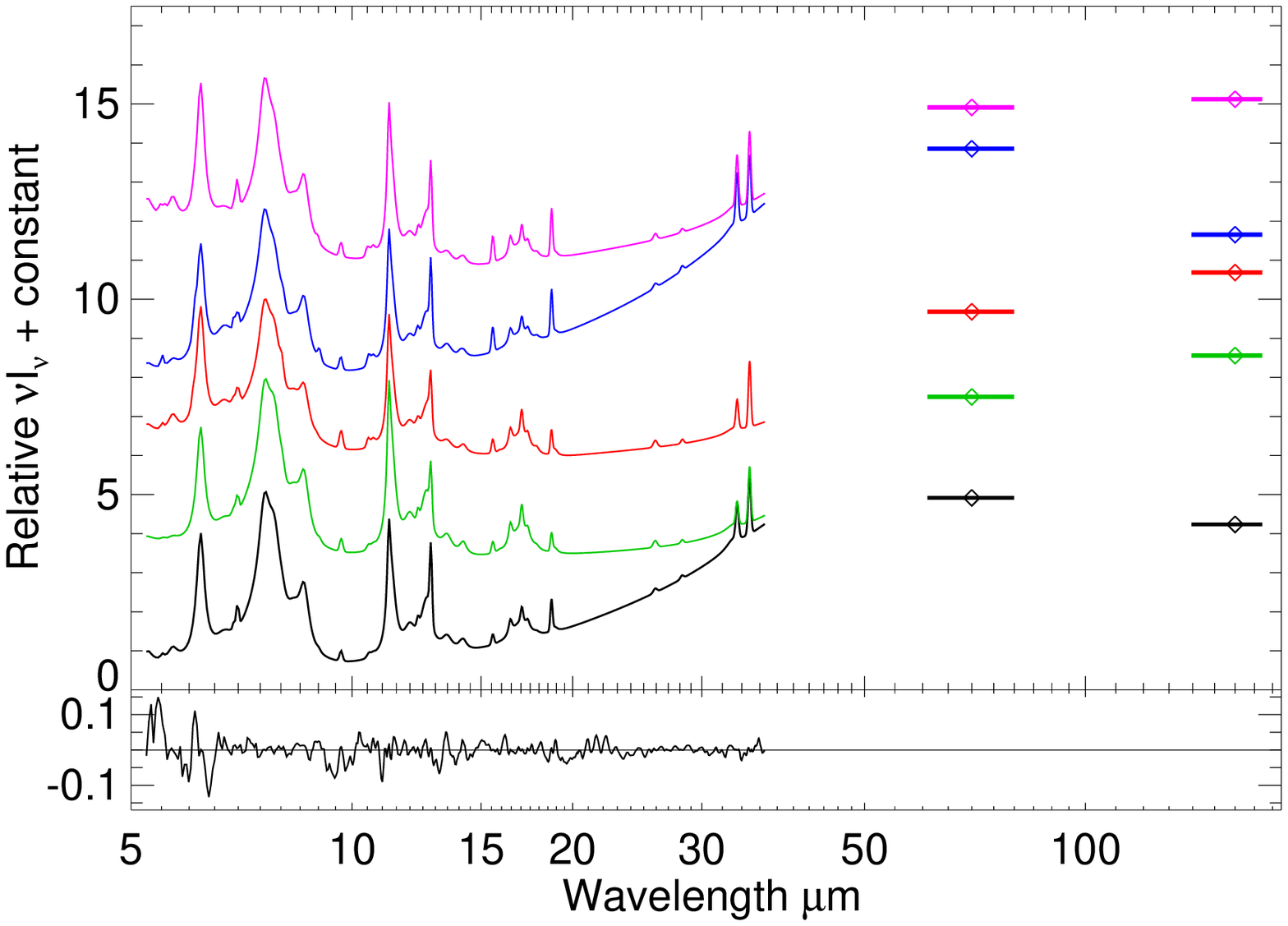}
  \caption{Noise-free MIR templates, produced by a TIR
    luminosity-weighted average of all spectra (black), and from
    sub-sample means for each region in Fig.~\ref{fig:main_band_rat}
    (color).  The units are arbitrary $\nu I_\nu$, with a constant
    offset of 2.5 units between spectra.  Mean matched 70\um\ and
    160\um\ intensities are also shown.  The lower panel contains the
    normalized residuals (model-spectrum)/model, averaged over the
    sample, with 20\% trimmed from the extremes of the distribution.}
\label{fig:templates}
\end{figure}

In Fig.~\ref{fig:templates}, five noiseless model templates from the
spectral decomposition technique of \S\,\ref{sec:decomp-mir-galaxy} are
shown --- an average spectrum weighted by the global total infrared
luminosity, and four sub-sample averages drawn from the
empirically-defined regions in Fig.~\ref{fig:main_band_rat}, formed by
lines in the $L(6.2\um)/L(7.7\um)$ vs. $L(11.3\um)/L(17\um)$ space.
These boundaries were chosen to ensure adequate sampling in each region,
and to bracket a modest range of variation within the sample.  Only
galaxies with reliable detection in all four main PAH bands are included
(eliminating all sources dominated by stellar continuum).  Emission
lines are not removed.  The region-averaged spectra corresponding to the
lower two quadrants in Fig.~\ref{fig:main_band_rat} have the largest
contribution from weak AGN sources.  The normalized residuals are also
shown, and are typically only a few percent, with some structure near
known artifacts (e.g. fringing at 20--23\um).  The luminosity-weighted
average has the steepest 30\um\ continuum, an indication that the most
luminous sources have the warmest thermal SEDs.  The 70\um\ and 160\um\
matched intensities (as described in \S\,\ref{sec:mips-data-reduction})
are shown for each template spectrum.

\begin{figure}
  \plotone{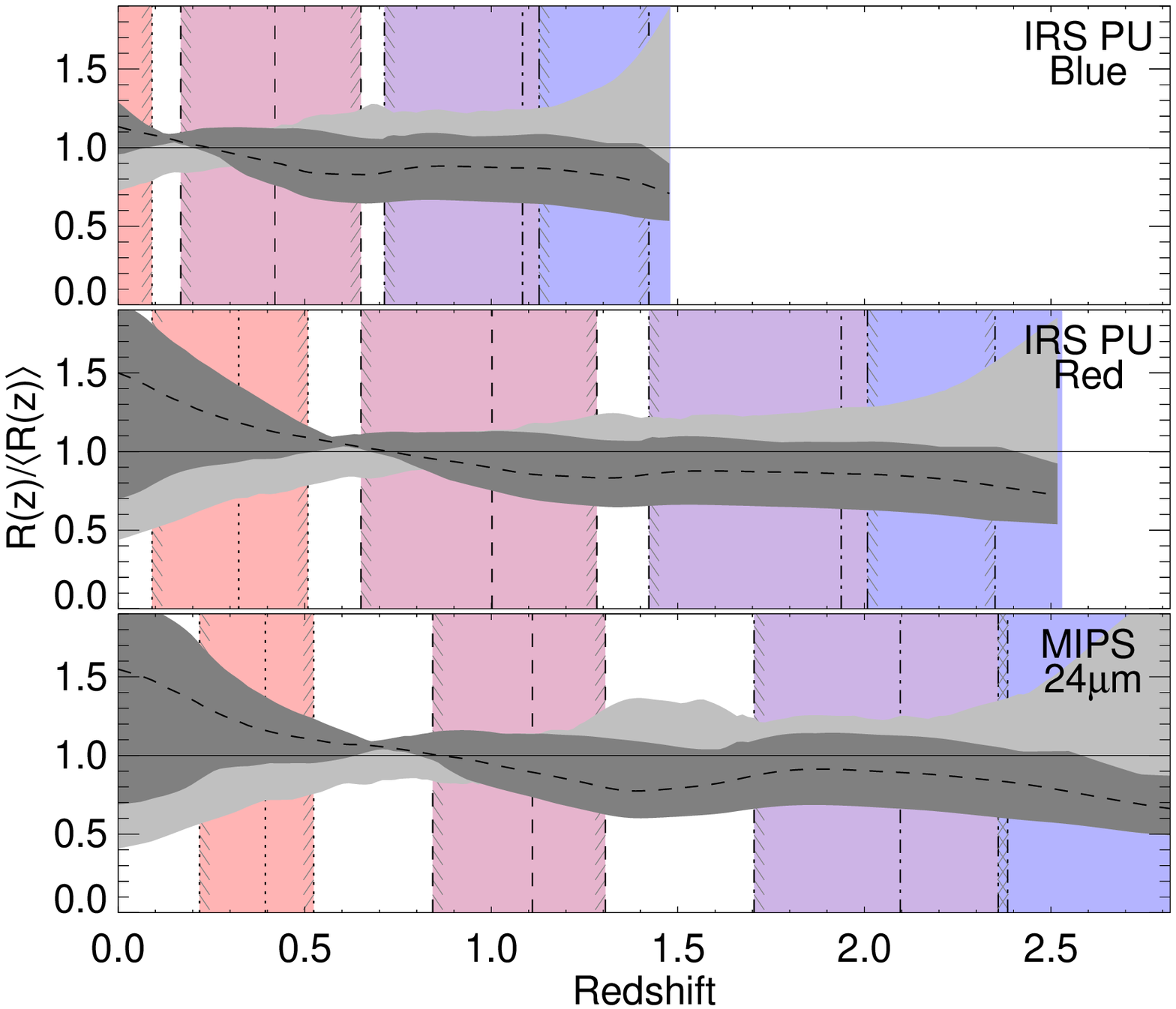}
  \caption{The redshift response of three broadband Spitzer filters (IRS
    PU Blue at 16\um, IRS PU Red at 22\um, and MIPS 24\um) to the range
    of sample spectra is shown, relative to the mean response at each
    redshift.  The global luminosity-weighted mean relative response is
    shown as the dashed line (the mean PAH template spectrum would
    produce the flat line $R(z)=1.0$).  The light gray region encloses
    the 10\%--90\% range of variation within the full sample, and the
    dark gray region encloses a similar 10\%--90\% range for the most
    luminous quartile of the sample ($L_{\TIR}>2.6\times10^{10}L_\sun$).
    The redshifts where the peak wavelengths of the main PAH features
    are shifted to the 50\% power and transmission-weighted central
    wavelengths of each broadband filter are indicated by shaded regions
    for the main PAH bands at (from left to right, red to blue): 17\um\
    (dotted, red), 11.3\um\ (long dash, red-blue), 7.7\um\ (dash-dot,
    blue-red), and 6.2\um\ (dash-dot-dot-dot, blue).}
\label{fig:redshift_response}
\end{figure}

The full sample was used to investigate the redshift response of the
Spitzer bands, defined as

\begin{equation}
  \label{eq:6}
  R(z)=\frac{\int g_\lambda I_\lambda(z)\: d\lambda}{\int_{5\mu m}^{20\mu m}
    I_\lambda(z)\: d\lambda}
\end{equation}

\noindent where $g_\lambda$ is the effective filter transmission
function for the three broadband Spitzer imaging filters of interest:
IRS Peak-Up Blue (16\um) and Red (22\um), and MIPS 24\um.  $R(z)$
quantifies the expected broadband response to a galaxy of a given
spectral appearance.  The normalization limits were chosen to ensure
that the PAH spectrum does not suffer artificial variations due to the
large range of slopes in the 20--30\um\ dust continuum.

In Fig.~\ref{fig:redshift_response}, $R(z)$ is plotted relative to its
sample mean at each redshift, indicating a wide range in broad
photometric response arising from normal variations in the PAH spectrum,
and, to a lesser degree, the underlying continuum.  In the highest
luminosity quartile, the range of variation is reduced by roughly 30\%,
and there is a waist of diminished variation (a factor of 1.4) at
$z\sim0.6$, in the relatively stable rest-frame continuum between the
11\um\ and 17\um\ bands. At the low and high redshifts ends, variations
in the continuum shape dominate the response.  The median range in the
10\%--90\% trimmed sample is roughly a factor of two (1.6 in the most
luminous quartile).  At redshifts where the region bracketing the
7.7\um\ and 11.3\um\ features shifts through the bands, the range of
variation increases significantly.

At luminosities above $L_{\TIR}$$\,\sim\,$$10^{11}L_\sun$, other MIR
emission components can begin to dominate over PAHs.  In particular, the
hot dust emission of embedded AGN can contribute the bulk of the
mid-infrared power and dilute away any PAH emission present, and very
strong silicate absorption can have a profound effect on the shape of
the MIR continuum.  A fundamental question in the modeling of deep
infrared survey results is what combination of features constitute a
useful suite of galaxy template spectra.  There is no guarantee that
distant galaxies forming in the early universe will exhibit the same
infrared properties as local galaxies, even if carefully sub-divided by
total infrared luminosity, metallicity, or other systemic physical
properties.  Because PAH emission is detected or inferred at ever higher
redshifts in systems with a wide variety of infrared properties,
applying the complete range of locally observed PAH behavior is the only
option which fully brackets the known uncertainties which result from
natural variations in the PAH emission spectrum.

\section{Summary and Conclusions}
\label{sec:conclusions}

We have presented a new suite of 5--38\um\ spectra from the inner
$\sim\!1/2$ arcmin$^2$ (roughly 1--10 kpc$^2$) of a broadly selected
sample of nearby, predominantly star-forming galaxies, many of which
host weak AGN.  The spectra are dominated in most cases by PAH emission
bands limited to the range 5--19\um, stellar photospheric emission of
varying strength, smooth thermal dust continuum, and occasional
absorption by silicate grains.  

The low-resolution spectra of star-forming galaxies can be successfully
reproduced using a simple physically-motivated decomposition model with
a small number of components, including starlight, thermal dust
continuum, and line and dust emission features, all absorbed by single
parameter dust extinction.  The decomposition method is also successful
at recovering unresolved emission lines blended with dust features.

Weak dust emission features at 6.69\um\ at 18.92\um\ are reported for
the first time in extragalactic sources, and many of the weak features
formerly seen only in ISO spectra of one or two of the brightest
starburst galaxies are found commonly among our sample.  The 17\um\
complex correlates strongly with other PAH bands, supporting the
identification of this new feature with PAH.

Compared to spline-based or other simple techniques for estimating the
strengths and continuum underlying PAH features, the decomposition
method recovers significantly more of the extended feature power.
Spline methods agree reasonably well within a simple scalar factor for
the two main bands in clean continuum regions at 6.2\um\ and 11.3\um,
but recover much less of the total feature power, and with significantly
more scatter, for the blended complexes at 7.7\um\ and 8.6\um.

We have formulated a new extinction profile for use in the 5--38\um\
region.  Measurable dust absorption at mid-infrared wavelengths by
silicate grains is found to occur in $\sim$1/8 of the sample targets.
Most normal star-forming galaxies appear to suffer less than $A_V\sim 3$
magnitudes of foreground screen silicate extinction, averaged over their
central few square kiloparsecs.

Matched Spitzer photometry at 24, 70, and 160\um, tracing the bolometric
dust emission, highlights the remarkable power emitted solely in the PAH
bands, which in aggregate can contribute up to 20\% of the total
3--1100\um\ infrared luminosity, with a typical value of 10\%.  The PAH
features vary by a factor of 10 relative to 24\um, with indications that
both stochastically and thermally heated grains contribute to the
continuum there.

The individual PAH bands show considerable variability --- factors of
2--5 in internal strength ratios in the sample --- with some features
varying more than others.  The 17\um\ band in particular exhibits
variation which tracks with metallicity, such that systems with higher
oxygen abundance are more likely to have stronger 17\um\ emission
relative to the shorter features.  Since the longer wavelength PAH
emission is expected to arise from larger PAH grains, this could
indicate that the size distribution of the PAHs depends on the metal
content of the ISM, even at relatively high metallicities reaching
$1.5Z_\sun$.

Contrary to the effects of luminous AGN, which completely destroy PAH
grains or mask their emission bands with hot dust, the presence of
low-luminosity AGN in the nucleus of a galaxy has a profound effect on
the form of its PAH spectrum, leading in some cases to peculiar emission
with very weak or absent 6.2\um, 7.7\um\ and 8.6\um\ bands, suppressed
in strength by up to a factor of 10.  Only sample galaxies with
low-luminosity LINER or Seyfert classifications from their optical
nuclear spectra produce these exceedingly unusual band ratios, raising
the possibility that the weak AGN itself can directly modify the PAH
grain size distribution, and even serve as the excitation source for PAH
emission.  If so, the abnormal PAH strengths could be used to detect
such weak AGN systems in dusty galaxies where optical diagnostics are
unavailable or unreliable.  Alternatively, low star formation intensity
in the centers of many AGN may play a role in modifying the form of the
PAH emission.

Both higher metallicity and the presence of an AGN shift power from the
short to long wavelength PAH bands, most likely by rebalancing the
distribution of grain sizes by favoring formation of larger grains, and
selectively destroying the smaller grains which emit the bulk of the
shorter wavelength PAH features. The integrated luminosity of the PAH
bands, relative to the total infrared, also shows an overall increase
with increasing oxygen abundance, but this trend does \emph{not} hold
for galaxies hosting AGN, which emit a much smaller fraction of their
total infrared luminosity in the PAH bands (a few percent), even at
relatively high metallicity.  Both the unusual PAH spectrum, and the
diminished PAH luminosity in these sources caution against the direct
use of the absolute PAH band strength as an indicator of star formation
rate in systems known to harbor AGN.

Taking into account a trimmed range of variations in the PAH band ratios
and the shape of the mid-infrared continuum seen in the sample, we find
that, compared to the common assumption of an invariant PAH emission
template used to model high-redshift sources, at least a factor of two
variation in broadband photometric response is imparted at 16\um, 22\um,
and 24\um, at all redshifts $z\!\lesssim$2.5.  At higher total infrared
luminosities, above $2.6\times 10^{10}L_\sun$, the effective range is
smaller by roughly 30\%, but still substantial.  While it is likely that
as galaxy luminosities approach $10^{12} L_\sun$, other effects, such as
the growing contribution of obscured AGN to the typical infrared SED,
will dominate the uncertainties in modeling deep mid-infrared source
populations, utilizing a single, fixed PAH template will impart factor
of two or larger uncertainties due solely to natural variations in the
form of the mid-infrared PAH emission spectrum.

\acknowledgments 

The authors thank H. Spoon for assistance with spline-based techniques,
K. Masters for early access to her updated local flow model,
C. Markwardt for his indispensable IDL L-M fitting package, J. Chiar,
E. Peeters, C. Papovich, D. Elbaz, and R.  Chary, for helpful
discussions, and an anonymous referee, for many useful suggestions.
This work made use of the NASA/IPAC Extragalactic Database (NED)
operated by JPL/Caltech, under contract with the NASA.  Support for this
work, part of the \textit{Spitzer Space Telescope} Legacy Science
Program, was provided by NASA through Contract \#1224769 issued by
JPL/Caltech under contract \#1407.  BTD has been partially supported by
NSF grant AST-0406883.

\bibliographystyle{apj}
\bibliography{general,myref,inprep}

\begin{thebibliography}{96}
\expandafter\ifx\csname natexlab\endcsname\relax\def\natexlab#1{#1}\fi

\bibitem[{{Allamandola} {et~al.}(1999){Allamandola}, {Hudgins}, \&
  {Sandford}}]{Allamandola1999}
{Allamandola}, L.~J., {Hudgins}, D.~M., \& {Sandford}, S.~A. 1999, \apjl, 511,
  L115

\bibitem[{{Allamandola} {et~al.}(1985){Allamandola}, {Tielens}, \&
  {Barker}}]{Allamandola1985}
{Allamandola}, L.~J., {Tielens}, A.~G.~G.~M., \& {Barker}, J.~R. 1985, \apjl,
  290, L25

\bibitem[{{Armus} {et~al.}(2004){Armus}, {Charmandaris}, {Spoon}, {Houck},
  {Soifer}, {Brandl}, {Appleton}, {Teplitz}, {Higdon}, {Weedman}, {Devost},
  {Morris}, {Uchida}, {van Cleve}, {Barry}, {Sloan}, {Grillmair}, {Burgdorf},
  {Fajardo-Acosta}, {Ingalls}, {Higdon}, {Hao}, {Bernard-Salas}, {Herter},
  {Troeltzsch}, {Unruh}, \& {Winghart}}]{Armus2004}
{Armus}, L., {et~al.} 2004, \apjs, 154, 178

\bibitem[{{Barvainis} {et~al.}(1999){Barvainis}, {Antonucci}, \&
  {Helou}}]{Barvainis1999}
{Barvainis}, R., {Antonucci}, R., \& {Helou}, G. 1999, \aj, 118, 645

\bibitem[{{Beintema} {et~al.}(1996){Beintema}, {van den Ancker}, {Molster},
  {Waters}, {Tielens}, {Waelkens}, {de Jong}, {de Graauw}, {Justtanont},
  {Yamamura}, {Heras}, {Lahuis}, \& {Salama}}]{Beintema1996}
{Beintema}, D.~A., {et~al.} 1996, \aap, 315, L369

\bibitem[{{Boulanger} {et~al.}(1998){Boulanger}, {Boisssel}, {Cesarsky}, \&
  {Ryter}}]{Boulanger1998}
{Boulanger}, F., {Boisssel}, P., {Cesarsky}, D., \& {Ryter}, C. 1998, \aap,
  339, 194

\bibitem[{{Brandl} {et~al.}(2006){Brandl}, {Bernard-Salas}, {Spoon}, {Devost},
  {Sloan}, {Guilles}, {Wu}, {Houck}, {Armus}, {Weedman}, {Charmandaris},
  {Appleton}, {Soifer}, {Hao}, {Marshall}, {Higdon}, \& {Herter}}]{Brandl2006}
{Brandl}, B.~R., {et~al.} 2006, ArXiv Astrophysics e-prints

\bibitem[{{Bregman} \& {Temi}(2005)}]{Bregman2005}
{Bregman}, J., \& {Temi}, P. 2005, \apj, 621, 831

\bibitem[{{Bregman} {et~al.}(2006){Bregman}, {Bregman}, \&
  {Temi}}]{Bregman2006}
{Bregman}, J.~D., {Bregman}, J.~N., \& {Temi}, P. 2006, ArXiv Astrophysics
  e-prints

\bibitem[{{Bressan} {et~al.}(2006){Bressan}, {Panuzzo}, {Buson}, {Clemens},
  {Granato}, {Rampazzo}, {Silva}, {Valdes}, {Vega}, \& {Danese}}]{Bressan2006}
{Bressan}, A., {et~al.} 2006, \apjl, 639, L55

\bibitem[{{Calzetti} {et~al.}(2005){Calzetti}, {Kennicutt}, {Bianchi},
  {Thilker}, {Dale}, {Engelbracht}, {Leitherer}, {Meyer}, {Sosey}, {Mutchler},
  {Regan}, {Thornley}, {Armus}, {Bendo}, {Boissier}, {Boselli}, {Draine},
  {Gordon}, {Helou}, {Hollenbach}, {Kewley}, {Madore}, {Martin}, {Murphy},
  {Rieke}, {Rieke}, {Roussel}, {Sheth}, {Smith}, {Walter}, {White}, {Yi},
  {Scoville}, {Polletta}, \& {Lindler}}]{Calzetti2005}
{Calzetti}, D., {et~al.} 2005, \apj, 633, 871

\bibitem[{{Cannon} {et~al.}(2006){Cannon}, {Smith}, {Walter}, {Bendo},
  {Calzetti}, {Dale}, {Draine}, {Engelbracht}, {Gordon}, {Helou}, {Kennicutt},
  {Jr.}, {Leitherer}, {Armus}, {Buckalew}, {Hollenbach}, {Jarrett}, {Li},
  {Meyer}, {Murphy}, {Regan}, {Rieke}, {Rieke}, {Roussel}, {Sheth}, \&
  {Thornley}}]{Cannon2006}
{Cannon}, J.~M., {et~al.} 2006, ArXiv Astrophysics e-prints

\bibitem[{{Caputi} {et~al.}(2006){Caputi}, {Dole}, {Lagache}, {McLure},
  {Puget}, {Rieke}, {Dunlop}, {Le Floc'h}, {Papovich}, \&
  {P{\'e}rez-Gonz{\'a}lez}}]{Caputi2006}
{Caputi}, K.~I., {et~al.} 2006, \apj, 637, 727

\bibitem[{{Cesarsky} {et~al.}(1996){Cesarsky}, {Lequeux}, {Abergel}, {Perault},
  {Palazzi}, {Madden}, \& {Tran}}]{Cesarsky1996}
{Cesarsky}, D., {Lequeux}, J., {Abergel}, A., {Perault}, M., {Palazzi}, E.,
  {Madden}, S., \& {Tran}, D. 1996, \aap, 315, L305

\bibitem[{{Cesarsky} {et~al.}(1998){Cesarsky}, {Lequeux}, {Pagani}, {Ryter},
  {Loinard}, \& {Sauvage}}]{Cesarsky1998}
{Cesarsky}, D., {Lequeux}, J., {Pagani}, L., {Ryter}, C., {Loinard}, L., \&
  {Sauvage}, M. 1998, \aap, 337, L35

\bibitem[{{Chiar} \& {Tielens}(2006)}]{Chiar2006}
{Chiar}, J.~E., \& {Tielens}, A.~G.~G.~M. 2006, \apj, 637, 774

\bibitem[{{Contursi} {et~al.}(2000){Contursi}, {Lequeux}, {Cesarsky},
  {Boulanger}, {Rubio}, {Hanus}, {Sauvage}, {Tran}, {Bosma}, {Madden}, \&
  {Vigroux}}]{Contursi2000}
{Contursi}, A., {et~al.} 2000, \aap, 362, 310

\bibitem[{{Dale} {et~al.}(2005){Dale}, {Bendo}, {Engelbracht}, {Gordon},
  {Regan}, {Armus}, {Cannon}, {Calzetti}, {Draine}, {Helou}, {Joseph},
  {Kennicutt}, {Li}, {Murphy}, {Roussel}, {Walter}, {Hanson}, {Hollenbach},
  {Jarrett}, {Kewley}, {Lamanna}, {Leitherer}, {Meyer}, {Rieke}, {Rieke},
  {Sheth}, {Smith}, \& {Thornley}}]{Dale2005}
{Dale}, D.~A., {et~al.} 2005, \apj, 633, 857

\bibitem[{{Dale} \& {Helou}(2002)}]{Dale2002}
{Dale}, D.~A., \& {Helou}, G. 2002, \apj, 576, 159

\bibitem[{{Dale} {et~al.}(2001){Dale}, {Helou}, {Contursi}, {Silbermann}, \&
  {Kolhatkar}}]{Dale2001}
{Dale}, D.~A., {Helou}, G., {Contursi}, A., {Silbermann}, N.~A., \&
  {Kolhatkar}, S. 2001, \apj, 549, 215

\bibitem[{{Dale} {et~al.}(2006){Dale}, {Smith}, {Armus}, {Buckalew}, {Helou},
  {Kennicutt}, {Moustakas}, {Roussel}, {Sheth}, {Bendo}, {Calzetti}, {Draine},
  {Engelbracht}, {Gordon}, {Hollenbach}, {Jarrett}, {Kewley}, {Leitherer},
  {Li}, {Malhotra}, {Murphy}, \& {Walter}}]{Dale2006}
{Dale}, D.~A., {et~al.} 2006, \apj, 646, 161

\bibitem[{{Draine} \& {Anderson}(1985)}]{Draine1985}
{Draine}, B.~T., \& {Anderson}, N. 1985, \apj, 292, 494

\bibitem[{{Draine} \& {Li}(2001)}]{Draine2001a}
{Draine}, B.~T., \& {Li}, A. 2001, \apj, 551, 807

\bibitem[{{Draine} \& {Li}(2006)}]{Draine2006}
---. 2006, ArXiv Astrophysics e-prints

\bibitem[{{Elbaz} {et~al.}(2002){Elbaz}, {Cesarsky}, {Chanial}, {Aussel},
  {Franceschini}, {Fadda}, \& {Chary}}]{Elbaz2002}
{Elbaz}, D., {Cesarsky}, C.~J., {Chanial}, P., {Aussel}, H., {Franceschini},
  A., {Fadda}, D., \& {Chary}, R.~R. 2002, \aap, 384, 848

\bibitem[{{Engelbracht} {et~al.}(2005){Engelbracht}, {Gordon}, {Rieke},
  {Werner}, {Dale}, \& {Latter}}]{Engelbracht2005}
{Engelbracht}, C.~W., {Gordon}, K.~D., {Rieke}, G.~H., {Werner}, M.~W., {Dale},
  D.~A., \& {Latter}, W.~B. 2005, \apjl, 628, L29

\bibitem[{{Fitzpatrick} \& {Massa}(1986)}]{Fitzpatrick1986}
{Fitzpatrick}, E.~L., \& {Massa}, D. 1986, \apj, 307, 286

\bibitem[{{Genzel} {et~al.}(1998){Genzel}, {Lutz}, {Sturm}, {Egami}, {Kunze},
  {Moorwood}, {Rigopoulou}, {Spoon}, {Sternberg}, {Tacconi-Garman}, {Tacconi},
  \& {Thatte}}]{Genzel1998}
{Genzel}, R., {et~al.} 1998, \apj, 498, 579+

\bibitem[{{Gillett} {et~al.}(1973){Gillett}, {Forrest}, \&
  {Merrill}}]{Gillett1973}
{Gillett}, F.~C., {Forrest}, W.~J., \& {Merrill}, K.~M. 1973, \apj, 183, 87

\bibitem[{{Guhathakurta} \& {Draine}(1989)}]{Guhathakurta1989}
{Guhathakurta}, P., \& {Draine}, B.~T. 1989, \apj, 345, 230

\bibitem[{{Hao} {et~al.}(2005){Hao}, {Spoon}, {Sloan}, {Marshall}, {Armus},
  {Tielens}, {Sargent}, {van Bemmel}, {Charmandaris}, {Weedman}, \&
  {Houck}}]{Hao2005}
{Hao}, L., {et~al.} 2005, \apjl, 625, L75

\bibitem[{{Helou} {et~al.}(2000){Helou}, {Lu}, {Werner}, {Malhotra}, \&
  {Silbermann}}]{Helou2000}
{Helou}, G., {Lu}, N.~Y., {Werner}, M.~W., {Malhotra}, S., \& {Silbermann}, N.
  2000, \apjl, 532, L21

\bibitem[{{Helou} {et~al.}(2004){Helou}, {Roussel}, {Appleton}, {Frayer},
  {Stolovy}, {Storrie-Lombardi}, {Hurt}, {Lowrance}, {Makovoz}, {Masci},
  {Surace}, {Gordon}, {Alonso-Herrero}, {Engelbracht}, {Misselt}, {Rieke},
  {Rieke}, {Willner}, {Pahre}, {Ashby}, {Fazio}, \& {Smith}}]{Helou2004}
{Helou}, G., {et~al.} 2004, \apjs, 154, 253

\bibitem[{{Ho} {et~al.}(1995){Ho}, {Filippenko}, \& {Sargent}}]{Ho1995}
{Ho}, L.~C., {Filippenko}, A.~V., \& {Sargent}, W.~L. 1995, \apjs, 98, 477

\bibitem[{{Ho} {et~al.}(1997){Ho}, {Filippenko}, \& {Sargent}}]{Ho1997a}
{Ho}, L.~C., {Filippenko}, A.~V., \& {Sargent}, W.~L.~W. 1997, \apj, 487, 579

\bibitem[{{Hogg} {et~al.}(2005){Hogg}, {Tremonti}, {Blanton}, {Finkbeiner},
  {Padmanabhan}, {Quintero}, {Schlegel}, \& {Wherry}}]{Hogg2005}
{Hogg}, D.~W., {Tremonti}, C.~A., {Blanton}, M.~R., {Finkbeiner}, D.~P.,
  {Padmanabhan}, N., {Quintero}, A.~D., {Schlegel}, D.~J., \& {Wherry}, N.
  2005, \apj, 624, 162

\bibitem[{{Hony} {et~al.}(2001){Hony}, {Van Kerckhoven}, {Peeters}, {Tielens},
  {Hudgins}, \& {Allamandola}}]{Hony2001}
{Hony}, S., {Van Kerckhoven}, C., {Peeters}, E., {Tielens}, A.~G.~G.~M.,
  {Hudgins}, D.~M., \& {Allamandola}, L.~J. 2001, \aap, 370, 1030

\bibitem[{{Houck} {et~al.}(2004){Houck}, {Roellig}, {van Cleve}, {Forrest},
  {Herter}, {Lawrence}, {Matthews}, {Reitsema}, {Soifer}, {Watson}, {Weedman},
  {Huisjen}, {Troeltzsch}, {Barry}, {Bernard-Salas}, {Blacken}, {Brandl},
  {Charmandaris}, {Devost}, {Gull}, {Hall}, {Henderson}, {Higdon}, {Pirger},
  {Schoenwald}, {Sloan}, {Uchida}, {Appleton}, {Armus}, {Burgdorf},
  {Fajardo-Acosta}, {Grillmair}, {Ingalls}, {Morris}, \&
  {Teplitz}}]{Houck2004a}
{Houck}, J.~R., {et~al.} 2004, \apjs, 154, 18

\bibitem[{{Houck} {et~al.}(2005){Houck}, {Soifer}, {Weedman}, {Higdon},
  {Higdon}, {Herter}, {Brown}, {Dey}, {Jannuzi}, {Le Floc'h}, {Rieke}, {Armus},
  {Charmandaris}, {Brandl}, \& {Teplitz}}]{Houck2005}
---. 2005, \apjl, 622, L105

\bibitem[{{Huang} {et~al.}(2006){Huang}, {Rigopoulou}, {Papovich}, {Ashby},
  {Willner}, {Ivison}, {Laird}, {Webb}, {Wilson}, {Barmby}, {Chapman},
  {Conselice}, {Mcleod}, {Shu}, {Smith}, {Le Floc'h}, {Egami}, {Willmer}, \&
  {Fazio}}]{Huang2006}
{Huang}, J.~., {et~al.} 2006, ArXiv Astrophysics e-prints

\bibitem[{{Joblin} {et~al.}(1996){Joblin}, {Tielens}, {Geballe}, \&
  {Wooden}}]{Joblin1996}
{Joblin}, C., {Tielens}, A.~G.~G.~M., {Geballe}, T.~R., \& {Wooden}, D.~H.
  1996, \apjl, 460, L119+

\bibitem[{{Kaneda} {et~al.}(2005){Kaneda}, {Onaka}, \& {Sakon}}]{Kaneda2005}
{Kaneda}, H., {Onaka}, T., \& {Sakon}, I. 2005, \apjl, 632, L83

\bibitem[{{Kauffmann} {et~al.}(2003){Kauffmann}, {Heckman}, {Tremonti},
  {Brinchmann}, {Charlot}, {White}, {Ridgway}, {Brinkmann}, {Fukugita}, {Hall},
  {Ivezi{\'c}}, {Richards}, \& {Schneider}}]{Kauffmann2003}
{Kauffmann}, G., {et~al.} 2003, \mnras, 346, 1055

\bibitem[{{Kemper} {et~al.}(2004){Kemper}, {Vriend}, \& {Tielens}}]{Kemper2004}
{Kemper}, F., {Vriend}, W.~J., \& {Tielens}, A.~G.~G.~M. 2004, \apj, 609, 826

\bibitem[{{Kennicutt} {et~al.}(2003){Kennicutt}, {Armus}, {Bendo}, {Calzetti},
  {Dale}, {Draine}, {Engelbracht}, {Gordon}, {Grauer}, {Helou}, {Hollenbach},
  {Jarrett}, {Kewley}, {Leitherer}, {Li}, {Malhotra}, {Regan}, {Rieke},
  {Rieke}, {Roussel}, {Smith}, {Thornley}, \& {Walter}}]{Kennicutt2003}
{Kennicutt}, R.~C., {et~al.} 2003, \pasp, 115, 928

\bibitem[{{Lagache} {et~al.}(2004){Lagache}, {Dole}, {Puget}, {P{\'
  e}rez-Gonz{\' a}lez}, {Le Floc'h}, {Rieke}, {Papovich}, {Egami},
  {Alonso-Herrero}, {Engelbracht}, {Gordon}, {Misselt}, \&
  {Morrison}}]{Lagache2004}
{Lagache}, G., {et~al.} 2004, \apjs, 154, 112

\bibitem[{{Laurent} {et~al.}(2000){Laurent}, {Mirabel}, {Charmandaris},
  {Gallais}, {Madden}, {Sauvage}, {Vigroux}, \& {Cesarsky}}]{Laurent2000}
{Laurent}, O., {Mirabel}, I.~F., {Charmandaris}, V., {Gallais}, P., {Madden},
  S.~C., {Sauvage}, M., {Vigroux}, L., \& {Cesarsky}, C. 2000, \aap, 359, 887

\bibitem[{{Le Floc'h} {et~al.}(2005){Le Floc'h}, {Papovich}, {Dole}, {Bell},
  {Lagache}, {Rieke}, {Egami}, {P{\'e}rez-Gonz{\'a}lez}, {Alonso-Herrero},
  {Rieke}, {Blaylock}, {Engelbracht}, {Gordon}, {Hines}, {Misselt}, {Morrison},
  \& {Mould}}]{LeFloch2005}
{Le Floc'h}, E., {et~al.} 2005, \apj, 632, 169

\bibitem[{{Leger} {et~al.}(1989){Leger}, {D'Hendecourt}, {Boissel}, \&
  {Desert}}]{Leger1989}
{Leger}, A., {D'Hendecourt}, L., {Boissel}, P., \& {Desert}, F.~X. 1989, \aap,
  213, 351

\bibitem[{{Leger} \& {Puget}(1984)}]{Leger1984}
{Leger}, A., \& {Puget}, J.~L. 1984, \aap, 137, L5

\bibitem[{{Leitherer} {et~al.}(1999){Leitherer}, {Schaerer}, {Goldader},
  {Delgado}, {Robert}, {Kune}, {de Mello}, {Devost}, \&
  {Heckman}}]{Leitherer1999}
{Leitherer}, C., {et~al.} 1999, \apjs, 123, 3

\bibitem[{{Li}(2004)}]{Li2004}
{Li}, A. 2004, in ASP Conf. Ser. 309: Astrophysics of Dust, ed. A.~N. {Witt},
  G.~C. {Clayton}, \& B.~T. {Draine}, 417--+

\bibitem[{{Li} \& {Draine}(2001)}]{Li2001}
{Li}, A., \& {Draine}, B.~T. 2001, \apj, 554, 778

\bibitem[{{Lu} {et~al.}(2003){Lu}, {Helou}, {Werner}, {Dinerstein}, {Dale},
  {Silbermann}, {Malhotra}, {Beichman}, \& {Jarrett}}]{Lu2003}
{Lu}, N., {et~al.} 2003, \apj, 588, 199

\bibitem[{{Lutz} {et~al.}(1998){Lutz}, {Spoon}, {Rigopoulou}, {Moorwood}, \&
  {Genzel}}]{Lutz1998a}
{Lutz}, D., {Spoon}, H.~W.~W., {Rigopoulou}, D., {Moorwood}, A.~F.~M., \&
  {Genzel}, R. 1998, \apjl, 505, L103

\bibitem[{{Lutz} {et~al.}(2005){Lutz}, {Valiante}, {Sturm}, {Genzel},
  {Tacconi}, {Lehnert}, {Sternberg}, \& {Baker}}]{Lutz2005}
{Lutz}, D., {Valiante}, E., {Sturm}, E., {Genzel}, R., {Tacconi}, L.~J.,
  {Lehnert}, M.~D., {Sternberg}, A., \& {Baker}, A.~J. 2005, \apjl, 625, L83

\bibitem[{{Madden} {et~al.}(2006){Madden}, {Galliano}, {Jones}, \&
  {Sauvage}}]{Madden2006}
{Madden}, S.~C., {Galliano}, F., {Jones}, A.~P., \& {Sauvage}, M. 2006, \aap,
  446, 877

\bibitem[{{Mart{\'{\i}}n-Hern{\'a}ndez}
  {et~al.}(2002){Mart{\'{\i}}n-Hern{\'a}ndez}, {Vermeij}, {Tielens}, {van der
  Hulst}, \& {Peeters}}]{Martin-Hernandez2002}
{Mart{\'{\i}}n-Hern{\'a}ndez}, N.~L., {Vermeij}, R., {Tielens}, A.~G.~G.~M.,
  {van der Hulst}, J.~M., \& {Peeters}, E. 2002, \aap, 389, 286

\bibitem[{{Meijerink} \& {Spaans}(2005)}]{Meijerink2005}
{Meijerink}, R., \& {Spaans}, M. 2005, \aap, 436, 397

\bibitem[{{Moutou} {et~al.}(2000){Moutou}, {Verstraete}, {L{\' e}ger},
  {Sellgren}, \& {Schmidt}}]{Moutou2000}
{Moutou}, C., {Verstraete}, L., {L{\' e}ger}, A., {Sellgren}, K., \& {Schmidt},
  W. 2000, \aap, 354, L17

\bibitem[{{O'Halloran} {et~al.}(2006){O'Halloran}, {Satyapal}, \&
  {Dudik}}]{OHalloran2006}
{O'Halloran}, B., {Satyapal}, S., \& {Dudik}, R.~P. 2006, \apj, 641, 795

\bibitem[{{Papovich} {et~al.}(2004){Papovich}, {Dole}, {Egami}, {Le Floc'h},
  {P{\'e}rez-Gonz{\'a}lez}, {Alonso-Herrero}, {Bai}, {Beichman}, {Blaylock},
  {Engelbracht}, {Gordon}, {Hines}, {Misselt}, {Morrison}, {Mould},
  {Muzerolle}, {Neugebauer}, {Richards}, {Rieke}, {Rieke}, {Rigby}, {Su}, \&
  {Young}}]{Papovich2004}
{Papovich}, C., {et~al.} 2004, \apjs, 154, 70

\bibitem[{{Papovich} {et~al.}(2006){Papovich}, {Moustakas}, {Dickinson}, {Le
  Floc'h}, {Rieke}, {Daddi}, {Alexander}, {Bauer}, {Brandt}, {Dahlen}, {Egami},
  {Eisenhardt}, {Elbaz}, {Ferguson}, {Giavalisco}, {Lucas}, {Mobasher},
  {P{\'e}rez-Gonz{\'a}lez}, {Stutz}, {Rieke}, \& {Yan}}]{Papovich2006}
---. 2006, \apj, 640, 92

\bibitem[{{Peeters} {et~al.}(2002){Peeters}, {Hony}, {Van Kerckhoven},
  {Tielens}, {Allamandola}, {Hudgins}, \& {Bauschlicher}}]{Peeters2002}
{Peeters}, E., {Hony}, S., {Van Kerckhoven}, C., {Tielens}, A.~G.~G.~M.,
  {Allamandola}, L.~J., {Hudgins}, D.~M., \& {Bauschlicher}, C.~W. 2002, \aap,
  390, 1089

\bibitem[{{Peeters} {et~al.}(2004){Peeters}, {Mattioda}, {Hudgins}, \&
  {Allamandola}}]{Peeters2004b}
{Peeters}, E., {Mattioda}, A.~L., {Hudgins}, D.~M., \& {Allamandola}, L.~J.
  2004, \apjl, 617, L65

\bibitem[{{Pilyugin} \& {Thuan}(2005)}]{Pilyugin2005}
{Pilyugin}, L.~S., \& {Thuan}, T.~X. 2005, \apj, 631, 231

\bibitem[{{Puget} {et~al.}(1985){Puget}, {Leger}, \& {Boulanger}}]{Puget1985}
{Puget}, J.~L., {Leger}, A., \& {Boulanger}, F. 1985, \aap, 142, L19

\bibitem[{{Reddy} {et~al.}(2006){Reddy}, {Steidel}, {Fadda}, {Yan}, {Pettini},
  {Shapley}, {Erb}, \& {Adelberger}}]{Reddy2006}
{Reddy}, N.~A., {Steidel}, C.~C., {Fadda}, D., {Yan}, L., {Pettini}, M.,
  {Shapley}, A.~E., {Erb}, D.~K., \& {Adelberger}, K.~L. 2006, \apj, 644, 792

\bibitem[{{Rieke} \& {Lebofsky}(1985)}]{Rieke1985}
{Rieke}, G.~H., \& {Lebofsky}, M.~J. 1985, \apj, 288, 618

\bibitem[{{Rieke} {et~al.}(2004){Rieke}, {Young}, {Engelbracht}, {Kelly},
  {Low}, {Haller}, {Beeman}, {Gordon}, {Stansberry}, {Misselt}, {Cadien},
  {Morrison}, {Rivlis}, {Latter}, {Noriega-Crespo}, {Padgett}, {Stapelfeldt},
  {Hines}, {Egami}, {Muzerolle}, {Alonso-Herrero}, {Blaylock}, {Dole}, {Hinz},
  {Le Floc'h}, {Papovich}, {P{\'e}rez-Gonz{\'a}lez}, {Smith}, {Su}, {Bennett},
  {Frayer}, {Henderson}, {Lu}, {Masci}, {Pesenson}, {Rebull}, {Rho}, {Keene},
  {Stolovy}, {Wachter}, {Wheaton}, {Werner}, \& {Richards}}]{Rieke2004}
{Rieke}, G.~H., {et~al.} 2004, \apjs, 154, 25

\bibitem[{{Rigopoulou} {et~al.}(1999){Rigopoulou}, {Spoon}, {Genzel}, {Lutz},
  {Moorwood}, \& {Tran}}]{Rigopoulou1999}
{Rigopoulou}, D., {Spoon}, H. W.~W., {Genzel}, R., {Lutz}, D., {Moorwood}, A.
  F.~M., \& {Tran}, Q.~D. 1999, \aj, 118, 2625

\bibitem[{{Rosenberg} {et~al.}(2006){Rosenberg}, {Ashby}, {Salzer}, \&
  {Huang}}]{Rosenberg2006}
{Rosenberg}, J.~L., {Ashby}, M.~L.~N., {Salzer}, J.~J., \& {Huang}, J.-S. 2006,
  \apj, 636, 742

\bibitem[{{Roussel} {et~al.}(2006){Roussel}, {Helou}, {Smith}, {Draine},
  {Hollenbach}, {Moustakas}, {Spoon}, {Kennicutt}, {Rieke}, {Walter}, {Armus},
  {Dale}, {Sheth}, {Bendo}, {Engelbracht}, {Gordon}, {Meyer}, {Regan}, \&
  {Murphy}}]{Roussel2006}
{Roussel}, H., {et~al.} 2006, \apj, 646, 841

\bibitem[{{Schutte} {et~al.}(1993){Schutte}, {Tielens}, \&
  {Allamandola}}]{Schutte1993}
{Schutte}, W.~A., {Tielens}, A.~G.~G.~M., \& {Allamandola}, L.~J. 1993, \apj,
  415, 397

\bibitem[{{Siebenmorgen} {et~al.}(2005){Siebenmorgen}, {Haas}, {Kr{\"u}gel}, \&
  {Schulz}}]{Siebenmorgen2005}
{Siebenmorgen}, R., {Haas}, M., {Kr{\"u}gel}, E., \& {Schulz}, B. 2005, \aap,
  436, L5

\bibitem[{{Silva} {et~al.}(1998){Silva}, {Granato}, {Bressan}, \&
  {Danese}}]{Silva1998}
{Silva}, L., {Granato}, G.~L., {Bressan}, A., \& {Danese}, L. 1998, \apj, 509,
  103

\bibitem[{{Smith} {et~al.}(2004){Smith}, {Dale}, {Armus}, {Draine},
  {Hollenbach}, {Roussel}, {Helou}, {Kennicutt}, {Li}, {Bendo}, {Calzetti},
  {Engelbracht}, {Gordon}, {Jarrett}, {Kewley}, {Leitherer}, {Malhotra},
  {Meyer}, {Murphy}, {Regan}, {Rieke}, {Rieke}, {Thornley}, {Walter}, \&
  {Wolfire}}]{2004ApJS..154..199S}
{Smith}, J.~D.~T., {et~al.} 2004, \apjs, 154, 199

\bibitem[{{Spoon} {et~al.}(2004){Spoon}, {Armus}, {Cami}, {Tielens}, {Chiar},
  {Peeters}, {Keane}, {Charmandaris}, {Appleton}, {Teplitz}, \&
  {Burgdorf}}]{Spoon2004}
{Spoon}, H.~W.~W., {et~al.} 2004, \apjs, 154, 184

\bibitem[{{Spoon} {et~al.}(2005){Spoon}, {Tielens}, {Armus}, {Sloan},
  {Sargent}, {Cami}, {Charmandaris}, {Houck}, \& {Soifer}}]{Spoon2005}
---. 2005, ArXiv Astrophysics e-prints

\bibitem[{{Sturm} {et~al.}(1996){Sturm}, {Lutz}, {Genzel}, {Sternberg},
  {Egami}, {Kunze}, {Rigopoulou}, {Bauer}, {Feuchtgruber}, {Moorwood}, \& {de
  Graauw}}]{Sturm1996}
{Sturm}, E., {et~al.} 1996, \aap, 315, L133

\bibitem[{{Sturm} {et~al.}(2000){Sturm}, {Lutz}, {Tran}, {Feuchtgruber},
  {Genzel}, {Kunze}, {Moorwood}, \& {Thornley}}]{Sturm2000}
{Sturm}, E., {Lutz}, D., {Tran}, D., {Feuchtgruber}, H., {Genzel}, R., {Kunze},
  D., {Moorwood}, A.~F.~M., \& {Thornley}, M.~D. 2000, \aap, 358, 481

\bibitem[{{Sturm} {et~al.}(2005){Sturm}, {Schweitzer}, {Lutz}, {Contursi},
  {Genzel}, {Lehnert}, {Tacconi}, {Veilleux}, {Rupke}, {Kim}, {Sternberg},
  {Maoz}, {Lord}, {Mazzarella}, \& {Sanders}}]{Sturm2005}
{Sturm}, E., {et~al.} 2005, \apjl, 629, L21

\bibitem[{{Thornley} {et~al.}(2000){Thornley}, {Schreiber}, {Lutz}, {Genzel},
  {Spoon}, {Kunze}, \& {Sternberg}}]{Thornley2000}
{Thornley}, M.~D., {Schreiber}, N. M. F.~., {Lutz}, D., {Genzel}, R., {Spoon},
  H. W.~W., {Kunze}, D., \& {Sternberg}, A. 2000, \apj, 539, 641

\bibitem[{{Uchida} {et~al.}(2000){Uchida}, {Sellgren}, {Werner}, \&
  {Houdashelt}}]{Uchida2000}
{Uchida}, K.~I., {Sellgren}, K., {Werner}, M.~W., \& {Houdashelt}, M.~L. 2000,
  \apj, 530, 817

\bibitem[{{Van Kerckhoven} {et~al.}(2000){Van Kerckhoven}, {Hony}, {Peeters},
  {Tielens}, {Allamandola}, {Hudgins}, {Cox}, {Roelfsema}, {Voors}, {Waelkens},
  {Waters}, \& {Wesselius}}]{VanKerckhoven2000}
{Van Kerckhoven}, C., {et~al.} 2000, \aap, 357, 1013

\bibitem[{{Vermeij} {et~al.}(2002){Vermeij}, {Peeters}, {Tielens}, \& {van der
  Hulst}}]{Vermeij2002}
{Vermeij}, R., {Peeters}, E., {Tielens}, A.~G.~G.~M., \& {van der Hulst}, J.~M.
  2002, \aap, 382, 1042

\bibitem[{{Verstraete} {et~al.}(2001){Verstraete}, {Pech}, {Moutou},
  {Sellgren}, {Wright}, {Giard}, {L{\'e}ger}, {Timmermann}, \&
  {Drapatz}}]{Verstraete2001}
{Verstraete}, L., {et~al.} 2001, \aap, 372, 981

\bibitem[{{Verstraete} {et~al.}(1996){Verstraete}, {Puget}, {Falgarone},
  {Drapatz}, {Wright}, \& {Timmermann}}]{Verstraete1996}
{Verstraete}, L., {Puget}, J.~L., {Falgarone}, E., {Drapatz}, S., {Wright},
  C.~M., \& {Timmermann}, R. 1996, \aap, 315, L337

\bibitem[{{Voit}(1992)}]{Voit1992}
{Voit}, G.~M. 1992, \mnras, 258, 841

\bibitem[{{Waters} {et~al.}(1998){Waters}, {Beintema}, {Zijlstra}, {de Koter},
  {Molster}, {Bouwman}, {de Jong}, {Pottasch}, \& {de Graauw}}]{Waters1998}
{Waters}, L.~B.~F.~M., {et~al.} 1998, \aap, 331, L61

\bibitem[{{Webb} {et~al.}(2006){Webb}, {van Dokkum}, {Egami}, {Fazio}, {Franx},
  {Gawiser}, {Herrera}, {Huang}, {Labb{\'e}}, {Lira}, {Marchesini}, {Maza},
  {Quadri}, {Rudnick}, \& {van der Werf}}]{Webb2006}
{Webb}, T.~M.~A., {et~al.} 2006, \apjl, 636, L17

\bibitem[{{Weingartner} \& {Draine}(2001)}]{Weingartner2001}
{Weingartner}, J.~C., \& {Draine}, B.~T. 2001, \apjs, 134, 263

\bibitem[{{Werner} {et~al.}(2004){Werner}, {Uchida}, {Sellgren}, {Marengo},
  {Gordon}, {Morris}, {Houck}, \& {Stansberry}}]{Werner2004}
{Werner}, M.~W., {Uchida}, K.~I., {Sellgren}, K., {Marengo}, M., {Gordon},
  K.~D., {Morris}, P.~W., {Houck}, J.~R., \& {Stansberry}, J.~A. 2004, \apjs,
  154, 309

\bibitem[{{Wu} {et~al.}(2006){Wu}, {Charmandaris}, {Hao}, {Brandl},
  {Bernard-Salas}, {Spoon}, \& {Houck}}]{Wu2006}
{Wu}, Y., {Charmandaris}, V., {Hao}, L., {Brandl}, B.~R., {Bernard-Salas}, J.,
  {Spoon}, H.~W.~W., \& {Houck}, J.~R. 2006, \apj, 639, 157

\bibitem[{{Yan} {et~al.}(2005){Yan}, {Chary}, {Armus}, {Teplitz}, {Helou},
  {Frayer}, {Fadda}, {Surace}, \& {Choi}}]{Yan2005}
{Yan}, L., {et~al.} 2005, \apj, 628, 604

\bibitem[{{Yan} {et~al.}(2004){Yan}, {Helou}, {Fadda}, {Marleau}, {Lacy},
  {Wilson}, {Soifer}, {Drozdovsky}, {Masci}, {Armus}, {Teplitz}, {Frayer},
  {Surace}, {Storrie-Lombardi}, {Appleton}, {Chapman}, {Choi}, {Fan},
  {Heinrichsen}, {Im}, {Schmitz}, {Shupe}, \& {Squires}}]{Yan2004}
---. 2004, \apjs, 154, 60

\end{thebibliography}

\typeout{get arXiv to do 4 passes: Label(s) may have changed. Rerun}
\end{document}